%% file: dadamo_etal_2021_pitching_flexible_foil_rev1.tex
\documentclass[reprint,onecolumn,amsmath,amssymb,groupedaddress,notitlepage,longbibliography]{revtex4-1}
\usepackage{psfrag}
\usepackage{url}
\usepackage{tikz} 
\usepackage{pgfplotstable, environ,subcaption,caption,framed,xspace,hyperref,grffile,bm}
\usetikzlibrary{arrows,positioning}
\usetikzlibrary{plotmarks}
\usetikzlibrary{3d,calc,decorations,patterns}
\usepackage{bold-extra}
\usetikzlibrary{external}
\usepackage{amsmath,amssymb,mathrsfs,bm,cancel}
\usepackage{MnSymbol}
\usepackage{pgfplots}
\pgfplotsset{compat=1.11}
\usepgfplotslibrary{groupplots}
\usetikzlibrary{shapes}
\usepackage[utf8x]{inputenc}
\usetikzlibrary{decorations.pathreplacing}
\usepackage{pgfplotstable}
\usetikzlibrary{positioning,arrows}
\usetikzlibrary{patterns}
\usetikzlibrary{babel}
\usetikzlibrary{arrows.meta}

\usetikzlibrary{shadows.blur}
\usetikzlibrary{shapes.symbols}
\makeatletter
\newsavebox{\measure@tikzpicture}
\NewEnviron{scaletikzpicturetowidth}[1]{%
	\def\tikz@width{#1}%
	\begin{lrbox}{\measure@tikzpicture}%
		\BODY
	\end{lrbox}%
	\pgfmathparse{#1/\wd\measure@tikzpicture}%
	\BODY
}
\makeatother
\definecolor{color1}{rgb}{1,0.498039215686275,0.0549019607843137}
\definecolor{color0}{rgb}{0.12156862745098,0.466666666666667,0.705882352941177}
\definecolor{color3}{rgb}{0.83921568627451,0.152941176470588,0.156862745098039}
\definecolor{color2}{rgb}{0.172549019607843,0.627450980392157,0.172549019607843}
\definecolor{color5}{rgb}{0.549019607843137,0.337254901960784,0.294117647058824}
\definecolor{color4}{rgb}{0.580392156862745,0.403921568627451,0.741176470588235}

\graphicspath{{figures/}{tikzs/}}

\def\centerarc[#1](#2)(#3:#4:#5)
{ \draw[#1] ($(#2)+({#5*cos(#3)},{#5*sin(#3)})$) arc (#3:#4:#5); }

\def\centerarct[#1](#2)(#3:#4:#5)(#6)
{ \draw[#1] ($(#2)+({#5*cos(#3)},{#5*sin(#3)})$) arc (#3:#4:#5) node[midway,fill=white,inner sep=1pt]{#6}; }

\newcounter{tmp}
\tikzset{line0/.style={
		preaction={draw=none},
		preaction={decoration={contour lineto closed, contour distance=3pt},
			decorate,
		},
		postaction={
			insert path={%
				\pgfextra{%
					\pgfinterruptpath
					\path[pattern=north west lines, pattern color=color0,even odd rule] 
					\mySecondList \myList 
					;
					\endpgfinterruptpath}
		}},
}}
\tikzset{line1/.style={
		preaction={draw=none},
		preaction={decoration={contour lineto closed, contour distance=4pt},
			decorate,
		},
		postaction={
			insert path={%
				\pgfextra{%
					\pgfinterruptpath
					\path[pattern=north east lines, pattern color=color1,even odd rule] 
					\mySecondList \myList 
					;
					\endpgfinterruptpath}
		}},
}}

\makeatletter
\def\pgfdecoratedcontourdistance{0pt}
\pgfset{
	decoration/contour distance/.code=%
	\pgfmathsetlengthmacro\pgfdecoratedcontourdistance{#1}}
\pgfdeclaredecoration{contour lineto closed}{start}{%
	\state{start}[
	next state=draw,
	width=0pt,
	persistent precomputation=\let\pgf@decorate@firstsegmentangle\pgfdecoratedangle]{%
		\pgfextra{\xdef\myList{}\xdef\mySecondList{}}
		\pgfextra{\setcounter{tmp}{0}}
		\pgfpathmoveto{\pgfpointlineattime{.5}
			{\pgfqpoint{0pt}{\pgfdecoratedcontourdistance}}
			{\pgfqpoint{\pgfdecoratedinputsegmentlength}{\pgfdecoratedcontourdistance}}}%
	}%
	\state{draw}[next state=draw, width=\pgfdecoratedinputsegmentlength]{%
		\ifpgf@decorate@is@closepath@%
		\pgfmathsetmacro\pgfdecoratedangletonextinputsegment{%
			-\pgfdecoratedangle+\pgf@decorate@firstsegmentangle}%
		\fi
		\pgfmathsetlengthmacro\pgf@decoration@contour@shorten{%
			-\pgfdecoratedcontourdistance*cot(-\pgfdecoratedangletonextinputsegment/2+90)}%
		\pgfpathlineto
		{\pgfpoint{\pgfdecoratedinputsegmentlength+\pgf@decoration@contour@shorten}
			{\pgfdecoratedcontourdistance}}%
		\stepcounter{tmp}
		\pgfcoordinate{muemmel\thetmp}{\pgfpoint{\pgfdecoratedinputsegmentlength+\pgf@decoration@contour@shorten}
			{\pgfdecoratedcontourdistance}}
		\pgfcoordinate{feep\thetmp}{\pgfpoint{\pgfdecoratedinputsegmentlength}{0pt}}      
		\pgfextra{\xdef\myList{\myList (muemmel\thetmp) -- }%
			\xdef\mySecondList{\mySecondList (feep\thetmp) -- }}
		\ifpgf@decorate@is@closepath@%
		\pgfpathclose
		\pgfextra{\xdef\myList{\myList cycle}%
			\xdef\mySecondList{\mySecondList cycle}}
		\fi
	}%
	\state{final}{\pgfextra{
	}}%
}
\makeatother
\tikzset{
	contour/.style={
		decoration={
			name=contour lineto closed,
			contour distance=#1
		},
		decorate}}

\begin{document}
	
	\title{Wake and aeroelasticity of a flexible pitching foil}
	\author{Juan D'Adamo\textsuperscript{1}}
	\author{Manuel Collaud\textsuperscript{1}}
	\author{Roberto Sosa\textsuperscript{1}}
	\author{Ramiro Godoy-Diana\textsuperscript{2}}
	\affiliation{\textsuperscript{1}Laboratorio de Fluidodin\'amica, Facultad de Ingenier\'{\i}a, Universidad de Buenos Aires, CONICET, Av. Paseo Col\'on 850, C1063ACV, Buenos Aires, Argentina}
	\affiliation{\textsuperscript{2} Laboratoire de Physique et M\'ecanique des Milieux H\'et\'erog\`enes (PMMH), CNRS UMR 7636, ESPCI Paris---Universit\'e PSL, Sorbonne Universit\'e, Universit\'e de Paris, F-75005 Paris, France}
	
	\begin{abstract}
		A flexible foil undergoing pitching oscillations is studied experimentally in a wind tunnel with different imposed free stream velocities. The chord-based Reynolds number is in the range 1600--4000, such that the dynamics of the system is governed by inertial forces and the wake behind the foil exhibits the reverse B\'enard-von K\'arm\'an vortex street characteristic of flapping-based propulsion. Particle Image Velocimetry (PIV) measurements are performed to examine the flow around the foil, whilst the deformation of the foil is also tracked. The first natural frequency of vibration of the foil is within the range of flapping frequencies explored, determining a strongly-coupled dynamics between the elastic foil deformation and the vortex shedding.
		Cluster-based reduced order modelling is applied on the PIV data in order to identify the coherent flow structures. Analysing the foil kinematics and using a control-volume calculation of the average drag forces from the corresponding velocity fields, we determine the optimal flapping configurations for thrust generation. We show that propulsive force peaks occur at dimensionless frequencies shifted with respect to the elastic resonances that are marked by maximum trailing edge oscillation amplitudes. The thrust peaks are better explained by a wake resonance, which we examine using the tools of classic hydrodynamic stability on the mean propulsive jet profiles.
	\end{abstract}
	\maketitle

	\section{Introduction}
	
	An oscillating elastic plate constitutes probably the simplest model of a bio-inspired propulsive mechanism: the back and forth motion of a structure subject to deformation as it interacts with the surrounding fluid is the basic fluid-dynamical problem behind animal swimming and flying \citep[see e.g.][and references therein]{Daniel:2002,Shyy:2010}, regardless of the broad collection of different geometrical and material properties, as well as the specificities of different kinematics with different degrees of complexity involved in insect \citep{Wootton:1992,Bomphrey:2018}, bird \citep{Biewener:1995,Tobalske:2007} and bat \citep{Song:2008} wings \citep{Chin:2016}, and fish fins and tails \citep{McHenry:1995,Blake:2004}. 
	
	From an engineering perspective, flapping-based propulsion has come back to the center stage in the past two decades especially with the development of micro-air-vehicles (MAVs)---fuelled by miniaturisation---see e.g. \citep{Whitney:2012,TheDelfly2016,Chen:2019} for a review. With respect to conventional aerodynamics, the main difference is rooted in the unsteady mechanisms of force production by a flapping wing. Average thrust and lift are the outcome of the periodic flapping motion of a structure that accelerates and decelerates, as opposed to the case of fixed-wing aircraft, where thrust is produced by a jet engine or propeller and lift is the result of the flow around the static wing. Moreover, these unsteady mechanisms are intimately linked to the problem of vortex shedding, especially the forced vortex shedding that occurs on the time scale clocked by the flapping characteristic frequency. The distinctive feature of the wake of a flapping foil is a sequence of counter-rotating vortices that bears resemblance to the vortex street behind a bluff body but where, in the regimes of interest for propulsion, the sign of vorticity is inverted: the reverse Bénard-von K\'arm\'an (BvK) vortex street \citep{Freymuth:1988,Anderson:1998,GodoyDiana:2008}. The average flow field behind a flapping foil producing thrust is a jet flow. In the case of a body in self-propulsion, as the body accelerates towards a cruising regime, the jet profile behind the flapping body tends to a momentum-less wake profile (typically an inner jet surrounded by a drag wake) where the average thrust is balanced by the global drag \citep{Afanasyev:2004,Arbie:2016}. In the case of flapping flight, the aerodynamic force produced by the flapping wings is directed obliquely downwards, and it is partly used for thrust and partly as lift to counter the weight of the flyer, the limit case being that of hovering, where all the thrust is directed downwards and there is no cruising speed.
	
	In addition to the imposed flapping motion, the other main ingredient governing the dynamics of wings, fins and bio-inspired appendages is their structural deformation response \citep[see e.g.][]{Marais:2012,HueraHuarte:2017}. The latter is determined by the material and geometric properties of the body and it is the basic solid mechanics problem of forced vibrations. Without the surrounding fluid, i.e. for a body flapping in vacuum, the whole problem is defined by the interplay between the solid inertial forces that drive the deformation and the elastic restoring force. Adding the fluid to this forced vibration problem means, from the point of view of the structure, new fluid forces (viscous and inertial). But from the point of view of the fluid there is a whole new set of equations of motion to be considered, where the elastic solid determines the boundary conditions. Physically, the flapping foil is thus a fluid-structure interaction problem governed by a few dimensionless parameters that measure the respective importance of inertial, viscous and elastic forces in both the solid and the fluid. We can be more specific considering a flat plate of chord length $L$, span $H$, mass per unit surface $\mu$, bending stiffness $B=EI$ as a model, immersed in a fluid of density $\rho$ and kinematic viscosity $\nu$, and actuated in pitching oscillations of amplitude $\theta_R$ and frequency $f_f$. The main dimensionless parameters are then: the {mass ra}tio $\mathcal{M}=\rho / \rho_s$, which compares the density of the fluid to that of the solid; the Reynolds number $Re=UL/\nu$, which compares fluid inertial vs. viscous forces---where $U$ is a characteristic velocity that can be the average flying or swimming speed in a cruising regime, or the flapping characteristic velocity defined in terms of $f_f$ and $\theta_R$); and the dimensionless frequency $f^+ = f_f /f_n$, that compares the forcing frequency $f_f$ to the natural frequency of the first mode of vibration of the structure $f_n$ and thus contains the information about the bending stiffness.
	
	The parameter space defined by $\mathcal{M}$, $Re$ and $f^+$ encompasses a wide variety of problems, ranging from relatively stiff flapping wings of some insects \citep{Combes:2003} to the very flexible parapodia of swimming snails \citep{Mohaghar:2019}. In this paper we focus on an aerodynamic problem with $\mathcal{M}\sim 10^{-3}$ and $Re\sim 10^{3}$, and we examine the case of a rectangular foil of span-to-chord aspect ratio of $\sim4.6$ bending chord-wise mostly on its first mode of deformation as in \citep{Thiria:2010,Ramananarivo:2011}. We use two-dimensional (2D) Particle Image Velocimetry (PIV) on the symmetry plane at mid-span to examine the flow in the wake and cluster-based reduced order modelling to identify the coherent flow structures. Simultaneously, the elastic deformation of the foil is tracked. The first natural frequency of vibration of the foil is within the range of frequencies of the forcing pitching oscillations, which determines a strong coupling between the structural dynamics of the elastic foil and the vortex shedding in the wake.
	We use a control-volume calculation based on the PIV data in order to compute the average drag force for each flapping configuration, which allows us to determine optimal points in the parameter space in terms of thrust generation. We show that propulsive force peaks occur at dimensionless frequencies shifted with respect to the elastic resonances, which are marked by maximum trailing edge oscillation amplitudes. The thrust peaks are better explained by a wake resonance \citep[][]{Triantafyllou:1993,moored2012hydrodynamic,moored2014linear}, which we examine using the tools of classic hydrodynamic stability on the mean propulsive jet profiles \citep[]{trianta1986}. 
	\section{Configuration and experimental Setup}
	\begin{figure}
	\begin{subfigure}[t]{.4\textwidth}
		\resizebox{\columnwidth}{!}{%
			\tikzsetnextfilename{tikzs/flow_config_vista_3D}
			\input{tikzs/flow_config_vista_3D.tikz}
		}
		\subcaption{}\label{fig:setup_config}
	\end{subfigure}	\hfill
	\begin{subfigure}[t]{.46\textwidth}
		\resizebox{\columnwidth}{!}{%
			\tikzsetnextfilename{tikzs/flow_config_f}
			\input{tikzs/flow_config_f.tikz}
		}
		\subcaption{}\label{fig:displacement_config}
	\end{subfigure}
	
	\begin{subfigure}[t]{.5\textwidth}
		\resizebox{\columnwidth}{!}{%
			\tikzsetnextfilename{tikz_matplotlib/cluster_snapshot}	
			\input{tikz_matplotlib/cluster_snapshot.tikz}
		}
		\caption{}\label{fig:instantaneous_piv}
	\end{subfigure}\hfill
	\begin{subfigure}[t]{.5\textwidth}
		\resizebox{\columnwidth}{!}{%
			\tikzsetnextfilename{tikz_matplotlib/mean_contour_scheme}	
			\input{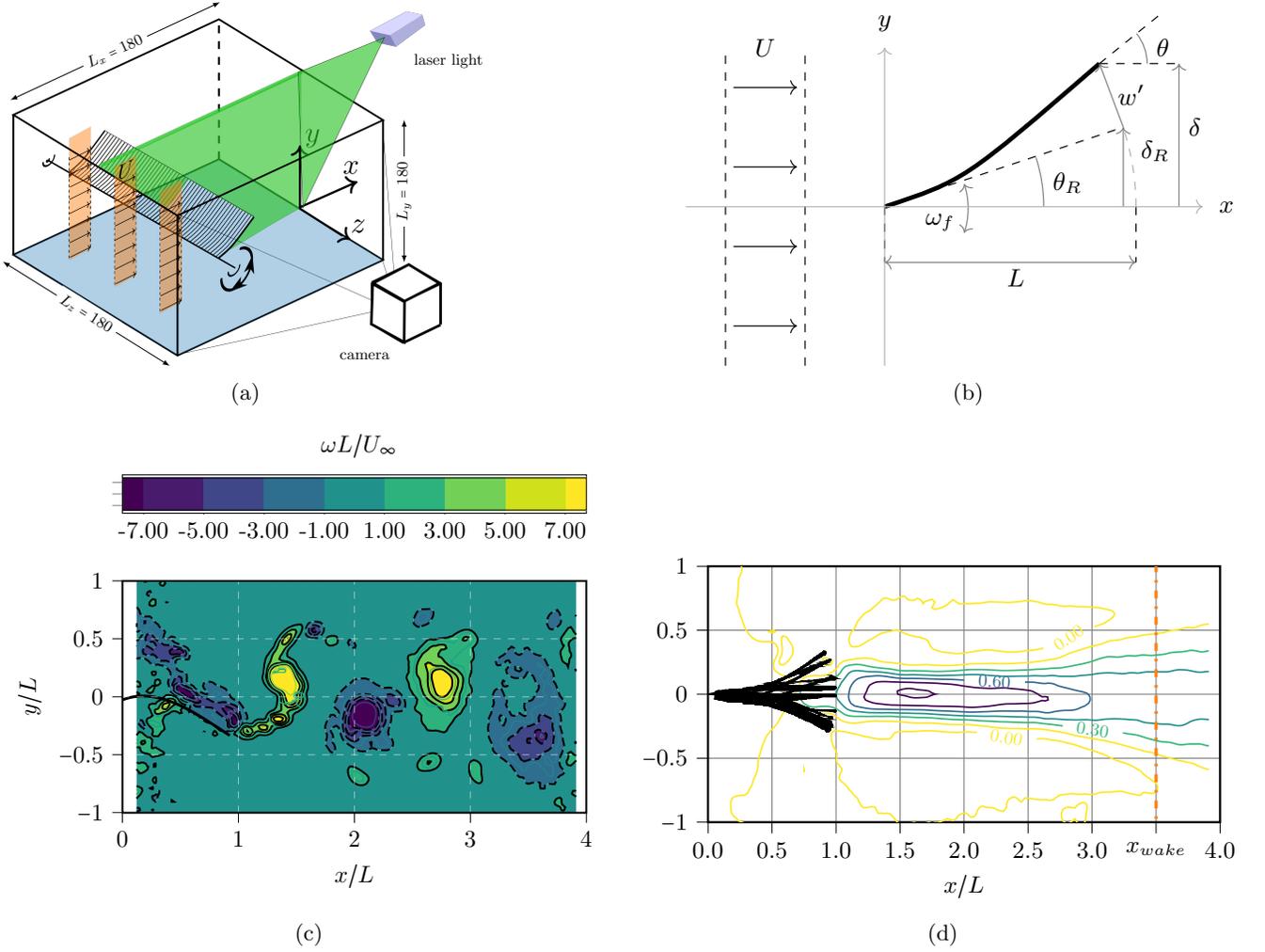}
		}
		\caption{}\label{fig:mean_contour_scheme}
	\end{subfigure}

	\caption{(a) General view of the experimental setup of the pitching flapping foil in the wind tunnel. A laser sheet defines the field of vision. (b) 2D view at midspan describing the flow configuration  and geometrical description. (c) PIV instantaneous vorticity field corresponding to $U_R=0.29$, $f^+=1.47$. (d) Contours of the time-averaged net velocity flow $u(x)=u_x-U$ showing a jet-like flow structure.}
\end{figure}
	
	The flow configuration is described in Fig. \ref{fig:setup_config} for a closed loop wind tunnel that has a test section of dimensions $L_x=L_y=L_z=180$ mm. A  flexible foil of chord length $L=35\,$mm and span $s=160\,$mm  undergoes an oscillatory rotation,  animated by a DC motor through a crank mechanism. The parameters of the movement are a forcing angular frequency $\omega_f = 2\pi f_f$ and a fixed angular amplitude $\theta_R=10^\circ$. We choose an aspect ratio $s:L\simeq 4.6$ that favors 2D flow structures even though the wake is 3D. Concerning flow blockage, for rigid motions, the frontal length of the foil $L_y:2 L\sin(\theta)\simeq 14.8$ assures that the wind tunnel walls should not affect the wake flow dynamics.
	
	As described by Figure \ref{fig:displacement_config}, the total displacement of the foil $w(x,t)$  can be written as the sum of a rigid motion $x\theta$ and an elastic deformation $\tilde w$, i.e. $w = \tilde w + x\theta$, as illustrated on Fig.\ref{fig:displacement_config}. Considering forced oscillations included as $\theta = \theta_0 \sin(\omega_f t)$, it is possible to characterize the dynamics using an Euler-Bernoulli beam framework for the displacements $w$ with times and spatial derivatives correspondingly $\dot w$, $w'$. We can write the conservation of momentum for the elastic foil as:
	\begin{equation}\label{eq:euler_bernoulli}
	\mu \ddot w + B w^{iv} + C_{Dp} L \left|\dot w + w'U\right|(\dot w + w'U)=0
	\end{equation}
	where $\mu$ is the mass per unit length, $B=EI$ is the flexural rigidity composed of the Young modulus E and the moment of inertia I. $C_{Dp}$ is the drag coefficient for the flow normal to a plate, a resistive contribution as proposed by \cite{taylor1952analysis} and later revisited by \cite{ramananarivo2011a,eloy2012origin,pineirua2017modelling}. In our case, we have contributions from the flapping motion $\dot w$ and from the projection of the free stream flow  $w'U$ on the same direction. Reynolds numbers based on the chord length are in the range $Re\sim [1600,4000]$, corresponding to the working velocities of the wind tunnel of $U\in[0.7,1.8]$m/s. Because viscous friction and added mass have small contributions, fluid forces on the foil are in practice given by the last term of Eq.~\ref{eq:euler_bernoulli}. The expression can be further simplified if we consider that the first mode of elastic deformation is dominant. It is thus sufficient to study the projection of the trailing edge displacements $\delta=w(L,t)$. 
	
	We have chosen the physical parameters of the foil and the kinematic parameter space in order to  define a problem with strong coupling between the fluid and solid elastic dynamics. On the one hand, considering a frontal area of the flapping object given by a characteristic length $\ell$  times the span $s$, a vortex shedding frequency can be roughly estimated using the Strouhal number  
	\begin{equation}\label{eq:strouhal}
	\text{St}=f_{vs}  \ell / U = f_{vs}  2\delta_R / U 
	\end{equation} 
	and a reference value for bluff bodies $\text{St}\approx 0.2$. For the range of wind tunnel speeds of the present experiment, the vortex shedding frequencies---which are here driven by the frequency of the forcing pitching oscillation (i.e. $f_{vs}\equiv f_f$)  are thus in the range $[10\ldots 30]$Hz. On the other hand, considering the foil as a cantilever beam, its elastic natural frequency is   $f_n =  k^2\sqrt{B / \mu}/2\pi$, where the wavenumber $k$ is such that $kL=1.875$ for the first elastic bending mode. For a  50$\mu$m polyethylene foil, the elastic natural frequency is  $f_n\simeq 12.27$ Hz. This value was confirmed experimentally through image processing of free damped vibrations of the foil, which gave $f_{n}=11.73$ Hz. 
	
	For image acquisition, we used a monochromatic high-speed video camera (Weinberger SpeedCam in a mode of 400 frames/s with a spatial resolution of 512$\times$512 pixels$^2$). We also performed  Particle Image Velocimetry measurements  using a LaVision system, composed of an ImagerPro $1600\times1200$ CCD camera with 14-bit dynamic range capable of recording double-frame pairs of images at 14 Hz and a two-rod Nd-YAG (15mJ) pulsed laser. The field of view was  170mm$\times$60mm (about 5$L\times$ 2$L$), with a spatial resolution of $0.8$mm (0.023$L$) and the number of snapshots for each case was 680, in order to assure statistical convergence. A typical instantaneous vorticity field $\mathbf \omega(\mathbf x,t)$ is presented in Fig. \ref{fig:instantaneous_piv}, where its contour levels follow the foil deformation. Two rows of eddies exhibit the characteristic arrangement of the reverse BvK vortex street, with positive (negative) eddies placed in regions where $y>0$ ($y<0$). This configuration produces a net momentum injection in the wake, typical of  propulsion produced by flapping motion, with a jet-like structure noticeable by averaging in time the corresponding component of the velocity field, $u_x(x)$.
	
	Assuming ergodicity,  time-averaged velocity fields $\langle\mathbf{u}(\mathbf x)\rangle$ and  fluctuations intensity  of the $j$ velocity component $\langle {{u_j'}^2}(\mathbf x)\rangle$ were determined respectively as:
	\begin{eqnarray}
	\langle{\mathbf u}({\mathbf x})\rangle = \frac{1}{N}\sum_{i=1}^N \mathbf{u}(\mathbf{x},t_i) ~~~~~~~~~
	\langle {u_j^{\prime}}^2(\mathbf x)\rangle = \frac{1}{N}\left\lbrace\sum_{i=1}^N (u_j(\mathbf{x},t_i)-\langle {u_j}\rangle)^2\right\rbrace \label{fluct_eq}
	\end{eqnarray}
	
	Then, with this definition, mean net momentum induced by flapping, $u =  u_x- U$ is illustrated as contour levels on Fig. \ref{fig:mean_contour_scheme}.
	
	The PIV measurements in the present experiment are not time resolved, since the maximum sampling frequency of the system $f_S\sim f_{nat}\sim f_{vs}$. However, the flow temporal dynamics was accurately recovered using a \textit{clustering} method. The algorithm is  based on the $k$-means algorithm (see Appendix \ref{app_cluster} for details) and was used to classify the acquired PIV fields and obtain a number of different representative states. As \textit{clustering} is similar to phase averaging, the velocity and vorticity fields are more clear than direct snapshots, and the whole period of flapping can be reconstructed.

	\section{Results}
	\subsection{Flexible foil dynamics}
	We proceed  to characterize the response of the  elastic foil by varying the forcing frequency $f_f$ or its non-dimensionalized form $f^+ = f_f /f_n$ in a range $f^+ \in[0.5,1.8]$.  We define the reduced velocity $U_R = U/(2\pi f_n L)$ that expresses the coupling between the free flow $U$ and the elastic foil dynamics. 
	
	The results are presented in Fig. \ref{fig:displacements1}, where the non-dimensional displacements of the trailing edge of the foil $\delta^+$ are obtained by comparing the total  to the rigid  displacements $\delta^+ =\delta / \delta_R$.  Similar curves are obtained for each reduced velocity $U_R$, and the data points are fitted with a third order polynomial around the maximum  of displacement $\delta^+_{\max}$. 
	The corresponding maxima $(f^+_{\max},\delta^+_{\max})$ are plotted on Fig. \ref{fig:maxima1} their relationship, together with a second-order polynomial fit. In the inset of the same Figure, a linear dependence between the reduced velocity $U_R$ and the maximum of amplitude $\delta^+_{\max}$ is displayed. It can be seen that the maximum deformation $\delta^+_{max}$ decreases and $f^+_{\max}$ shifts with $U_R$. Modeling the foil dynamics as a nonlinear oscillator with damping contributions from $\dot w$ and $U_R$, it is reasonable to find such behavior. This could be done by reducing eq.(\ref{eq:euler_bernoulli}) to a  system of equations regarding trailing edge dynamics. Through these results it is possible the approximation of displacements for different flow conditions by interpolation, criterion supported by Fig. \ref{fig:delta_f_adim}, where a scaling can be performed showing that the regime is similar for each reduced velocity,

	Another way to scrutinize the kinematics is to compare the changes on the Strouhal number defined in eq.\ref{eq:strouhal} with respect to a different Strouhal number   defined using the elastic displacement $2\delta$, St$^*=2\delta f_f / U$. Fig. \ref{fig_strouhal1} shows a family of similar curves for the different reduced velocities. More insight on the fluid--structure coupling can be gained recalling the different terms in eq. (\ref{eq:euler_bernoulli}) for the Euler--Bernoulli model of the elastic displacements of the foil. We define a Cauchy number 
	\begin{equation}C^*_Y=(\dot w+Uw^\prime)^2 / (\omega_n L)^2 \label{cauchy}\end{equation}
	
	\noindent that compares the kinetic energy of the  fluid velocities (we can identify the  flapping velocity $\dot w$ that is responsible for a major component of damping drag, and also the contribution from the  free stream velocity coupled by the angle, $Uw^\prime$) to that of the solid elastic vibrations given by $(\omega_n L)^2$, for each reduced velocity $U_R$. Both $\dot w$ and $w'$ are evaluated at the trailing edge of the foil, therefore, $C_Y^*(U,f_f,w')$ is a function that changes for every measured case. The inset of Fig. \ref{fig_strouhal1} shows that the Cauchy number $C_Y^*$ provides a satisfactory scaling for the Strouhal numbers relationship.

	\subsection{Flow field results}
	
	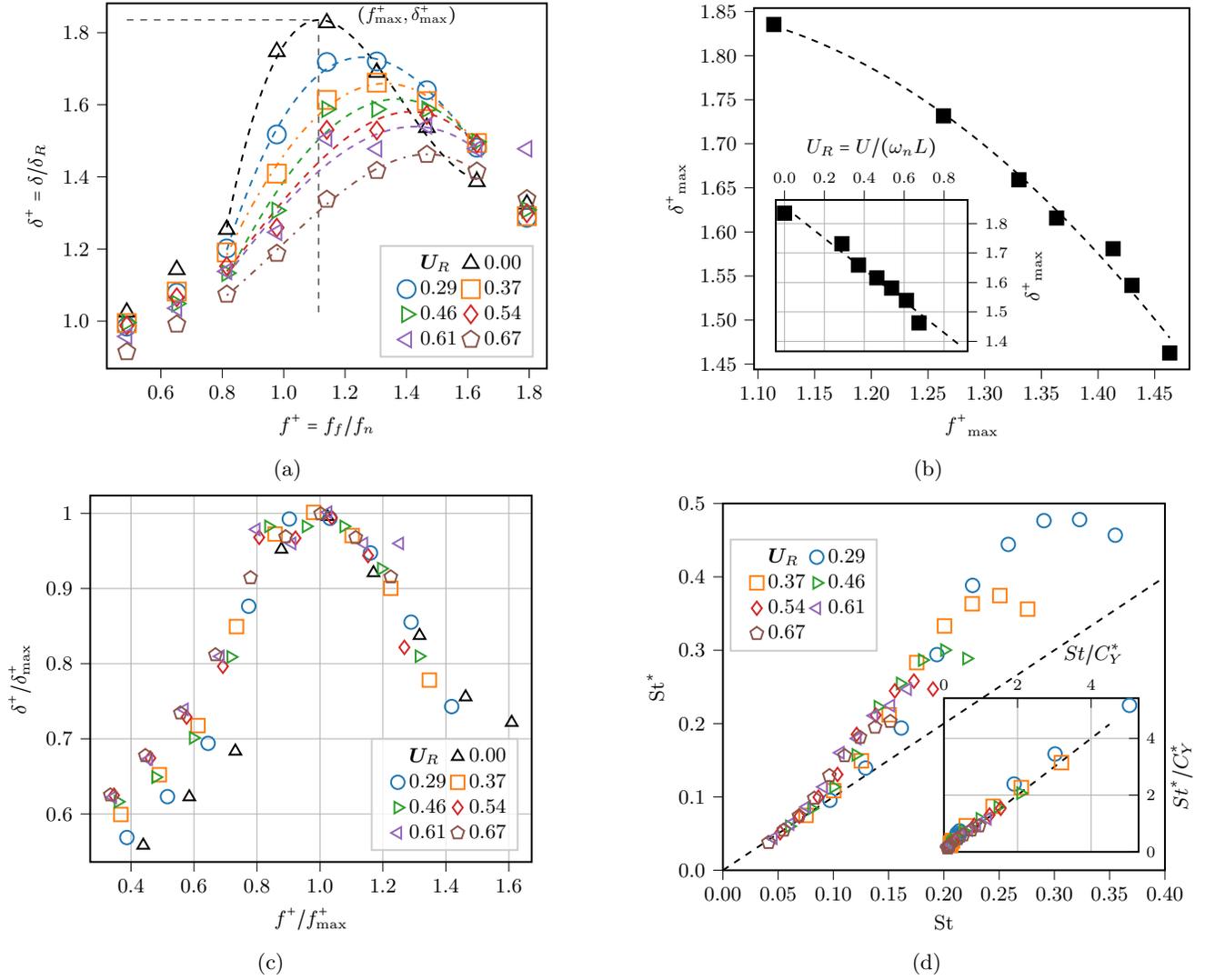
\begin{figure}[t]
		\begin{subfigure}[t]{.45\textwidth}
			\resizebox{\columnwidth}{!}{%
				\tikzsetnextfilename{tikz_matplotlib/resumen_escala_1}
				\input{tikz_matplotlib/resumen_escala1.tikz}
			}
			\caption{}\label{fig:displacements1}
		\end{subfigure}\hspace{1.3cm}
		\begin{subfigure}[t]{.45\textwidth}
			\resizebox{\columnwidth}{!}{%
				\tikzsetnextfilename{tikz_matplotlib/delta_freq_kynematics}			
				\input{tikz_matplotlib/delta_freq_kynematics.tikz}
			}
			\caption{}\label{fig:maxima1}
		\end{subfigure}
		
		\begin{subfigure}[t]{.45\textwidth}
			\resizebox{\columnwidth}{!}{%
				\tikzsetnextfilename{tikz_matplotlib/resumen_escala_2}			
				\input{tikz_matplotlib/resumen_escala2.tikz}
			}
			\caption{}\label{fig:delta_f_adim}
		\end{subfigure}\hfill
		\begin{subfigure}[t]{.48\textwidth}
			\resizebox{\columnwidth}{!}{%
				\resizebox{\columnwidth}{!}{%
					\tikzsetnextfilename{tikz_matplotlib/strouhal_scale}			
					\input{tikz_matplotlib/strouhal_scale.tikz}
				}
			}
			\caption{}\label{fig_strouhal1}
		\end{subfigure}
		\caption{a) Amplitude of non-dimensional displacements $\delta^+$ as a function of the non-dimensional forcing frequency $f^+$ under various reduced velocities $U_R$. The maxima of ($f^+_{\max},\delta^+_{\max}$)   are determined for each curve and plotted on b), showing the amplitude attenuation and frequency shift. In the inset, a quasi-linear dependence with respect to $U_R=U/(\omega_n L)$ is shown. c) Renormalized version of a) $\delta^+/\delta^+_{\max}$ vs. $f^+/f^+_{\max}$. d) Strouhal numbers defined from rigid and from total elastic displacements. The inset shows a  scaling through the Cauchy-like number $C_Y^*$.
			 }	
	\end{figure}
	
	\begin{figure}[t]
		\begin{subfigure}[b]{.48\textwidth}
			\resizebox{\columnwidth}{!}{%
				\tikzsetnextfilename{tikz_matplotlib/media_ux_029}				
				\input{tikz_matplotlib/media_ux_0.29.tikz}
			}
			\subcaption{Profiles of the net mean velocity $\langle u_x\rangle-U $ at $y=0$ for $U_R=0.29$ and different forcing frequencies $f^+$.}\label{fig:jets}
		\end{subfigure}\hfill
		\begin{subfigure}[b]{.46\textwidth}
			\resizebox{\columnwidth}{!}{%
				\tikzsetnextfilename{tikz_matplotlib/media_ux_max}				
				\input{tikz_matplotlib/media_ux_max.tikz}
			}
			\subcaption{Maxima of velocity profiles $\langle u_x(y=0)\rangle-U$ for each $(f^+,U_R)$ case.}\label{fig:jets_max}
			
		\end{subfigure}\hfill
		\begin{subfigure}[b]{.48\textwidth}
			\resizebox{\columnwidth}{!}{%
				\tikzsetnextfilename{tikz_matplotlib/media_uystd_029}				
				\input{tikz_matplotlib/media_uystd_0.29.tikz}
			}
			\subcaption{Profiles of the intensity of velocity fluctuations   $\langle {u'}^2_y\rangle^{1/2} $ at $y=0$ for $U_R=0.29$ and different forcing frequencies $f^+$.}\label{fig:fluctuations}
		\end{subfigure}\hfill
		\begin{subfigure}[b]{.46\textwidth}
			\resizebox{\columnwidth}{!}{%
				\tikzsetnextfilename{tikz_matplotlib/media_uystd_max}				
				\input{tikz_matplotlib/media_uystd_max.tikz}
			}
			\subcaption{Maxima of velocity fluctuations profiles $\langle {u'}^2_y(y=0)\rangle^{1/2}$ for each $(f^+,U_R)$ case.}\label{fig:fluctuations_max}
			
		\end{subfigure}
		\caption{Forced wake flow characteristics.}\label{fig:mean_flow}
	\end{figure}
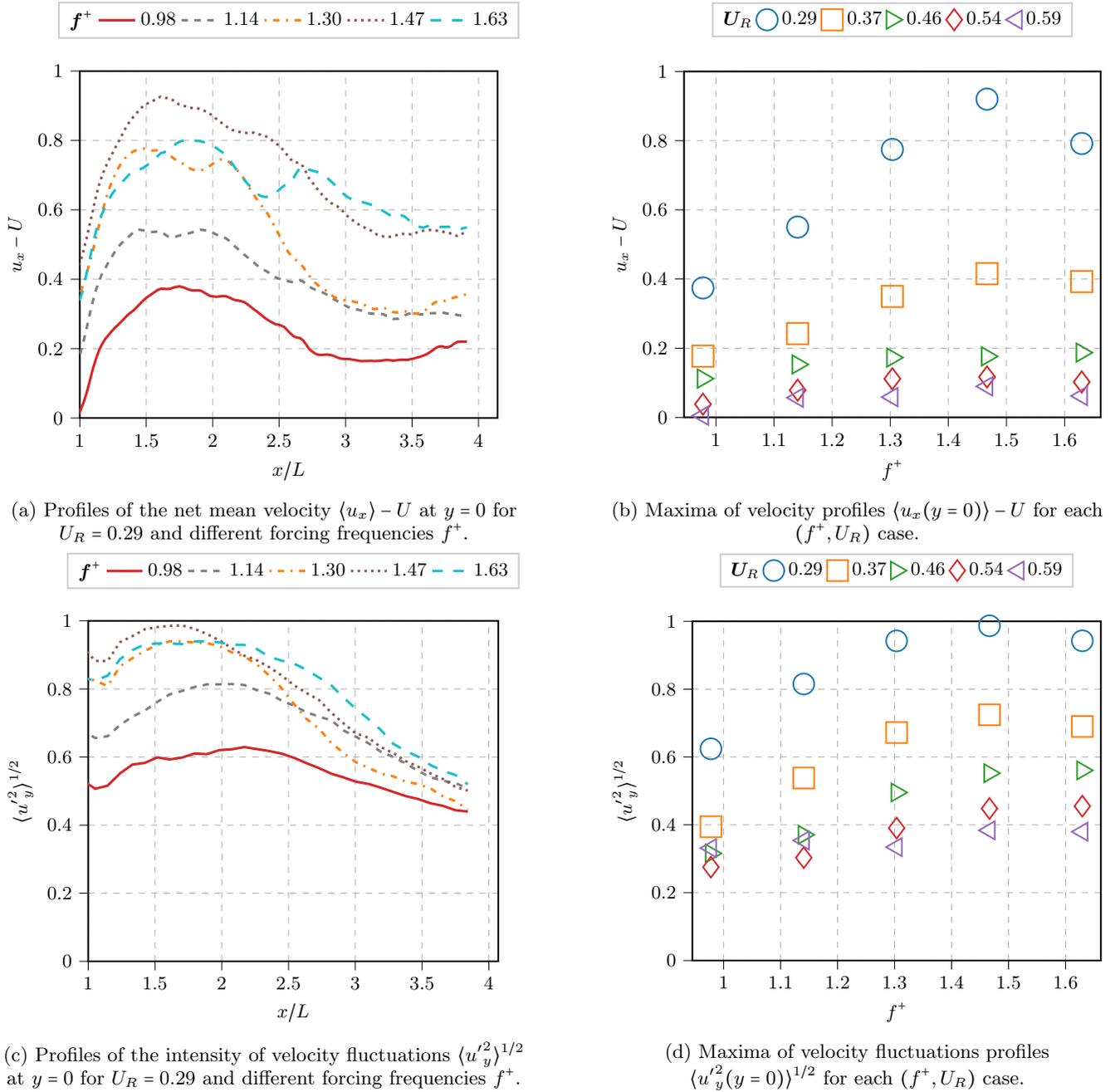
	
	The flow dynamics in the wake of the foil was inspected for different parameters $(f^+,U_R)$ through PIV measurements. A representative state of the flow was already presented in Fig. \ref{fig:instantaneous_piv} (c) for the case of $U_R = 0.29$ and $f^+=1.47$, where the well-known reverse BvK vortex street regime is established.  A first characterization of the induced velocity field can be obtained by representing contour lines of the difference between  time mean averaged  and  free flow  as shown in Fig. \ref{fig:mean_contour_scheme}. A jet-like flow structure is the result in each forced case, and we can compare them conveniently by plotting a profile at $y = 0$ as displayed on Fig.\ref{fig:jets}. For $U_R=0.29$ an increment in the overall magnitude of the velocity  profiles is observed up to $f^+=1.47$.  This is expected as $\omega_f = 2\pi f_f L$ gets higher, but  effective flapping amplitudes reduce for $f^+>1.3$ recalling Fig. \ref{fig:displacements1}. Maxima for the curves are summarized in Fig. \ref{fig:jets_max}, where it is clear that for increasing $U_R$, induced motion  diminishes monotonically.   
	On the other hand, flapping  also produces transversal velocity fluctuations $\langle {u'_y}^2\rangle^{1/2} $ which are displayed in Fig. \ref{fig:fluctuations}. These augment with $f^+$ and attain a maximum for a value in the range $1.30<f^+<1.63]$.  
	
	Furthermore, Fig. \ref{fig:fluctuations_max}  outlines the evolution of transversal velocity fluctuations maxima for each forced case. We shall analyze how these components constrain effective streamwise momentum production or, in a more general context, how they are related to the determination of drag forces. We will also later discuss this aspect in terms of the wake receptivity or its hydrodynamical stability properties.
	
	\subsection{Drag forces}
	\begin{figure}
		\begin{subfigure}[t]{.63\textwidth}
			\tikzsetnextfilename{tikz_matplotlib/U=0.83_f18_scheme_integral_f}
			\resizebox{\columnwidth}{!}{%
				\input{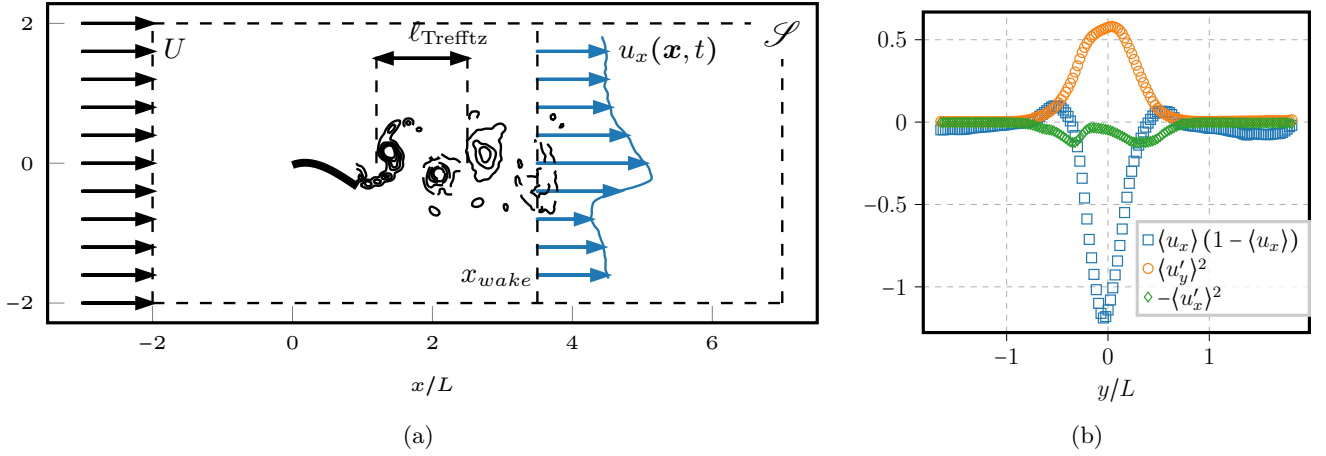}
			}
			\subcaption{}\label{fig:scheme_integral}
		\end{subfigure}	
		\begin{subfigure}[t]{.35\textwidth}
			\resizebox{\columnwidth}{!}{%
				\tikzsetnextfilename{tikz_matplotlib/profiles_UR029fr146}				
				\input{tikz_matplotlib/profiles_UR029fr146.tikz}
			}	
			\subcaption{}\label{fig:drag_terms}
		\end{subfigure}		
		\caption{(a) Schematic representation of a control volume approach to determine forces from velocity fields. (b) Momentum fluxes for drag estimation in equation (\ref{eq_scheme2}) for $U_R=0.29$, $f^+=1.47$ at $x_{\text{wake}}= 2.5L$.}\label{fig:drag_analysis}
	\end{figure}
	Time-averaged aerodynamic forces, particularly drag/thrust in the streamwise direction can be estimated from PIV 2D-velocity field data. 
	There are different approaches to achieve this goal in the literature (see for instance \cite{kurtulus2007}). Here we start by considering the general framework depicted in Fig., \ref{fig:scheme_integral} that consists in integrating both the momentum budget in a control volume $\mathscr V$ enclosing the flapping foil and the pressure forces on the corresponding boundary surface $\mathscr S$. The expression for the force reads: 
	
	\begin{equation}\label{eq:momentum1}
	\mathbf F(t)=-\frac{\mathrm{D}}{\mathrm{D}t}\int_\mathscr V \rho\mathbf u\, \mathrm{d}v + \int_\mathscr S \left(-p\mathbf{I}+\mathbf{T}\right)\cdot \mathbf n\, \mathrm{d}s,
	\end{equation}
	\noindent where $p$ is the pressure field, $\mathbf{I}$  the unit tensor, $\mathbf T=\mu(\nabla\mathbf u+\nabla^T\mathbf u)$ is the viscous stress tensor (with viscosity $\mu$) and $\mathbf n$ is a unit vector orthogonal to the boundary $\mathscr S$.
	
	Neglecting the effect of inhomogeneities in the flow along the spanwise direction, the projection of the time-averaged Eq.~\eqref{eq:momentum1} in the streamwise direction yields

	\begin{equation}\label{eq_scheme2}
	F_D= \int_{\infty}^{+\infty}u_x(1-u_x)dy +\frac{1}{\rho} \int_{\infty}^{+\infty}\Delta p dy- \int_{\infty}^{+\infty}\langle{u_x^\prime}^2\rangle dy
	\end{equation}
	where, for the sake of simpler notation $F_D\equiv\langle F_D\rangle$, $u_x \equiv \langle u_x\rangle$ and in the  equation  the integral is evaluated on a fixed wake coordinate $x=x_{\text{wake}}$. Viscous terms are therefore very small compared with convective and pressure $\Delta p = \left(p(x_{\text{wake}},y)-p_\infty\right)$ terms. The main difficulty of this expression is the estimation of the pressure field from the 2D velocity field.  This is achieved either by means of the Poisson equation (see e.g. \cite{fujisawa2005}), or by integrating the Navier-Stokes (NS) equation along the control surface \cite{unal1997}. The time-averaged pressure gradient on $x=x_{\text{wake}}$ is given by:
	\begin{equation}
	\frac{1}{\rho}\frac{\partial \langle p\rangle}{\partial y} = \langle u_x\rangle\frac{\partial \langle u_y\rangle}{\partial x}+\langle u_y\rangle\frac{\partial \langle u_y\rangle}{\partial y}+\frac{\partial \langle u_x^\prime u_y^\prime\rangle}{\partial x} + \boxed{\frac{\partial \langle u_y^\prime u_y^\prime\rangle}{\partial y}}
	\end{equation}
	where, for wakes, the last term is much larger than the rest, so  $\Delta p\approx \rho \langle {u^\prime_y}^2\rangle$ accounts for the change of pressure in the wake due to transverse velocity fluctuations gradients. Figure \ref{fig:drag_terms} presents an example of how the terms in Eq.~(\ref{eq_scheme2}) can be evaluated at a position $x_{\text{wake}}$ and integrated in order to obtain drag forces. It can be seen that the  transverse velocity fluctuations (orange circles) bring a contribution of the order of the momentum flux  in the stream direction (blue squares). 
	
	A possible approach to explicitly compute the force was proposed by \cite{Noca:1997p249} and later revisited by \cite{wang2013lift}, where pressure terms can be replaced by products of velocity and vorticity impulse contributions (for a formulation in the case of a cylinder wake, see e.g. \cite{dadamo2011}).
	
	However, in order to obtain accurate results, the evaluation of the flow field at a position  $x_{\text{wake}}$ far enough from the body is needed, which is out of reach of the PIV measurement window of the present experiments. 

	To circumvent this problem, we used an alternative method  following the study presented in \cite{Hall:1996} (see also \citep{Minotti:2011}). Through a control volume scheme, under the hypothesis of  periodicity, for a wake flow expressed as $\mathbf{u} = U \mathbf i + \mathbf\nabla \phi=(U+u)\mathbf i+v\mathbf j + w\mathbf k$, a superposition of a potential flow $\mathbf\nabla\phi$ and a far field velocity $U$, it is possible to estimate time-averaged forces on a flapping body. We recall, that in the context of wake flows, the Reynolds numbers of these experiments $Re\sim 10^3$ allow us to consider that starting at the very near wake, inertia is  dominant with respect to viscous forces.
	
\begin{figure}
		\begin{subfigure}[t]{.46\textwidth}
			\tikzsetnextfilename{tikz_matplotlib/drag_Trefftz_1}
			\resizebox{\columnwidth}{!}{%
				\input{tikz_matplotlib/drag_Trefftz_1.tikz}
			}
			\subcaption{}\label{fig:drag_ur}
		\end{subfigure}\hfill
		\begin{subfigure}[t]{.46\textwidth}
			\tikzsetnextfilename{tikz_matplotlib/CD_trefftz_2}	
			\resizebox{\columnwidth}{!}{%
				\input{tikz_matplotlib/CD_trefftz_2.tikz}
			}
			\subcaption{}\label{trefftz_plot} 
		\end{subfigure}

		\caption{ a) Drag Coefficients as a function of the non-dimensional forcing frequency $f^+$ for different values of the reduced velocity $U_R$. b) Corresponding Trefftz lengths. In cases marked by black circles, $UT>\ell_{\text{PIV domain}}$ and  the force computation with eq. (\ref{eq_hall}) cannot be performed. }\label{drag_resume}
	\end{figure}
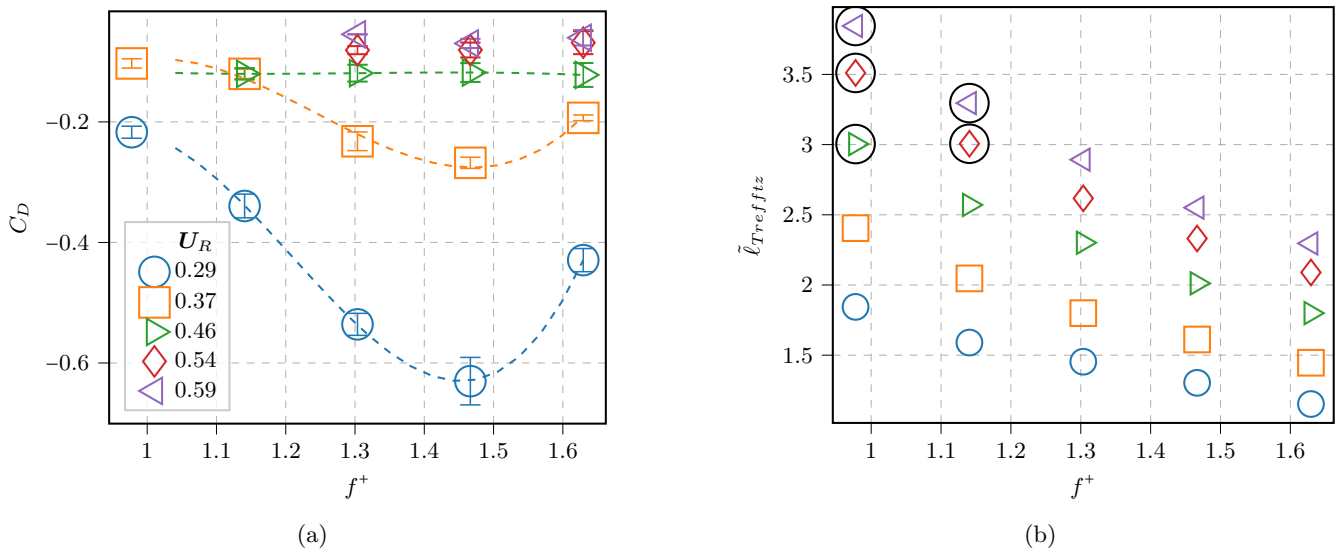

	A light loading hypothesis applies when the unsteady wake and its associated potential are simply convected in the $x$-direction with speed $\sim U$, given that $|\mathbf \nabla \phi|\ll U$. To leading order, the potential is $\phi = \phi\left(x-U t, y,z\right)$. Furthermore,  the unsteady flow is periodic in the $x$-direction so averaging over one temporal period is equivalent to averaging over one spatial period. It results that the momentum balance reduces to:
	\begin{equation}\label{eq_hall}
	\mathbf F = -\frac{1}{T}\iiint_\mathscr V  \rho\left(u\mathbf i + v\mathbf j + w\mathbf k\right) d \mathscr V
	\end{equation}
	where $\mathscr V$ is the Trefftz volume, which encloses an $x-$periodic control volume in the wake, characterised by a length $\ell_{\mathrm{Trefftz}}=UT = U/f_f$, as  showed in Fig.~\ref{fig:scheme_integral}. In non-dimensional form,
	\begin{equation*}
	\tilde \ell_{\mathrm{Trefftz}} = \ell_{\mathrm{Trefftz}} /L = U T / L = U / f_f L \; .
	\end{equation*}
	Recalling that $\mathrm{St}  = f_{f} 2\delta_R /U $ and $\delta_R = L\sin(\theta_R)$, it follows that
	\begin{equation}\label{eq:Trefftz}
	\tilde \ell_{\mathrm{Trefftz}} =  \frac{U}{f_f L} \frac{\delta_R}{\delta_R} = \frac{U}{f_f\delta_R} \sin(\theta_R)  = \frac{2\sin(\theta_R)}{\mathrm{St}}
	\end{equation}
	We can evaluate the integral in Eq. (\ref{eq_hall}) on the volume defined by this length and divide by the period $T$ or multiply equivalently by $f_f$ or $\mathrm{St}$ in non-dimensional form. The result of this computation expressed as a drag coefficient $C_D = 2 F_x / (\rho U^2 2\delta_R)$, considering homogeneity in the spanwise direction, is shown in Fig. \ref{drag_resume}.
	The computed forces are plotted as a function of the forcing frequency for all the different values of the reduced velocity $U_R$  on Fig. \ref{fig:drag_ur}. It is evident that as $U_R$ increases, the momentum produced from flapping is smaller compared to the incoming flow, so in $U_R=0.29$ we appreciate the highest thrust (or the minimum drag).
	Among these values, a minimum is attained at a forcing frequency around $f^+=1.47$. We note that this vaue does not match to the solid elastic resonance $f^+=1$ nor to the maximum elastic displacement observed in Fig. \ref{fig:displacements1} at $f^+=1.30$.
	
	Trefftz volumes, defined by the associated lengths $\ell_{\mathrm{Trefftz}}$ are determined from the experimental parameters and plotted on Fig. \ref{trefftz_plot}. For some cases,  $U T$ becomes higher than the PIV domain behind the foil, so we cannot use this scheme. Nonetheless, the relationship between drag and forcing frequency presents strong changes for low relative velocity values,  $U_R=[0.29;0.37]$, where Trefftz volumes are well defined. For the other cases, the induced momentum, as already depicted in Fig. \ref{fig:jets_max}, is just high enough to compensate the losses in the wake due to transversal fluctuations. 
	
	In this context, previous works have hypothesized on the causes of an optimal propulsion.  \citet{triantafyllou1991wake, Triantafyllou:1993}, for the case of wake dynamics of a pitching rigid foil postulated that the frequency of maximum spatial amplification in the wake provides optimal thrust production per input power. In the same sense, later, \citet{moored2012hydrodynamic,moored2014linear} experimentally showed that
	resonant peaks in thrust occurred for discrete values of the effective flexibility, a
	non-dimensional parameter measuring the ratio of added mass forces to internal bending
	forces. In their study, not only friction but added-mass contributions play a role on more complex dynamics. Their linear stability analysis used the Orr-Sommerfeld equation as viscous terms cannot be neglected and the flow finds more combinations of  resonant frequencies for the wakes. They found on the one hand that for flexible propulsors each peak in efficiency occurs when the driving frequency of motion is tuned to a wake resonant frequency, not a structural resonant frequency. On the other hand, panel flexibilities that attain global optimally efficient locomotion are those for which structural and wake resonant frequencies are tuned.  
	
	Although, the analysis is focused on a self-propelled condition, we found that some of these concepts can be applied to our present results, as we discuss in what follows.

	\subsection{Drag forces and Circulation}

\begin{figure}
		\begin{subfigure}[b]{.45\textwidth}
			\tikzsetnextfilename{tikz_matplotlib/circulacion_piv2}			
			\resizebox{\columnwidth}{!}{%
				\input{tikz_matplotlib/circulacion_piv2.tikz}
			}
			\subcaption{}\label{fig_circulation_piv}
		\end{subfigure}\hfill
		\begin{subfigure}[b]{.48\textwidth}
			\tikzsetnextfilename{tikzs/Circulation_Scheme}
			\resizebox{\columnwidth}{!}{%
				\input{tikzs/Circulation_Scheme.tikz}
			}
			\vspace{5mm}
			\subcaption{}\label{fig_circulation_scheme}
		\end{subfigure}
		
		\caption{a) Non-dimensional circulation as a function of forcing frequency. Filled symbols come from PIV measurements and non-filled symbols from the kinematic model represented schematically in b). The model predicts circulation from the parameters of the foil kinematics and the free flow (see text).}\label{fig_circulation}
	\end{figure}
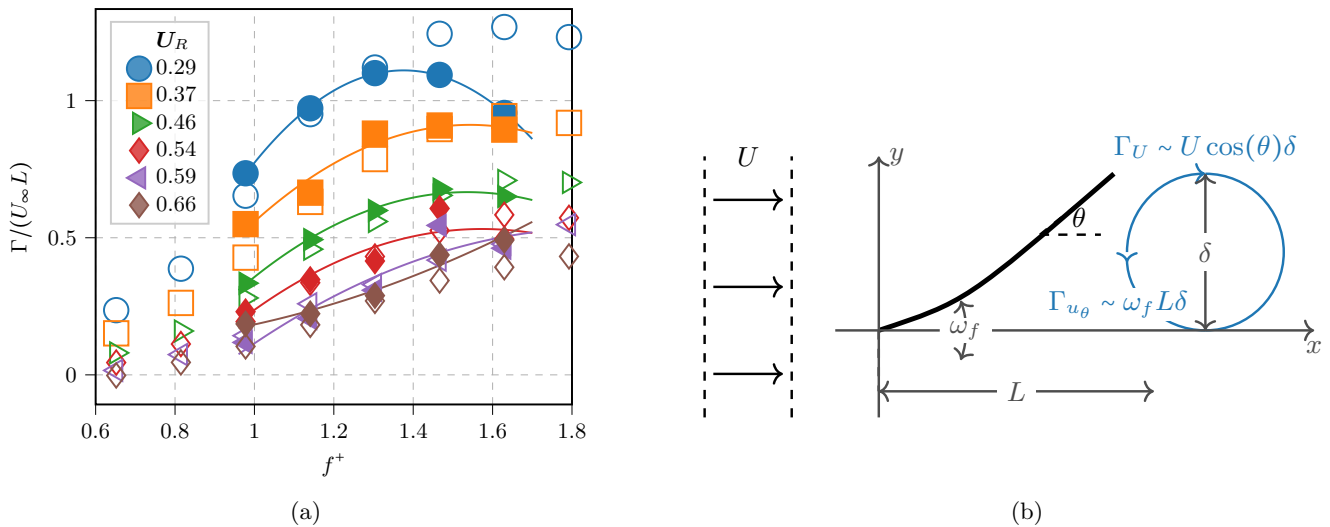

	Considering a potential flow theoretical framework, the forces induced on the body are proportional to each element of the vortex street circulation $\Gamma_i$ and to their distance $y_i$ respect the symmetry axis $y=0$,  as described by the impulse equation :
	\begin{equation}\label{eq:drag_inviscid}
	F = \frac{d}{dt}\sum \Gamma_i y_i
	\end{equation}
	The circulation  $\Gamma_i = \int_{\mathbf s} \omega \mathbf ds$ can be determined for each case as we have clear quantitative results of vorticity and  vortex definition  from the PIV cluster analysis. The results are displayed on Fig \ref{fig_circulation_piv}. An alternative estimate for the circulation can be obtained using the foil kinematics. Since there is a competition between vorticity production from the free flow and from flapping, as depicted by Fig. \ref{fig_circulation_scheme}, $\Gamma_i=\int_s \bar u \bar {dl}$, we express:
	\begin{equation}
	\Gamma_i =\Gamma_{u\theta}- \Gamma_U  = \omega_fL\delta-U\cos(\theta)\delta 
	\end{equation}
	The values computed from this kinematic approximation are also presented in Fig. \ref{fig_circulation_piv} and a reasonable agreement with the calculation from the PIV fields is found for most cases. Also, the kinematic computation brings an estimate of the circulation where no PIV data is available, thereby extending the range of frequencies explored. For $Ur=0.29$ the circulation prediction from the kinematic model exceeds the calculation from the PIV measurements for frequencies over the maximum, $f^+>1.30$. Maxima are delayed for increasing $U_R$ so differences in other cases are not so clearly displayed. We hypothesize that the wake is less sensitive to higher frequencies so forcing at $f^+>f^+_{\max}$ ceases to create effective vorticity.
	
	Examining  $U_R=0.29$, for $f^+=1.14$ and $f^+=1.47$ we observe that while the value of vortex circulation from PIV measurements does not change significantly, the drag coefficient on Fig.\ref{fig:drag_ur}  is about the double. Recalling Eq.~(\ref{eq:drag_inviscid}), this fact confirms that the change in the resulting force must be due to a modification in vortex arrangements. 
	
\begin{figure}
		\begin{subfigure}[t]{.48\textwidth}
			\includegraphics[width=\textwidth]{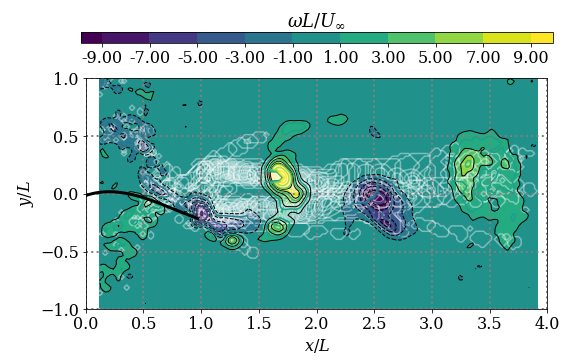}
			\subcaption{$f^+=1.14$}\label{fig:vortex_config_1}
		\end{subfigure}\hfill
			\begin{subfigure}[t]{.48\textwidth} 
		\includegraphics[width=\textwidth]{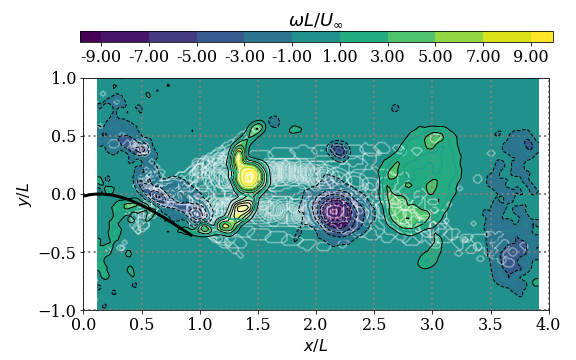}
		\subcaption{$f^+=1.30$}\label{fig:vortex_config_optim}
	\end{subfigure}
		\begin{subfigure}[t]{.48\textwidth} 
			\includegraphics[width=\textwidth]{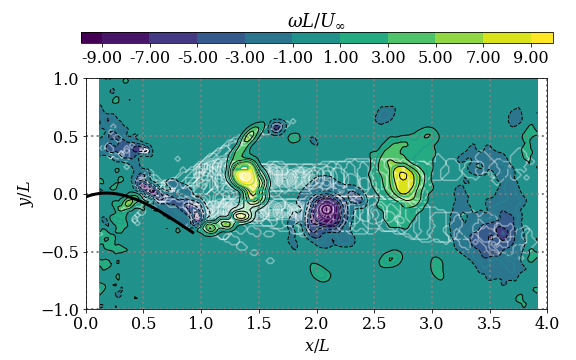}
			\subcaption{$f^+=1.47$}\label{fig:vortex_config_optim}
		\end{subfigure}
		\caption{Contours of vorticity for a given phase of the flapping period at $U_R=0.29$ at three different forcing frequencies. White contours shows the path of vortices during the whole period of oscillation.  Full animated trajectories are available in Supplementary Material. }\label{fig:vortex_config}
	\end{figure}

	The spatial development of the vorticity field is depicted in Fig. \ref{fig:vortex_config}, for a similar forcing phase period under three different forcing frequencies. We observe that the intensity and position of vortices in the near wake does no differ from one case to the other but for larger distances ($x/L>2.5$) $f^+=1.47$ shows a more ordered structure. Traces for every phase of the period are also plotted for both cases, describing crossing ($f^+=1.14; 1.30$) and parallel ($f^+=1.47$) vortex streets. 
	In an air flow, for $Re\sim 2000$, where we can neglect friction and added mass contributions on the wake dynamics, instability of these trajectories can be attributed to an inviscid mechanism. We study this particular aspect of forced wakes   by means of a hydrodynamical stability analysis based on previous approaches \cite{trianta1986,thiria2007} and hypotheses \cite{moored2012hydrodynamic,moored2014linear}.

	\subsection{Hydrodynamical stability}
	\begin{figure}
		\tikzsetnextfilename{tikz_matplotlib/freq_globales_f}
		\resizebox{.35\columnwidth}{!}{%
			\input{tikz_matplotlib/freq_globales_f.tikz}		
		}	
		\caption{The bottom plot presents Strouhal numbers ($\text{St}$) calculated in the wakes through linear stability as a function of $f^+$ for $U_R$=0.29; 0.37. The shaded horizontal lines correspond to the elastic non-dimensional frequencies of the foil for each case. Resonance between the wake and the elastic foil occurs when they intersect the Strouhal curves.
			 The top plot reproduces the $C_D$ curves of Fig. \ref{fig:drag_ur} for the same reduced velocities. Shaded vertical lines mark the frequencies for the drag minima to highlight that these coincide with  the aforementioned wake-foil resonance. }\label{fig_freqs}
	\end{figure}
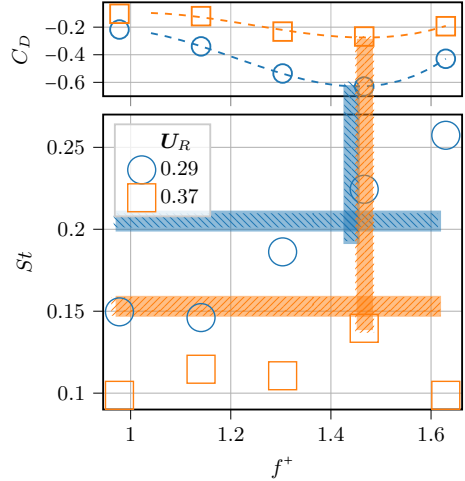
	In order to analyze the stability properties of the wake,  we explore its receptivity  to disturbances in the region $x/L>1$. Neglecting viscosity effects, time mean average velocity profiles $u_x(y)$ are analyzed with the inviscid Rayleigh equation. Global frequencies for the wake are obtained for two forcing cases $U_R=0.29$ and 0.37 where clear propulsive force peaks were observed. Details on the method and implementation are included in Appendix \ref{app:hydro}.
	
	Figure \ref{fig_freqs} summarizes the results of the analysis. Both panels in the figure share the same abscissa $f^+$. The top plot reproduces the drag coefficient curves from  Fig.~\ref{fig:drag_ur} whilst the bottom panel presents the non-dimensional frequencies selected by the wake (written as Strouhal numbers), as obtained by the stability analysis. The horizontal bars represent Strouhal numbers calculated as $f_n2\delta_R/U$ while vertical bars mark values of drag minima for each reduced velocity. These lines coincide in the vicinity of non-dimensional frequencies issued from linear stability analysis.
	
	It is thus resonance (e.g. $U_R=0.29, f^+=1.47$) between the stability properties from the wake and the elastic vibrations from the foil the criteria that leads to maximum propulsion (or drag minima). Such condition is in agreement with parallel trajectories of vortices as already depicted on Fig. \ref{fig:vortex_config_optim}. Again, $y_i$ in eq. (\ref{eq:drag_inviscid} remains constant and assures the maximum impulse. On the contrary, when foil elastic and wake frequencies are not synchronized (i.e. ($U_R=0.29,f^+=1.14$)), even for similar values of circulation produced in the near wake as measured and predicted on Fig. \ref{fig_circulation_piv}, vortex paths result unstable producing higher transversal fluctuations and less added net momentum. From these facts, we found that the interaction between the foil induced motion and the free flow (through $U_R$) is not only in terms of non-linear damping but also in setting the wake stability conditions.

	\section{Conclusions}
	We characterized the dynamics of a flexible foil performing pitch oscillations immerse in an uniform airflow. For small to moderate elastic deformations, scale laws can describe the relationship between the maximum of amplitude, and the corresponding forcing frequency of the foil and the reduced velocity. In the same sense,  inertial drag resistive contribution importance is depicted through a Cauchy number.
	
	Study of the flow main structures was performed through PIV measurements and applying clustering methods. Propulsion forces were determined from velocities by means of an integral momentum scheme. Maximum propulsion was not obtained for the forcing and elastic resonance condition $f^+=1$ but rather when the  wake resonant frequency is tuned with the foil dynamics. A convective instability associated with a \textit{jet}-like structure that develops in the wake was studied by hydrodynamic linear stability analysis. Reduced velocity, forcing oscillations and the corresponding elastic deformations condition the shape of such  \textit{jet} velocity profiles, defining the resulting wake global resonant frequency. In agreement with these findings, we observe that vortex trajectories corresponding to propulsion maxima are more stable and remain parallel in the reverse B\'enard-von K\'arm\'an configuration.

	\appendix
	\section{Appendix}
	\subsection{Hydrodynamical stability}\label{app:hydro}
	Wake flow stability properties are surveyed through a linear inviscid 1D approach, using the Rayleigh equation:
	\begin{equation}\label{eq:rayleigh}
	(kU_y-\omega)\left(\frac{d^2\psi}{dy^2}-k^2\psi\right)-k\frac{d^2U_y}{dy^2}\psi=0
	\end{equation}
	being  $U_y$ a time-averaged velocity profile at a fixed $x-$coordinate, $k$ the wavenumber, $\omega$ the frequency and $\psi$ the stream function. 
	
	An analytical fit for $U_y$ to avoid numerical noise is used, as in the study by \cite{trianta1986} for cylinder wakes:
	\begin{equation}\label{eq:fit}
	U_y  = 1 - A + A \tanh\left[a\left(y^2-b\right)\right]
	\end{equation}
	
	The boundary conditions are on the one hand, symmetry about $y=0$, corresponding to BvK vortex streets. This also is the case of reverse BvK, which applies directly to our measurements. On the other hand, for $y\gg 1$, $U_y\simeq U$ the free flow velocity, then $\psi\longrightarrow0$.
	
	A generalized eigenvalue problem solves (\ref{eq:rayleigh})
	$$\omega B(k)\psi=A(k)\psi$$ 
	We use NumPy linear algebra libraries in order to calculatie the pseudoinverse of B, for a $k=k_r+ik_i$ and then computing the most unstables eigenvalues $\omega_0=\omega_{0r}+i\omega_{0i}$.
	
\begin{figure}
		\tikzsetnextfilename{tikz_matplotlib/stab_hidro_appendix}	
		\resizebox{\columnwidth}{!}{%
			\input{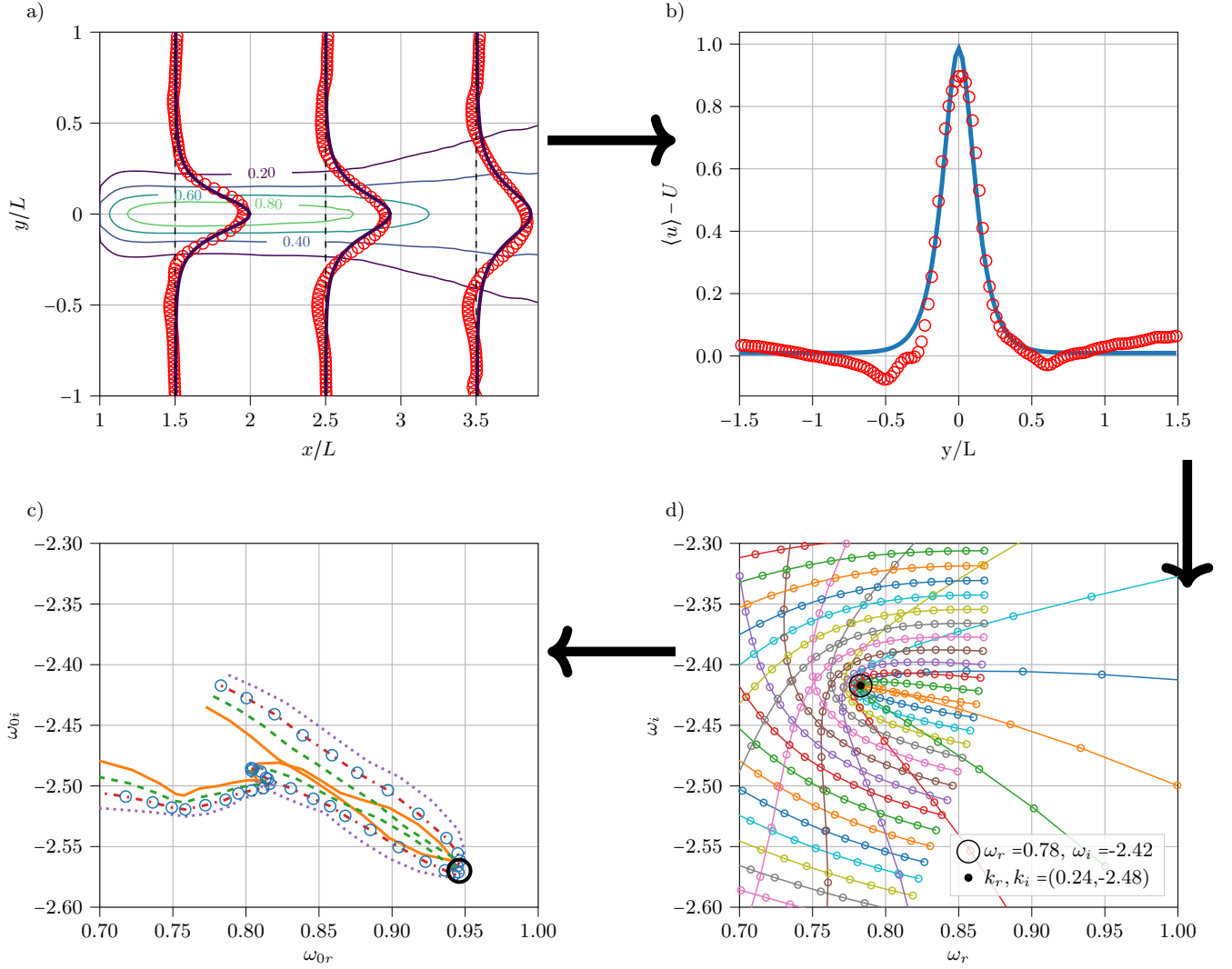}		
		}
		\caption{Schematic representation of the necessary steps involved in the linear stability analysis: (a) Time-averaged net mean velocity flow contours for $u_x$ with three profiles (circle symbols) fitted by eq. (\ref{eq:fit}) . (b) A mean velocity profile at $\langle U_y(y,x=x_c)\rangle$,  for a fixed $x-$coordinate $x_c$. Solving  the Rayleigh equation for this profile, for a range of wavenumbers $k= k_r+i k_i$ produces   (d)  a map for $k_i=$ constant in the $\omega$-plane. At a critical point, a cusp on the  curve, the point $(\omega_{0r},\omega_{0i})$ is determined for $x_c$. In (c) the results for each $x-$position (symbols) are plotted in a $\omega-$plane. Criteria for the prediction of a global frequency in a spatially developing flow is  given by the condition ${\partial\omega/\partial x|}_{x=x_s}=0 $ \cite{chomaz1991frequency}. The saddle point $\omega_s=\omega(x=x_s)$ (big black circle) is obtained through an analytic continuation of the complex function $\omega_0(x)$ considering  $x = x_r + i x_i$ (lines). $\omega_s$ corresponds to the appearance of   a cusp point. (dashed curve) \cite{Hammond:1997p387}.  Considering all the cases, the points $\mathbf{\omega_s}=(\omega,\sigma)$ provide the data for Fig.\ref{fig_freqs}}\label{fig:stability_procedure}
	\end{figure}

	The global frequency for the spatially developing flow is determined using the criterion proposed by \cite{chomaz1991frequency} which consists in finding a saddle point, $\partial\omega_0/\partial x\|_{x=x_s}=0$ of the complex function $\omega_0(x)$ through use of the Cauchy--Riemann equations and analytic continuation to complex values of $x = x_r + ix_i$. Figure \ref{fig:stability_procedure} summarizes the procedure, Hence, we could  select the onset frequency $f=\omega_0(x_s)/2\pi$ for this spatially developing flow to characterize the wake resonance.
	\subsection{Clustering}\label{app_cluster}
	A clustering technique, named k-means algorithm \cite{steinhaus1956,lloyd1982}, was applied in order to obtain clear and representative states of the flow.  \citet{burkardt2006} used the method for the first time in a fluid mechanics context. Later, \citet{kaiser2014} developed Cluster-based Reduced Order Modeling (CROM) which constitutes and alternative to classical POD-Galerkin reduced model schemes. 
	
	Strictly,  k-means   is a classification algorithm which minimizes the average distance between a specified number $N_c$ of points (called centroids) and the data in a phase space. For example, a point $\omega_i$ in the phase space representing one PIV snapshot can be defined by its vorticity values in the spatial domain of interest, in our case, $x\in[1\ 3.5]$, $y\in[-1\ 1D]$. Other components of the flow field may be chosen or even projections on some suitable base functions. Taking into considerations our PIV measurements, given their spatial resolution, $\omega_i$ corresponds to 10836 values.  We can use i.e. the  euclidean norm $\mathrm{d}(\omega_i,\omega_j)=\sqrt{\sum_{s=1}^{10836} (\omega_i(s)-\omega_j(s))^2}$ as a proper distance between two points of the phase space.
	
\begin{figure}
		\resizebox{\columnwidth}{!}{%
			\tikzsetnextfilename{tikz_matplotlib/cluster_method_f}				
			\input{tikz_matplotlib/cluster_method_f.tikz}
		}	
		
		\caption{Left: Steps for the construction of Clusters. Right: application example from our PIC data. Three diferent snapshots of voricity fields corresponding to the same cluster contributes to its determination through ensamble average.}\label{fig:cluster_method}
	\end{figure}
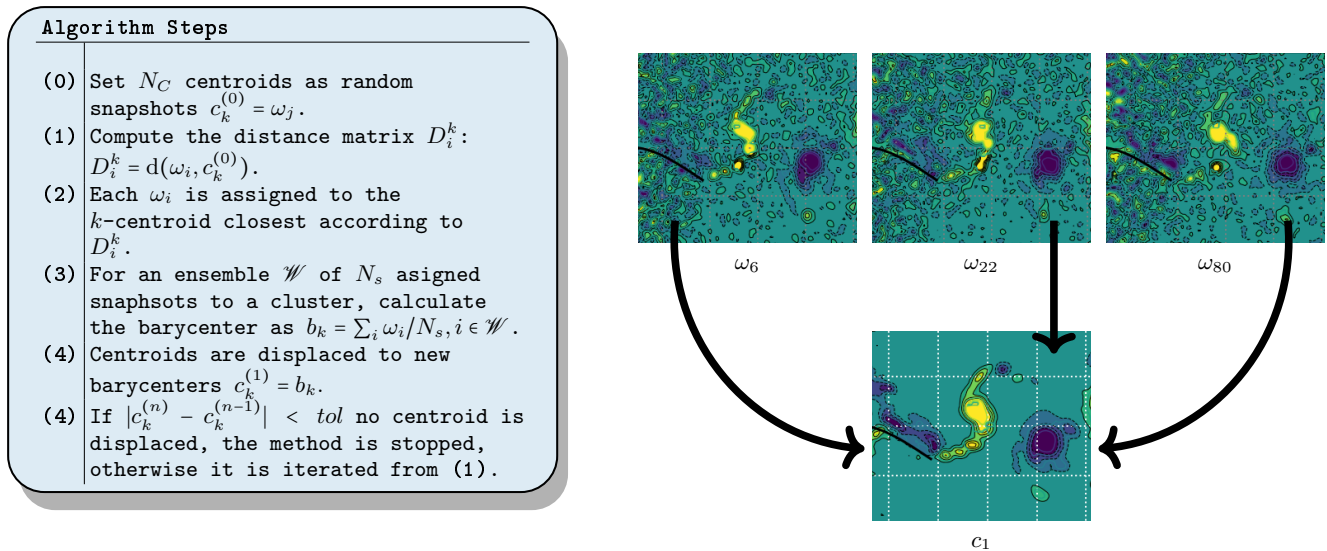

	The $N_c$ centroids $c_k$, $k\in[1,N_c]$ can be  initialized by picking $N_c$ random PIV fields. Figure \ref{fig:cluster_method} presents the  algorithm calculations and  illustrate how coherent structures are clearer and more identifiable.
	
	When the method is stopped, $N_c$ centroids are obtained, each one of them representing average states issued from the data and as separated as possible one from the other. A representation of a characteristic state of the flow can then be obtained by averaging all snapshots belonging to the same cluster.  This method can be considered as a generalized phase averaging technique without the necessity of a strict periodicity of the phenomenon. An object oriented, open source k-mean algorithm is available in the  Scikit-Learn \cite{scikit-learn} Python library .
	
	\bigskip
	
	We acknowledge support of the French-Argentinian International Research Project IVMF,
	CNRS--INSIS (France), CONICET (Argentina).

\end{document}

%% file: tikzs/flow_config_vista_3D.tikz
\tikzset{grid/.style={gray,very thin,opacity=1}}
\begin{tikzpicture}
\node at(0,-5cm){\includegraphics[height=7cm]{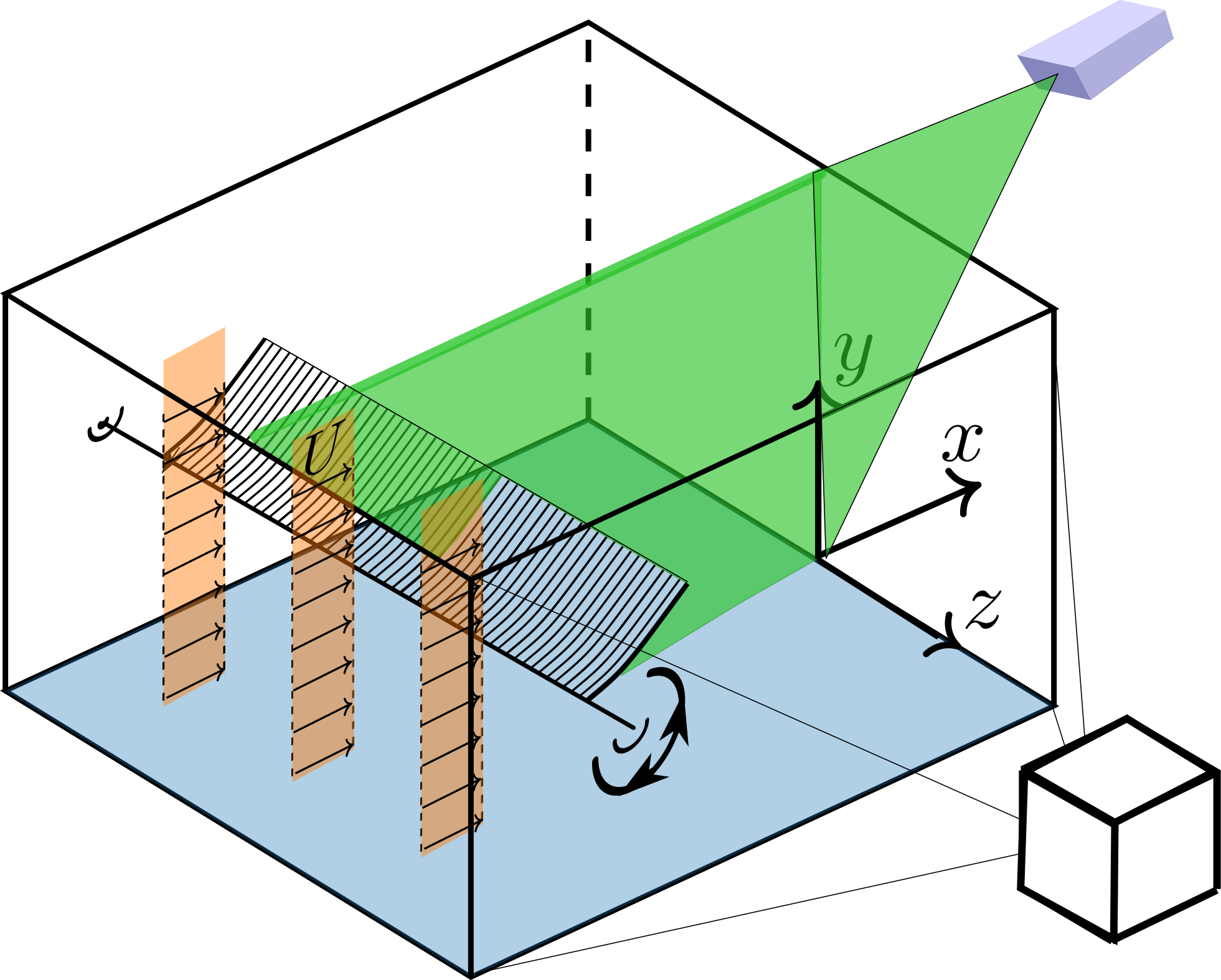}};
\node at(4.5,-2.5cm){laser light};
\node at(2.8,-8.5cm){camera};
\draw[latex-latex](-4.5,-6.5)++(58:-.1)--++(-32:4)node[midway,fill=white,rotate=-32]{$L_z=180$};
\draw[latex-latex](-4.4,-3.3)--++(25:4.7)node[midway,fill=white,rotate=25]{$L_x=180$};
\draw[latex-latex](3.6,-3.7)--++(-90:2.9)node[midway,fill=white,rotate=90]{$L_y=180$};
\end{tikzpicture}
		

%% file: tikzs/flow_config_f.tikz
\begin{tikzpicture}

\draw[dashed] (-2,-2) -- (-2,2);
\draw[dashed] (-1,-2) -- (-1,2);
\draw[->] (-1.9,-1.5) -- (-1.1,-1.5);
\draw[->] (-1.9,-0.5) -- (-1.1,-0.5);
\draw[->] (-1.9,0.5) -- (-1.1,0.5);
\draw[->] (-1.9,1.5) -- (-1.1,1.5);
\node () at (-1.5,2) {$U$};

\draw[dashed](0,0)--(3,1); 
\draw[ultra thick](0,0).. controls(1,0.33) ..(2.7,1.8);

\draw[black!50,->](3,1)->(2.7,1.8); 
\node[]at (3.1,1.4) {$w^\prime$} ;
\draw[black!50,->](3,0)->(3,1); 
\draw[black!50,->](3.7,0)->(3.7,1.8); 
\node[]at (3.4,.7) {$\delta_R$} ;
\node[]at (3.9,1) {$\delta$} ;

\draw[dashed](2.7,1.8)--++(39:10mm);
\draw[dashed](2.7,1.8)--++(0:10mm);
\draw[black!50] (3.3,1.8) arc (0:39:6mm);
\draw[black!50] (2,0) arc (0:18:20mm);

\draw[black!50,<->] (1,-.33) arc(-18:18:10mm);
\node[]at (0.7,-.2) {$\omega_f$} ;
\node[]at (2.3,.3) {$\theta_R$} ;
\node[]at (3.5,2) {$\theta$} ;

\draw[dashed](3.16,0)--(3.16,-.7);
\draw[dashed](0,0)--(0,-.7);
\draw[black!50,<->] (0,-.7)--++(3.16,0);
\draw[black!30,dashed] (3.16,0) arc(0:18:31.6mm);
\node[]at (1.65,-.9) {$L$} ;



\draw[black!30][->] (-2.5,0) -- (4,0);
\node () at (4.3,0) {$x$};
\draw[black!30][->] (0,-2) -- (0,2);
\node () at (0,2.3) {$y$};

\end{tikzpicture}
		

%% file: tikz_matplotlib/cluster_snapshot.tikz
\begin{tikzpicture}
\Large
\node at(0,-5cm){\includegraphics[height=7cm,trim=0cm 0 0 0,clip]{{tikz_matplotlib/U=0.83_f18_vort_cluster}.pdf}};
\end{tikzpicture}

%% file: tikz_matplotlib/resumen_escala1.tikz

\begin{tikzpicture}
\definecolor{color0}{rgb}{0.12156862745098,0.466666666666667,0.705882352941177}
\definecolor{color1}{rgb}{1,0.498039215686275,0.0549019607843137}
\definecolor{color2}{rgb}{0.172549019607843,0.627450980392157,0.172549019607843}
\definecolor{color3}{rgb}{0.83921568627451,0.152941176470588,0.156862745098039}
\definecolor{color4}{rgb}{0.580392156862745,0.403921568627451,0.741176470588235}
\definecolor{color5}{rgb}{0.549019607843137,0.337254901960784,0.294117647058824}

\begin{axis}[ thick,
legend cell align={left},
legend columns=2,
legend style={fill opacity=0.8, draw opacity=1, text opacity=1, at={(0.97,0.03)}, anchor=south east, draw=white!80!black},
tick align=outside,
tick pos=left,
xlabel={\(\displaystyle f^+=f_f/f_n\)},
xmin=0.423659768616588, xmax=1.85758513931889,
xtick style={color=black},
xtick={0.4,0.6,0.8,1,1.2,1.4,1.6,1.8,2},
xticklabels={0.4,0.6,0.8,1.0,1.2,1.4,1.6,1.8,2.0},
y grid style={white!69.0196078431373!black},
ylabel={\(\displaystyle \delta^+=\delta /\delta_R\)},
ymin=0.868409348665972, ymax=1.8813887505007,
ytick style={color=black},
ytick={0.8,1,1.2,1.4,1.6,1.8,2},
yticklabels={0.8,1.0,1.2,1.4,1.6,1.8,2.0}
]
\addlegendimage{empty legend}
\addlegendentry{\hspace{-0cm}\textbf{$\bm U_R$}}
\addplot [thick, black, mark=triangle, mark size=4, mark options={solid,fill opacity=0}, only marks]
table {%
0.488838194557601 1.02457134828939
0.651784259410135 1.14266835728225
0.814730324262669 1.25438174416739
0.977676389115203 1.74751655198892
1.14062245396774 1.82731182833544
1.30356851882027 1.69006395301942
1.4665145836728 1.53685702243409
1.62946064852534 1.38684190290262
1.79240671337787 1.32460158735233
};
\addlegendentry{$0.00$}
\addplot [thick, black, forget plot,dashed]
table {%
0.814730324262669 1.25996741351165
0.831357473737417 1.32952469593021
0.847984623212166 1.39381385146114
0.864611772686914 1.45299342072821
0.881238922161663 1.50722194435519
0.897866071636411 1.55665796296586
0.914493221111159 1.60146001718399
0.931120370585908 1.64178664763334
0.947747520060656 1.67779639493768
0.964374669535404 1.70964779972079
0.981001819010153 1.73749940260644
0.997628968484901 1.7615097442184
1.01425611795965 1.78183736518044
1.0308832674344 1.79864080611633
1.04751041690915 1.81207860764984
1.06413756638389 1.82230931040473
1.08076471585864 1.8294914550048
1.09739186533339 1.83378358207379
1.11401901480814 1.83534423223549
1.13064616428289 1.83433194611366
1.14727331375764 1.83090526433208
1.16390046323238 1.82522272751451
1.18052761270713 1.81744287628473
1.19715476218188 1.8077242512665
1.21378191165663 1.7962253930836
1.23040906113138 1.78310484235981
1.24703621060613 1.76852113971887
1.26366336008087 1.75263282578458
1.28029050955562 1.7355984411807
1.29691765903037 1.717576526531
1.31354480850512 1.69872562245926
1.33017195797987 1.67920426958923
1.34679910745462 1.65917100854469
1.36342625692936 1.63878437994943
1.38005340640411 1.61820292442719
1.39668055587886 1.59758518260176
1.41330770535361 1.5770896950969
1.42993485482836 1.5568750025364
1.44656200430311 1.53709964554401
1.46318915377785 1.5179221647435
1.4798163032526 1.49950110075866
1.49644345272735 1.48199499421325
1.5130706022021 1.46556238573104
1.52969775167685 1.45036181593579
1.5463249011516 1.43655182545129
1.56295205062634 1.4242909549013
1.57957920010109 1.41373774490959
1.59620634957584 1.40505073609994
1.61283349905059 1.39838846909611
1.62946064852534 1.39390948452188
};
\addplot [thin, black, dashed, forget plot]
table {%
0.488838194557601 1.83534423223549
1.11401901480814 1.83534423223549
};
\addplot [very thin, black, dashed, forget plot]
table {%
1.11401901480814 1.02457134828939
1.11401901480814 1.83534423223549
};
\addplot [thick, color0, mark=*, mark size=4, mark options={solid,fill opacity=0}, only marks]
table {%
0.488838194557601 0.984673710116129
0.651784259410135 1.07883213620503
0.814730324262669 1.20171686177868
0.977676389115203 1.51770615611092
1.14062245396774 1.71879025250417
1.30356851882027 1.7203861580311
1.4665145836728 1.64059088168457
1.62946064852534 1.48100032899152
1.79240671337787 1.28647452979585
};
\addlegendentry{$0.29$}
\addplot [thick, color0, dashed, forget plot]
table {%
0.814730324262669 1.19737245228871
0.831357473737417 1.24025310342012
0.847984623212166 1.28109672890891
0.864611772686914 1.3199346035543
0.881238922161663 1.3567980021555
0.897866071636411 1.39171819951174
0.914493221111159 1.42472647042224
0.931120370585908 1.45585408968624
0.947747520060656 1.48513233210294
0.964374669535404 1.51259247247157
0.981001819010153 1.53826578559135
0.997628968484901 1.56218354626152
1.01425611795965 1.58437702928128
1.0308832674344 1.60487750944986
1.04751041690915 1.62371626156649
1.06413756638389 1.64092456043039
1.08076471585864 1.65653368084078
1.09739186533339 1.67057489759688
1.11401901480814 1.68307948549792
1.13064616428289 1.69407871934311
1.14727331375764 1.70360387393169
1.16390046323238 1.71168622406287
1.18052761270713 1.71835704453588
1.19715476218188 1.72364761014994
1.21378191165663 1.72758919570427
1.23040906113138 1.7302130759981
1.24703621060613 1.73155052583064
1.26366336008087 1.73163282000113
1.28029050955562 1.73049123330878
1.29691765903037 1.72815704055281
1.31354480850512 1.72466151653246
1.33017195797987 1.72003593604693
1.34679910745462 1.71431157389546
1.36342625692936 1.70751970487727
1.38005340640411 1.69969160379158
1.39668055587886 1.69085854543761
1.41330770535361 1.68105180461458
1.42993485482836 1.67030265612172
1.44656200430311 1.65864237475826
1.46318915377785 1.64610223532341
1.4798163032526 1.63271351261639
1.49644345272735 1.61850748143643
1.5130706022021 1.60351541658276
1.52969775167685 1.58776859285459
1.5463249011516 1.57129828505114
1.56295205062634 1.55413576797165
1.57957920010109 1.53631231641533
1.59620634957584 1.5178592051814
1.61283349905059 1.49880770906909
1.62946064852534 1.47918910287763
};
\addplot [thick, color1, mark=square, mark size=4, mark options={solid,fill opacity=0}, only marks]
table {%
0.488838194557601 0.994249143277713
0.651784259410135 1.08202394725889
0.814730324262669 1.19054552309017
0.977676389115203 1.40918458027965
1.14062245396774 1.61346048772675
1.30356851882027 1.66133765353467
1.4665145836728 1.61026867667289
1.62946064852534 1.49376757320696
1.79240671337787 1.29108757128679
};
\addlegendentry{$0.37$}
\addplot [thick, color1, dash pattern=on 1pt off 3pt on 3pt off 3pt, forget plot]
table {%
0.814730324262669 1.18338928005274
0.831357473737417 1.21276380740252
0.847984623212166 1.24125048841894
0.864611772686914 1.26884442463947
0.881238922161663 1.29554071760159
0.897866071636411 1.32133446884277
0.914493221111159 1.34622077990047
0.931120370585908 1.37019475231218
0.947747520060656 1.39325148761535
0.964374669535404 1.41538608734747
0.981001819010153 1.436593653046
0.997628968484901 1.45686928624841
1.01425611795965 1.47620808849218
1.0308832674344 1.49460516131478
1.04751041690915 1.51205560625367
1.06413756638389 1.52855452484633
1.08076471585864 1.54409701863022
1.09739186533339 1.55867818914283
1.11401901480814 1.57229313792163
1.13064616428289 1.58493696650407
1.14727331375764 1.59660477642764
1.16390046323238 1.6072916692298
1.18052761270713 1.61699274644803
1.19715476218188 1.6257031096198
1.21378191165663 1.63341786028257
1.23040906113138 1.64013209997383
1.24703621060613 1.64584093023103
1.26366336008087 1.65053945259166
1.28029050955562 1.65422276859318
1.29691765903037 1.65688597977307
1.31354480850512 1.65852418766879
1.33017195797987 1.65913249381781
1.34679910745462 1.65870599975762
1.36342625692936 1.65723980702567
1.38005340640411 1.65472901715944
1.39668055587886 1.65116873169641
1.41330770535361 1.64655405217403
1.42993485482836 1.64088008012979
1.44656200430311 1.63414191710116
1.46318915377785 1.6263346646256
1.4798163032526 1.61745342424058
1.49644345272735 1.60749329748358
1.5130706022021 1.59644938589208
1.52969775167685 1.58431679100353
1.5463249011516 1.57109061435541
1.56295205062634 1.5567659574852
1.57957920010109 1.54133792193036
1.59620634957584 1.52480160922836
1.61283349905059 1.50715212091669
1.62946064852534 1.48838455853279
};
\addplot [thick, color2, mark=triangle, mark size=4, mark options={solid,rotate=270,fill opacity=0}, only marks]
table {%
0.488838194557601 0.995845048804643
0.651784259410135 1.04850993119335
0.814730324262669 1.13309292412067
0.977676389115203 1.30704662655609
1.14062245396774 1.58792599929586
1.30356851882027 1.58792599929586
1.4665145836728 1.58792599929586
1.62946064852534 1.49695938426083
1.79240671337787 1.30864253208302
};
\addlegendentry{$0.46$}
\addplot [thick, color2, dashed, forget plot]
table {%
0.814730324262669 1.11836246278876
0.831357473737417 1.14617664142069
0.847984623212166 1.17328410379139
0.864611772686914 1.19967618339022
0.881238922161663 1.22534421370657
0.897866071636411 1.25027952822981
0.914493221111159 1.2744734604493
0.931120370585908 1.29791734385442
0.947747520060656 1.32060251193454
0.964374669535404 1.34252029817903
0.981001819010153 1.36366203607727
0.997628968484901 1.38401905911862
1.01425611795965 1.40358270079247
1.0308832674344 1.42234429458817
1.04751041690915 1.44029517399511
1.06413756638389 1.45742667250265
1.08076471585864 1.47373012360016
1.09739186533339 1.48919686077703
1.11401901480814 1.50381821752262
1.13064616428289 1.5175855273263
1.14727331375764 1.53049012367744
1.16390046323238 1.54252334006542
1.18052761270713 1.55367650997961
1.19715476218188 1.56394096690938
1.21378191165663 1.5733080443441
1.23040906113138 1.58176907577315
1.24703621060613 1.5893153946859
1.26366336008087 1.59593833457171
1.28029050955562 1.60162922891996
1.29691765903037 1.60637941122003
1.31354480850512 1.61018021496128
1.33017195797987 1.61302297363309
1.34679910745462 1.61489902072483
1.36342625692936 1.61579968972586
1.38005340640411 1.61571631412557
1.39668055587886 1.61464022741333
1.41330770535361 1.6125627630785
1.42993485482836 1.60947525461046
1.44656200430311 1.60536903549859
1.46318915377785 1.60023543923224
1.4798163032526 1.5940657993008
1.49644345272735 1.58685144919364
1.5130706022021 1.57858372240012
1.52969775167685 1.56925395240963
1.5463249011516 1.55885347271153
1.56295205062634 1.54737361679519
1.57957920010109 1.53480571814999
1.59620634957584 1.5211411102653
1.61283349905059 1.50637112663049
1.62946064852534 1.49048710073494
};
\addplot [thick, color3, mark=diamond, mark size=4, mark options={solid,fill opacity=0}, only marks]
table {%
0.488838194557601 0.98786552116999
0.651784259410135 1.06606489198959
0.814730324262669 1.15224379044383
0.977676389115203 1.25916946074818
1.14062245396774 1.53047340032637
1.30356851882027 1.52887749479944
1.4665145836728 1.57196694402656
1.62946064852534 1.49217166768003
1.79240671337787 1.29906709892144
};
\addlegendentry{$0.54$}
\addplot [thick, color3, dashed, forget plot]
table {%
0.814730324262669 1.13769065194825
0.831357473737417 1.15594213116034
0.847984623212166 1.1741777280497
0.864611772686914 1.19236889965196
0.881238922161663 1.21048710300279
0.897866071636411 1.22850379513782
0.914493221111159 1.24639043309272
0.931120370585908 1.26411847390312
0.947747520060656 1.28165937460468
0.964374669535404 1.29898459223304
0.981001819010153 1.31606558382386
0.997628968484901 1.33287380641278
1.01425611795965 1.34938071703546
1.0308832674344 1.36555777272754
1.04751041690915 1.38137643052467
1.06413756638389 1.39680814746251
1.08076471585864 1.41182438057669
1.09739186533339 1.42639658690287
1.11401901480814 1.4404962234767
1.13064616428289 1.45409474733383
1.14727331375764 1.4671636155099
1.16390046323238 1.47967428504057
1.18052761270713 1.49159821296148
1.19715476218188 1.50290685630829
1.21378191165663 1.51357167211664
1.23040906113138 1.52356411742218
1.24703621060613 1.53285564926057
1.26366336008087 1.54141772466744
1.28029050955562 1.54922180067845
1.29691765903037 1.55623933432925
1.31354480850512 1.56244178265549
1.33017195797987 1.56780060269282
1.34679910745462 1.57228725147688
1.36342625692936 1.57587318604333
1.38005340640411 1.57852986342781
1.39668055587886 1.58022874066597
1.41330770535361 1.58094127479347
1.42993485482836 1.58063892284595
1.44656200430311 1.57929314185906
1.46318915377785 1.57687538886845
1.4798163032526 1.57335712090976
1.49644345272735 1.56870979501866
1.5130706022021 1.56290486823078
1.52969775167685 1.55591379758178
1.5463249011516 1.5477080401073
1.56295205062634 1.538259052843
1.57957920010109 1.52753829282453
1.59620634957584 1.51551721708752
1.61283349905059 1.50216728266764
1.62946064852534 1.48745994660052
};
\addplot [thick, color4, mark=triangle, mark size=4, mark options={solid,rotate=90,fill opacity=0}, only marks]
table {%
0.488838194557601 0.957543316158311
0.651784259410135 1.03574268697791
0.814730324262669 1.13788064070146
0.977676389115203 1.24640221653273
1.14062245396774 1.50653481742241
1.30356851882027 1.47780851793766
1.4665145836728 1.54164473901488
1.62946064852534 1.47780851793766
1.79240671337787 1.47780851793766
};
\addlegendentry{$0.61$}
\addplot [thick, color4, dashed, forget plot]
table {%
0.814730324262669 1.12436610739007
0.831357473737417 1.14343483478633
0.847984623212166 1.16224087270937
0.864611772686914 1.18076933736918
0.881238922161663 1.19900534497577
0.897866071636411 1.21693401173916
0.914493221111159 1.23454045386936
0.931120370585908 1.25180978757636
0.947747520060656 1.26872712907019
0.964374669535404 1.28527759456085
0.981001819010153 1.30144630025834
0.997628968484901 1.31721836237268
1.01425611795965 1.33257889711387
1.0308832674344 1.34751302069192
1.04751041690915 1.36200584931685
1.06413756638389 1.37604249919865
1.08076471585864 1.38960808654735
1.09739186533339 1.40268772757293
1.11401901480814 1.41526653848543
1.13064616428289 1.42732963549483
1.14727331375764 1.43886213481116
1.16390046323238 1.44984915264442
1.18052761270713 1.46027580520462
1.19715476218188 1.47012720870176
1.21378191165663 1.47938847934586
1.23040906113138 1.48804473334693
1.24703621060613 1.49608108691496
1.26366336008087 1.50348265625998
1.28029050955562 1.51023455759198
1.29691765903037 1.51632190712099
1.31354480850512 1.521729821057
1.33017195797987 1.52644341561002
1.34679910745462 1.53044780699007
1.36342625692936 1.53372811140715
1.38005340640411 1.53626944507126
1.39668055587886 1.53805692419243
1.41330770535361 1.53907566498066
1.42993485482836 1.53931078364594
1.44656200430311 1.53874739639831
1.46318915377785 1.53737061944775
1.4798163032526 1.53516556900429
1.49644345272735 1.53211736127792
1.5130706022021 1.52821111247867
1.52969775167685 1.52343193881653
1.5463249011516 1.51776495650151
1.56295205062634 1.51119528174362
1.57957920010109 1.50370803075288
1.59620634957584 1.49528831973929
1.61283349905059 1.48592126491285
1.62946064852534 1.47559198248359
};
\addplot [thick, color5, mark=pentagon, mark size=4, mark options={solid,fill opacity=0}, only marks]
table {%
0.488838194557601 0.914453866931187
0.651784259410135 0.991057332223851
0.814730324262669 1.07404441962424
0.977676389115203 1.18735371203631
1.14062245396774 1.33736883156777
1.30356851882027 1.4171641079143
1.4665145836728 1.46184946266835
1.62946064852534 1.41556820238737
1.79240671337787 1.3389647370947
};
\addlegendentry{$0.67$}
\addplot [thick, color5, dash pattern=on 1pt off 3pt on 3pt off 3pt, forget plot]
table {%
0.814730324262669 1.07129591566119
0.831357473737417 1.08327026760176
0.847984623212166 1.09553059555282
0.864611772686914 1.10804543631274
0.881238922161663 1.12078332667988
0.897866071636411 1.13371280345262
0.914493221111159 1.14680240342934
0.931120370585908 1.1600206634084
0.947747520060656 1.17333612018817
0.964374669535404 1.18671731056704
0.981001819010153 1.20013277134336
0.997628968484901 1.21355103931552
1.01425611795965 1.22694065128189
1.0308832674344 1.24027014404084
1.04751041690915 1.25350805439073
1.06413756638389 1.26662291912995
1.08076471585864 1.27958327505687
1.09739186533339 1.29235765896985
1.11401901480814 1.30491460766727
1.13064616428289 1.31722265794751
1.14727331375764 1.32925034660892
1.16390046323238 1.3409662104499
1.18052761270713 1.3523387862688
1.19715476218188 1.36333661086401
1.21378191165663 1.37392822103389
1.23040906113138 1.38408215357682
1.24703621060613 1.39376694529116
1.26366336008087 1.4029511329753
1.28029050955562 1.4116032534276
1.29691765903037 1.41969184344643
1.31354480850512 1.42718543983017
1.33017195797987 1.43405257937719
1.34679910745462 1.44026179888587
1.36342625692936 1.44578163515456
1.38005340640411 1.45058062498166
1.39668055587886 1.45462730516552
1.41330770535361 1.45789021250453
1.42993485482836 1.46033788379705
1.44656200430311 1.46193885584145
1.46318915377785 1.46266166543611
1.4798163032526 1.46247484937941
1.49644345272735 1.4613469444697
1.5130706022021 1.45924648750538
1.52969775167685 1.45614201528479
1.5463249011516 1.45200206460633
1.56295205062634 1.44679517226837
1.57957920010109 1.44048987506926
1.59620634957584 1.43305470980739
1.61283349905059 1.42445821328113
1.62946064852534 1.41466892228886
};
\draw[ultra thick] (axis cs:1.21401901480814,1.83534423223549) node[
  scale=0.9,
  anchor=base west,
  text=black,
  rotate=0.0
]{$\left(f^+_{\max},\delta_{\max}^+\right)$};
\end{axis}

\end{tikzpicture}

%% file: tikz_matplotlib/delta_freq_kynematics.tikz
\begin{tikzpicture}

\begin{axis}[name=principal,thick,
tick align=outside,
tick pos=left,
x grid style={white!69.0196078431373!black},
xlabel={\(\displaystyle {f^+}_{\max}\)},
xmin=1.09656050785965, xmax=1.48064766072634,
xtick style={color=black},
xtick={1.05,1.1,1.15,1.2,1.25,1.3,1.35,1.4,1.45,1.5},
xticklabels={1.05,1.10,1.15,1.20,1.25,1.30,1.35,1.40,1.45,1.50},
y grid style={white!69.0196078431373!black},
ylabel={\(\displaystyle {\delta^+}_{\max}\)},
ymin=1.44402753709615, ymax=1.85397836057547,
ytick style={color=black},
ytick={1.4,1.45,1.5,1.55,1.6,1.65,1.7,1.75,1.8,1.85,1.9},
yticklabels={1.40,1.45,1.50,1.55,1.60,1.65,1.70,1.75,1.80,1.85,1.90}
]
\addplot [thick, black, mark=square*, mark size=3, mark options={solid}, only marks]
table {%
	1.11401901480814 1.8353442322355
	1.26366336008087 1.73163282000112
	1.33017195797987 1.65913249381781
	1.36342625692936 1.61579968972586
	1.41330770535361 1.58094127479347
	1.42993485482836 1.53931078364594
	1.46318915377785 1.46266166543612
};
\addplot [thick, black, dashed]
table {%
	1.11401901480814 1.83361852674662
	1.1211449360116 1.83064686429962
	1.12827085721507 1.82749820407706
	1.13539677841853 1.82417254607894
	1.14252269962199 1.82066989030526
	1.14964862082546 1.81699023675601
	1.15677454202892 1.81313358543121
	1.16390046323238 1.80909993633084
	1.17102638443585 1.80488928945492
	1.17815230563931 1.80050164480343
	1.18527822684278 1.79593700237638
	1.19240414804624 1.79119536217377
	1.1995300692497 1.7862767241956
	1.20665599045317 1.78118108844187
	1.21378191165663 1.77590845491258
	1.22090783286009 1.77045882360772
	1.22803375406356 1.76483219452731
	1.23515967526702 1.75902856767133
	1.24228559647048 1.75304794303979
	1.24941151767395 1.7468903206327
	1.25653743887741 1.74055570045004
	1.26366336008087 1.73404408249182
	1.27078928128434 1.72735546675804
	1.2779152024878 1.72048985324869
	1.28504112369127 1.71344724196379
	1.29216704489473 1.70622763290333
	1.29929296609819 1.6988310260673
	1.30641888730166 1.69125742145572
	1.31354480850512 1.68350681906857
	1.32067072970858 1.67557921890586
	1.32779665091205 1.66747462096759
	1.33492257211551 1.65919302525376
	1.34204849331897 1.65073443176437
	1.34917441452244 1.64209884049942
	1.3563003357259 1.63328625145891
	1.36342625692936 1.62429666464283
	1.37055217813283 1.6151300800512
	1.37767809933629 1.605786497684
	1.38480402053976 1.59626591754125
	1.39192994174322 1.58656833962293
	1.39905586294668 1.57669376392905
	1.40618178415015 1.56664219045961
	1.41330770535361 1.55641361921461
	1.42043362655707 1.54600805019404
	1.42755954776054 1.53542548339792
	1.434685468964 1.52466591882624
	1.44181139016746 1.51372935647899
	1.44893731137093 1.50261579635619
	1.45606323257439 1.49132523845782
	1.46318915377785 1.47985768278389
};
\coordinate (insetPosition) at (rel axis cs:0.05,0.05);
\end{axis}
\begin{axis}[at={(insetPosition)},width=4.6cm, thick,
label style={font=\small},
tick label style={font=\scriptsize},
tick align=outside,
tick pos=right,
x grid style={white!69.0196078431373!black},
xlabel={\(\displaystyle U_R = U / (\omega_n L)\)},
xmajorgrids,
xmin=-0.0437137731822493, xmax=0.917989236827235,
xtick style={color=black},
xtick={-0.2,0,0.2,0.4,0.6,0.8,1},
xticklabels={−0.2,0.0,0.2,0.4,0.6,0.8,1.0},
y grid style={white!69.0196078431373!black},
ylabel={\(\displaystyle {\delta^+}_{\max}\)},
ymajorgrids,
ymin=1.36625304950016, ymax=1.88027446379534,
ytick style={color=black},
ytick={1.3,1.4,1.5,1.6,1.7,1.8,1.9},
yticklabels={1.3,1.4,1.5,1.6,1.7,1.8,1.9}
]
\addplot [thick, black, mark=square*, mark size=3, mark options={solid}, only marks]
table {%
	0 1.8353442322355
	0.287660588619901 1.73163282000112
	0.370481023980761 1.65913249381781
	0.463101279975952 1.61579968972586
	0.537197484772104 1.58094127479347
	0.611293689568256 1.53931078364594
	0.674275463644986 1.46266166543612
};
\addplot [thick, black, dashed]
table {%
	0 1.85690985405465
	0.0178423564009181 1.84737327865029
	0.0356847128018362 1.83783670324593
	0.0535270692027542 1.82830012784156
	0.0713694256036723 1.8187635524372
	0.0892117820045904 1.80922697703284
	0.107054138405508 1.79969040162847
	0.124896494806427 1.79015382622411
	0.142738851207345 1.78061725081975
	0.160581207608263 1.77108067541538
	0.178423564009181 1.76154410001102
	0.196265920410099 1.75200752460666
	0.214108276811017 1.74247094920229
	0.231950633211935 1.73293437379793
	0.249792989612853 1.72339779839357
	0.267635346013771 1.7138612229892
	0.285477702414689 1.70432464758484
	0.303320058815607 1.69478807218048
	0.321162415216525 1.68525149677611
	0.339004771617444 1.67571492137175
	0.356847128018362 1.66617834596739
	0.37468948441928 1.65664177056302
	0.392531840820198 1.64710519515866
	0.410374197221116 1.6375686197543
	0.428216553622034 1.62803204434993
	0.446058910022952 1.61849546894557
	0.46390126642387 1.60895889354121
	0.481743622824788 1.59942231813684
	0.499585979225706 1.58988574273248
	0.517428335626624 1.58034916732812
	0.535270692027542 1.57081259192375
	0.553113048428461 1.56127601651939
	0.570955404829379 1.55173944111503
	0.588797761230297 1.54220286571066
	0.606640117631215 1.5326662903063
	0.624482474032133 1.52312971490194
	0.642324830433051 1.51359313949757
	0.660167186833969 1.50405656409321
	0.678009543234887 1.49451998868885
	0.695851899635805 1.48498341328448
	0.713694256036723 1.47544683788012
	0.731536612437641 1.46591026247576
	0.749378968838559 1.45637368707139
	0.767221325239478 1.44683711166703
	0.785063681640396 1.43730053626267
	0.802906038041314 1.4277639608583
	0.820748394442232 1.41822738545394
	0.83859075084315 1.40869081004958
	0.856433107244068 1.39915423464521
	0.874275463644986 1.38961765924085
};
\end{axis}

\end{tikzpicture}

%% file: tikz_matplotlib/resumen_escala2.tikz
\begin{tikzpicture}

\definecolor{color0}{rgb}{0.12156862745098,0.466666666666667,0.705882352941177}
\definecolor{color1}{rgb}{1,0.498039215686275,0.0549019607843137}
\definecolor{color2}{rgb}{0.172549019607843,0.627450980392157,0.172549019607843}
\definecolor{color3}{rgb}{0.83921568627451,0.152941176470588,0.156862745098039}
\definecolor{color4}{rgb}{0.580392156862745,0.403921568627451,0.741176470588235}
\definecolor{color5}{rgb}{0.549019607843137,0.337254901960784,0.294117647058824}

\begin{axis}[thick,
legend cell align={left},
legend columns=2,
legend style={fill opacity=0.8, draw opacity=1, text opacity=1, at={(0.97,0.03)}, anchor=south east, draw=white!80!black},
tick align=outside,
tick pos=left,
x grid style={white!69.0196078431373!black},
xlabel={\(\displaystyle f^+/f^+_{\max}\)},
xmajorgrids,
xmin=0.270347693351425, xmax=1.67269843962008,
xtick style={color=black},
xtick={0.2,0.4,0.6,0.8,1,1.2,1.4,1.6,1.8},
xticklabels={0.2,0.4,0.6,0.8,1.0,1.2,1.4,1.6,1.8},
y grid style={white!69.0196078431373!black},
ylabel={\(\displaystyle \delta^+  /\delta^+_{\max}\)},
ymajorgrids,
ymin=0.536081213672941, ymax=1.02367980641164,
ytick style={color=black}
]
\addlegendimage{empty legend}
\addlegendentry{\hspace{-0cm}\textbf{$\bm U_R$}}
\addplot [thick, black, mark=triangle, mark size=3, mark options={solid,fill opacity=0}, only marks]
table {%
0.438805970149254 0.558244786070154
0.585074626865672 0.622590758296309
0.73134328358209 0.683458569861591
0.877611940298507 0.952146480914049
1.02388059701493 0.995623489174964
1.17014925373134 0.920843034966189
1.31641791044776 0.837367179105232
1.46268656716418 0.755630403574711
1.6089552238806 0.721718337131197
};
\addlegendentry{$0.00$}
\addplot [thick, color0, mark=*, mark size=3, mark options={solid,fill opacity=0}, only marks]
table {%
0.386842105263158 0.568638858505515
0.51578947368421 0.623014373338296
0.644736842105263 0.693979028289551
0.773684210526316 0.87645956959278
0.902631578947368 0.992583550422107
1.03157894736842 0.993505169317578
1.16052631578947 0.947424224544035
1.28947368421053 0.85526233499695
1.41842105263158 0.742925702802867
};
\addlegendentry{$0.29$}
\addplot [thick, color1, mark=square, mark size=3, mark options={solid,fill opacity=0}, only marks]
table {%
0.3675 0.599258435949172
0.49 0.652162471225584
0.6125 0.717571096658239
0.735 0.849350239074028
0.8575 0.972472357535495
0.98 1.0013291040499
1.1025 0.970548574434535
1.225 0.900330491249479
1.3475 0.778170264338492
};
\addlegendentry{$0.37$}
\addplot [thick, color2, mark=triangle, mark size=3, mark options={solid,rotate=270,fill opacity=0}, only marks]
table {%
0.358536585365854 0.616317143230544
0.478048780487805 0.648910838305236
0.597560975609756 0.70125828797065
0.717073170731707 0.808916250490088
0.836585365853659 0.982749290888447
0.95609756097561 0.982749290888447
1.07560975609756 0.982749290888447
1.19512195121951 0.926451090304888
1.31463414634146 0.809903938219624
};
\addlegendentry{$0.46$}
\addplot [thick, color3, mark=diamond, mark size=3, mark options={solid,fill opacity=0}, only marks]
table {%
0.345882352941176 0.624859086748204
0.461176470588235 0.674322891676576
0.576470588235294 0.728834023638455
0.691764705882353 0.796468205887453
0.807058823529412 0.968077325026701
0.92235294117647 0.96706785962
1.03764705882353 0.994323425600939
1.15294117647059 0.943850155265866
1.26823529411765 0.821704841054989
};
\addlegendentry{$0.54$}
\addplot [thick, color4, mark=triangle, mark size=3, mark options={solid,rotate=90,fill opacity=0}, only marks]
table {%
0.341860465116279 0.622059772679767
0.455813953488372 0.672861320781948
0.569767441860465 0.739214363201123
0.683720930232558 0.809714470771497
0.797674418604651 0.978707375682833
0.911627906976744 0.96004558250244
1.02558139534884 1.00151623401443
1.13953488372093 0.96004558250244
1.25348837209302 0.96004558250244
};
\addlegendentry{$0.61$}
\addplot [thick, color5, mark=pentagon, mark size=3, mark options={solid,fill opacity=0}, only marks]
table {%
0.334090909090909 0.625198491585906
0.445454545454545 0.67757114009572
0.556818181818182 0.734308175981352
0.668181818181818 0.811776051902119
0.779545454545455 0.914339155233838
0.890909090909091 0.968893997431561
1.00227272727273 0.999444709062286
1.11363636363636 0.967802900587606
1.225 0.915430252077792
};
\addlegendentry{$0.67$}
\end{axis}

\end{tikzpicture}

%% file: tikz_matplotlib/strouhal_scale.tikz
\begin{tikzpicture}

\definecolor{color0}{rgb}{0.12156862745098,0.466666666666667,0.705882352941177}
\definecolor{color1}{rgb}{1,0.498039215686275,0.0549019607843137}
\definecolor{color2}{rgb}{0.172549019607843,0.627450980392157,0.172549019607843}
\definecolor{color3}{rgb}{0.83921568627451,0.152941176470588,0.156862745098039}
\definecolor{color4}{rgb}{0.580392156862745,0.403921568627451,0.741176470588235}
\definecolor{color5}{rgb}{0.549019607843137,0.337254901960784,0.294117647058824}

\begin{axis}[name=principal,thick,
legend cell align={left},
legend columns=2,
legend style={fill opacity=0.8, draw opacity=1, text opacity=1,legend columns=-1,at={(0.35,0.75)},anchor=east,  draw=white!80!black},
tick align=outside,
tick pos=left,
x grid style={white!69.0196078431373!black},
xlabel={\(\displaystyle {\rm St} \)},
xmin=0, xmax=0.4,
xtick style={color=black},
xtick={0,0.05,0.1,0.15,0.2,0.25,0.3,0.35,0.4},
xticklabels={0.00,0.05,0.10,0.15,0.20,0.25,0.30,0.35,0.40},
y grid style={white!69.0196078431373!black},
ylabel={\(\displaystyle {\rm St^*}\)},
ymin=0, ymax=0.500086892625447,
ytick style={color=black},
ytick={0,0.1,0.2,0.3,0.4,0.5,0.6},
yticklabels={0.0,0.1,0.2,0.3,0.4,0.5,0.6}
]
\addlegendimage{empty legend}
\addlegendentry{\hspace{-0cm}\textbf{$\bm U_R$}}
\addplot [thick, color0, mark=*, mark size=3, mark options={solid,fill opacity=0}, only marks]
table {%
0.096841090403167 0.095356875778978
0.129121453870889 0.139300373909431
0.161401817338612 0.193959285417532
0.193682180806334 0.293952638138762
0.225962544274056 0.38838221852929
0.258242907741779 0.444277523888658
0.290523271209501 0.476629829663482
0.322803634677224 0.478072289156627
0.355083998144946 0.45680651955155
};
\addlegendentry{0.29}
\addplot [thick, color1, mark=square, mark size=3, mark options={solid,fill opacity=0}, only marks]
table {%
0.0751924208388461 0.07476
0.100256561118462 0.10848
0.125320701398077 0.1492
0.150384841677692 0.21192
0.175448981957308 0.28308
0.200513122236923 0.33312
0.225577262516538 0.36324
0.250641402796154 0.3744
0.275705543075769 0.35596
};
\addlegendentry{0.37}
\addplot [thick, color2, mark=triangle, mark size=3, mark options={solid,rotate=270,fill opacity=0}, only marks]
table {%
0.0601539366710769 0.059904
0.0802052488947692 0.084096
0.100256561118462 0.1136
0.120307873342154 0.157248
0.140359185565846 0.22288
0.160410497789538 0.25472
0.180461810013231 0.28656
0.200513122236923 0.30016
0.220564434460615 0.28864
};
\addlegendentry{0.46}
\addplot [thick, color3, mark=diamond, mark size=3, mark options={solid,fill opacity=0}, only marks]
table {%
0.0518568419578249 0.0512275862068966
0.0691424559437666 0.0737103448275862
0.0864280699297082 0.0995862068965517
0.10371368391565 0.130593103448276
0.120999297901592 0.185186206896552
0.138284911887533 0.211420689655172
0.155570525873475 0.244551724137931
0.172856139859416 0.257931034482759
0.190141753845358 0.247006896551724
};
\addlegendentry{0.54}
\addplot [thick, color4, mark=triangle, mark size=3, mark options={solid,rotate=90,fill opacity=0}, only marks]
table {%
0.0455711641447552 0.0436363636363636
0.060761552193007 0.0629333333333333
0.0759519402412587 0.0864242424242424
0.0911423282895105 0.1136
0.106332716337762 0.160193939393939
0.121523104386014 0.179587878787879
0.136713492434266 0.210763636363636
0.151903880482517 0.224484848484848
0.167094268530769 0.246933333333333
};
\addlegendentry{0.61}
\addplot [thick, color5, mark=pentagon, mark size=3, mark options={solid,fill opacity=0}, only marks]
table {%
0.041314516944421 0.0377802197802198
0.0550860225925613 0.0545934065934066
0.0688575282407016 0.073956043956044
0.0826290338888419 0.0981098901098901
0.0964005395369823 0.128923076923077
0.110172045185123 0.156131868131868
0.123943550833263 0.181186813186813
0.137715056481403 0.194945054945055
0.151486562129544 0.202835164835165
};
\addlegendentry{0.67}
\addplot [thick, black, dashed, forget plot]
table {%
0 0
0.0172413793103448 0.0172413793103448
0.0344827586206897 0.0344827586206897
0.0517241379310345 0.0517241379310345
0.0689655172413793 0.0689655172413793
0.0862068965517241 0.0862068965517241
0.103448275862069 0.103448275862069
0.120689655172414 0.120689655172414
0.137931034482759 0.137931034482759
0.155172413793103 0.155172413793103
0.172413793103448 0.172413793103448
0.189655172413793 0.189655172413793
0.206896551724138 0.206896551724138
0.224137931034483 0.224137931034483
0.241379310344828 0.241379310344828
0.258620689655172 0.258620689655172
0.275862068965517 0.275862068965517
0.293103448275862 0.293103448275862
0.310344827586207 0.310344827586207
0.327586206896552 0.327586206896552
0.344827586206897 0.344827586206897
0.362068965517241 0.362068965517241
0.379310344827586 0.379310344827586
0.396551724137931 0.396551724137931
0.413793103448276 0.413793103448276
0.431034482758621 0.431034482758621
0.448275862068966 0.448275862068966
0.46551724137931 0.46551724137931
0.482758620689655 0.482758620689655
0.5 0.5
};
\coordinate (insetPosition) at (rel axis cs:0.5,0.05);

\end{axis}
\begin{axis}[at={(insetPosition)},xshift=0cm,width=4.6cm, thick,
label style={font=\small},
tick label style={font=\scriptsize},
legend cell align={left},
legend columns=2,
legend style={fill opacity=0.8, draw opacity=1, text opacity=1,legend columns=-1,at={(0.35,0.75)},anchor=east,  draw=white!80!black},
tick align=inside,
tick pos=right,
x grid style={white!69.0196078431373!black},
xlabel={\(~~~~~~~~~~~~~~~\displaystyle St /C_Y^*\)},
xmajorgrids,
xmin=0, xmax=5.29239618980983,
xtick style={color=black},
y grid style={white!69.0196078431373!black},
ylabel={\(\displaystyle St^* /C_Y^*\)},
ymajorgrids,
ymin=0, ymax=5.42030030043181,
ytick style={color=black}
]
\addplot [thick, color0, mark=*, mark size=3, mark options={solid,fill opacity=0}, only marks]
table {%
	5.04486819105398 5.16882740445044
	3.0223413963961 3.45353387856608
	1.90836097320349 2.39381316606797
	0.424848932375529 0.742430541421057
	0.355884316394134 0.650311620866074
	0.28158404864691 0.475895050363408
	0.391744790702826 0.60205573259361
	0.406612947543162 0.563907873915602
	0.389764380699777 0.516282517368322
};

\addplot [thick, color1, mark=square, mark size=3, mark options={solid,fill opacity=0}, only marks]
table {%
	3.19635311949369 3.14736488501312
	2.10300380216019 2.26878808433178
	1.33667387931689 1.60630353947422
	0.608117638064367 0.922943882929924
	0.226775925107248 0.389780249576953
	0.230981363440657 0.397377140426456
	0.187487967684269 0.307591050208383
	0.190764311482734 0.282522008065769
	0.182283842210392 0.234503520196995
};
\addplot [thick, color2, mark=triangle, mark size=3, mark options={solid,rotate=270,fill opacity=0}, only marks]
table {%
	2.08246624553663 2.07049028052955
	1.46722600849003 1.58757367722729
	0.978784428425412 1.16528741933224
	0.530062711186624 0.746956199185415
	0.2696166008033 0.435015732231321
	0.211140823275506 0.350776199905907
	0.169400516515106 0.272780345556484
	0.185588456965028 0.277226018975876
	0.167130219131118 0.215779748706624
};
\addplot [thick, color3, mark=diamond, mark size=3, mark options={solid,fill opacity=0}, only marks]
table {%
	1.54469051500053 1.53827240129877
	1.24061750780909 1.30079977775017
	0.823798301473171 0.933440026301874
	0.267113362317823 0.349129619125567
	0.151343400115511 0.240322119865257
	0.140798744990202 0.22357798783817
	0.150418495575455 0.238853439899235
	0.151654147942485 0.227020099924583
	0.14752810743513 0.193061556067326
};
\addplot [thick, color4, mark=triangle, mark size=3, mark options={solid,rotate=90,fill opacity=0}, only marks]
table {%
	1.18035309503044 1.16603012538686
	0.891327662481479 0.950213128230648
	0.67028756083814 0.77233467978749
	0.459145064185749 0.578141442875957
	0.173666667516162 0.265792215156809
	0.115751315602599 0.17696958141824
	0.118763211098301 0.186691842012978
	0.105316189389063 0.157149833954385
	0.120586553093061 0.156650023695539
};
\addplot [thick, color5, mark=pentagon, mark size=3, mark options={solid,fill opacity=0}, only marks]
table {%
	0.957101443244318 0.91646608986407
	0.742771359436443 0.769320003632934
	0.524558690335876 0.596885178644905
	0.373900102864344 0.466029916971935
	0.158424955892882 0.238672712001236
	0.130244553738398 0.192476510929595
	0.120348782110658 0.185535066787744
	0.0994458032972689 0.146961855185857
	0.0943082159369332 0.139369484823104
};
\addplot [black, dashed, forget plot]
table {%
	0 0
	0.155172413793103 0.155172413793103
	0.310344827586207 0.310344827586207
	0.46551724137931 0.46551724137931
	0.620689655172414 0.620689655172414
	0.775862068965517 0.775862068965517
	0.931034482758621 0.931034482758621
	1.08620689655172 1.08620689655172
	1.24137931034483 1.24137931034483
	1.39655172413793 1.39655172413793
	1.55172413793103 1.55172413793103
	1.70689655172414 1.70689655172414
	1.86206896551724 1.86206896551724
	2.01724137931034 2.01724137931034
	2.17241379310345 2.17241379310345
	2.32758620689655 2.32758620689655
	2.48275862068966 2.48275862068966
	2.63793103448276 2.63793103448276
	2.79310344827586 2.79310344827586
	2.94827586206897 2.94827586206897
	3.10344827586207 3.10344827586207
	3.25862068965517 3.25862068965517
	3.41379310344828 3.41379310344828
	3.56896551724138 3.56896551724138
	3.72413793103448 3.72413793103448
	3.87931034482759 3.87931034482759
	4.03448275862069 4.03448275862069
	4.18965517241379 4.18965517241379
	4.3448275862069 4.3448275862069
	4.5 4.5
};
\end{axis}

\end{tikzpicture}

%% file: tikz_matplotlib/media_ux_0.29.tikz
\begin{tikzpicture}

\definecolor{color0}{rgb}{0.12156862745098,0.466666666666667,0.705882352941177}
\definecolor{color1}{rgb}{1,0.498039215686275,0.0549019607843137}
\definecolor{color2}{rgb}{0.172549019607843,0.627450980392157,0.172549019607843}
\definecolor{color3}{rgb}{0.83921568627451,0.152941176470588,0.156862745098039}
\definecolor{color4}{rgb}{0.580392156862745,0.403921568627451,0.741176470588235}
\definecolor{color5}{rgb}{0.549019607843137,0.337254901960784,0.294117647058824}
\definecolor{color6}{rgb}{0.890196078431372,0.466666666666667,0.76078431372549}
\definecolor{color7}{rgb}{0.737254901960784,0.741176470588235,0.133333333333333}
\definecolor{color8}{rgb}{0.0901960784313725,0.745098039215686,0.811764705882353}

\begin{axis}[thick,
legend cell align={left},
legend style={fill opacity=0.8, draw opacity=1, text opacity=1, draw=white!80!black,legend columns=-1,at={(0.5,1.2)},anchor=north},
tick align=outside,
tick pos=left,
x grid style={white!69.0196078431373!black},
xlabel={\(\displaystyle x/L\)},
xmajorgrids,
xmin=1, xmax=4.14356885084084,
xtick style={color=black},
y grid style={white!69.0196078431373!black},
ylabel={\(\displaystyle u_x-U\)},
ymajorgrids,
ymin=0, ymax=1,
ytick style={color=black}
]
\addlegendimage{empty legend}
\addlegendentry{\hspace{-0cm}\textbf{$\bm f^+$}}	
\addplot [thick, color0, forget plot]
table {%
0 0
};
\addplot [thick, color1, forget plot]
table {%
0 0
};
\addplot [thick, color2, forget plot]
table {%
0 0
};
\addplot [very thick, color3]
table {%
-0.715008099590029 -0.963895846575881
-0.691755757161549 -0.966966362598484
-0.668503414733069 -0.972275345003248
-0.645251072304589 -0.976292868469586
-0.621998729876109 -0.975175174166828
-0.59874638744763 -0.971483616833361
-0.57549404501915 -0.966555305967576
-0.55224170259067 -0.965616533115529
-0.52898936016219 -0.963474087744894
-0.50573701773371 -0.958277301655806
-0.48248467530523 -0.951827308790356
-0.45923233287675 -0.94835128399351
-0.435979990448271 -0.946414922976354
-0.412727648019791 -0.94749433668576
-0.389475305591311 -0.950335117581187
-0.366222963162831 -0.956641809745237
-0.342970620734351 -0.959124238645351
-0.319718278305871 -0.958385747816534
-0.296465935877391 -0.957172565573325
-0.273213593448911 -0.952530561078123
-0.249961251020432 -0.950977650499476
-0.226708908591952 -0.949634257151976
-0.203456566163472 -0.962136139042367
-0.180204223734992 -0.976501316447564
-0.156951881306512 -0.993391817326157
-0.133699538878032 -1.00658936558508
-0.110447196449552 -1.01450256819381
-0.0871948540210724 -1.01444515503926
-0.0639425115925925 -1.01023970157611
-0.0406901691641126 -1.01250293161446
-0.0174378267356328 -1.01703922618786
0.00581451569284712 -1.0200716574977
0.029066858121327 -1.01486078761601
0.0523192005498069 -1.01025364785918
0.0755715429782867 -1.00719352161305
0.0988238854067666 -1.00727815484044
0.122076227835246 -1.0040717294705
0.145328570263726 -1.00117426015857
0.168580912692206 -1.00080409701276
0.191833255120686 -1.00478727611259
0.215085597549166 -1.0056415216056
0.238337939977646 -1.00552694567502
0.261590282406126 -1.00564129082565
0.284842624834606 -1.00605311379421
0.308094967263085 -1.00584287176124
0.331347309691565 -1.00573027232966
0.354599652120045 -1.00747999296028
0.377851994548525 -1.00983992187765
0.401104336977005 -1.01264829373482
0.424356679405485 -1.01073103224914
0.447609021833965 -1.00337568776713
0.470861364262445 -0.990840238667184
0.494113706690925 -0.976863780611795
0.517366049119404 -0.965444092024907
0.540618391547884 -0.95158151358335
0.563870733976364 -0.927259075045274
0.587123076404844 -0.885295387342726
0.610375418833324 -0.825966271267613
0.633627761261804 -0.758705334322114
0.656880103690284 -0.681028519258641
0.680132446118764 -0.59355191646519
0.703384788547243 -0.506864114444586
0.726637130975723 -0.42959334109085
0.749889473404203 -0.361231325438838
0.773141815832683 -0.301819714580174
0.796394158261163 -0.25050855835674
0.819646500689643 -0.205954471907156
0.842898843118123 -0.158432293752873
0.866151185546603 -0.11344104814935
0.889403527975082 -0.0808441217082219
0.912655870403562 -0.0554386980985302
0.935908212832042 -0.0333268150414313
0.959160555260522 -0.0127672713532275
0.982412897689002 0.00518480801669052
1.00566524011748 0.0224877563056944
1.02891758254596 0.0452703015942826
1.05216992497444 0.0733698852862947
1.07542226740292 0.107397272138232
1.0986746098314 0.142121928224248
1.12192695225988 0.170855012778292
1.14517929468836 0.19574771795782
1.16843163711684 0.213591313595659
1.19168397954532 0.228223680641484
1.2149363219738 0.238390078176749
1.23818866440228 0.249182391480132
1.26144100683076 0.259103021072799
1.28469334925924 0.267332203504412
1.30794569168772 0.276064684651978
1.3311980341162 0.28545021324063
1.35445037654468 0.294798908340176
1.37770271897316 0.302479342747711
1.40095506140164 0.30895006311814
1.42420740383012 0.317726708223471
1.4474597462586 0.326968083269578
1.47071208868708 0.336025181157165
1.49396443111556 0.344429905964705
1.51721677354404 0.350888256362128
1.54046911597252 0.357217108342816
1.563721458401 0.363270036910837
1.58697380082948 0.369170581461344
1.61022614325796 0.372395060120355
1.63347848568644 0.372335350162243
1.65673082811492 0.371700040689501
1.6799831705434 0.372484925722829
1.70323551297188 0.375762873537072
1.72648785540036 0.37801239753771
1.74974019782884 0.379338436668698
1.77299254025732 0.377295368871661
1.7962448826858 0.374193210693482
1.81949722511428 0.370232615729507
1.84274956754276 0.368080044991558
1.86600190997124 0.367728093928072
1.88925425239972 0.368372247510013
1.9125065948282 0.36677336858911
1.93575893725668 0.361646501875607
1.95901127968516 0.356205597539825
1.98226362211364 0.351426467106133
2.00551596454212 0.350259952686987
2.0287683069706 0.349458094252538
2.05202064939908 0.350411330044016
2.07527299182756 0.351548800264154
2.09852533425604 0.351131786258752
2.12177767668452 0.347736225352856
2.145030019113 0.343260189489902
2.16828236154148 0.340012104342592
2.19153470396996 0.338891713578089
2.21478704639844 0.336759302086495
2.23803938882692 0.333701638656215
2.2612917312554 0.328013002194326
2.28454407368387 0.32089199809195
2.30779641611235 0.313735224436671
2.33104875854083 0.307383246334097
2.35430110096931 0.300499047395592
2.37755344339779 0.293017574884368
2.40080578582627 0.286026369493164
2.42405812825475 0.280398361419773
2.44731047068323 0.275379004699053
2.47056281311171 0.270500182935245
2.49381515554019 0.268956701724373
2.51706749796867 0.26639107602192
2.54031984039715 0.261714422141533
2.56357218282563 0.252250789306062
2.58682452525411 0.243231984495398
2.61007686768259 0.235303463423239
2.63332921011107 0.231909839872676
2.65658155253955 0.23052236059722
2.67983389496803 0.223630781166678
2.70308623739651 0.212134895106941
2.72633857982499 0.198752016236184
2.74959092225347 0.191341563242109
2.77284326468195 0.185440274239503
2.79609560711043 0.182201482682988
2.81934794953891 0.181904381729395
2.84260029196739 0.182576304167126
2.86585263439587 0.182936281047331
2.88910497682435 0.181336445974821
2.91235731925283 0.178967018302655
2.93560966168131 0.174941806031568
2.95886200410979 0.171698138671809
2.98211434653827 0.17051856994271
3.00536668896675 0.170212012593099
3.02861903139523 0.16983420752975
3.05187137382371 0.168356691841531
3.07512371625219 0.166995196144438
3.09837605868067 0.165129999916058
3.12162840110915 0.164011893397078
3.14488074353763 0.164401094918136
3.16813308596611 0.164381458831844
3.19138542839459 0.165099182535423
3.21463777082307 0.164103614868923
3.23789011325155 0.16449061640903
3.26114245568003 0.165191787451867
3.28439479810851 0.166386130583444
3.30764714053699 0.167455390824238
3.33089948296547 0.166928336332676
3.35415182539395 0.165319088720121
3.37740416782243 0.166237589337243
3.40065651025091 0.167384633814305
3.42390885267939 0.168160806697215
3.44716119510787 0.167713286936608
3.47041353753635 0.167985741078544
3.49366587996483 0.170266861301532
3.51691822239331 0.171792672554385
3.54017056482179 0.174720317819262
3.56342290725027 0.176742660324539
3.58667524967875 0.180244943077199
3.60992759210723 0.183975677267784
3.63317993453571 0.189730631867599
3.65643227696419 0.195623592314491
3.67968461939267 0.20234453532997
3.70293696182115 0.206180613116394
3.72618930424963 0.206679503068366
3.74944164667811 0.203800362833833
3.77269398910659 0.204129070973634
3.79594633153507 0.208126008189607
3.81919867396355 0.21517181718117
3.84245101639203 0.219807919076587
3.86570335882051 0.220163270992575
3.88895570124899 0.220449587098745
3.91220804367747 0.219974280293629
};
\addlegendentry{0.98}
\addplot [thick, color4, forget plot]
table {%
0 0
};
\addplot [thick, color5, forget plot]
table {%
0 0
};
\addplot [thick, color6, forget plot]
table {%
0 0
};
\addplot [very thick, white!49.8039215686275!black, dashed]
table {%
-0.715008099590029 -0.981785981978602
-0.691755757161549 -0.982486666060383
-0.668503414733069 -0.985307591296172
-0.645251072304589 -0.986686244392748
-0.621998729876109 -0.988356372076063
-0.59874638744763 -0.986504546779951
-0.57549404501915 -0.986595933892959
-0.55224170259067 -0.988432939594598
-0.52898936016219 -0.992188937801951
-0.50573701773371 -0.997115893299952
-0.48248467530523 -1.00202295296293
-0.45923233287675 -1.00001662372499
-0.435979990448271 -0.992030544049995
-0.412727648019791 -0.980387572719583
-0.389475305591311 -0.975716780423691
-0.366222963162831 -0.970255247321257
-0.342970620734351 -0.965157368059601
-0.319718278305871 -0.962957877148911
-0.296465935877391 -0.963199044599806
-0.273213593448911 -0.964063796828855
-0.249961251020432 -0.965206452714322
-0.226708908591952 -0.972313633998211
-0.203456566163472 -0.988544857057345
-0.180204223734992 -0.996675852439114
-0.156951881306512 -1.00062596825578
-0.133699538878032 -0.999184606789669
-0.110447196449552 -1.00022046594697
-0.0871948540210724 -1.000560170883
-0.0639425115925925 -0.999846530688531
-0.0406901691641126 -1.0052770153724
-0.0174378267356328 -1.01645501506946
0.00581451569284712 -1.01808252952032
0.029066858121327 -1.01558588932957
0.0523192005498069 -1.00646093041389
0.0755715429782867 -1.00670355716938
0.0988238854067666 -1.00732598514633
0.122076227835246 -1.00487135952643
0.145328570263726 -1.00183390752783
0.168580912692206 -1.00098897266374
0.191833255120686 -1.00471805139723
0.215085597549166 -1.00559316551383
0.238337939977646 -1.00523321496409
0.261590282406126 -1.00629844688427
0.284842624834606 -1.00760352429454
0.308094967263085 -1.0061928978545
0.331347309691565 -1.0025850290355
0.354599652120045 -1.00208287210039
0.377851994548525 -1.00573137939796
0.401104336977005 -1.00715743453353
0.424356679405485 -1.00341532893275
0.447609021833965 -0.992269339003155
0.470861364262445 -0.975754518011022
0.494113706690925 -0.95374649575209
0.517366049119404 -0.928614404583023
0.540618391547884 -0.89496341615726
0.563870733976364 -0.851417105117192
0.587123076404844 -0.788311570338941
0.610375418833324 -0.707792419170406
0.633627761261804 -0.614355966994244
0.656880103690284 -0.516246083989097
0.680132446118764 -0.421220043685195
0.703384788547243 -0.33199188992516
0.726637130975723 -0.251171056538357
0.749889473404203 -0.182011596311542
0.773141815832683 -0.113977426801609
0.796394158261163 -0.0512063971005404
0.819646500689643 0.000231069351320556
0.842898843118123 0.0364877499099048
0.866151185546603 0.0598533703314714
0.889403527975082 0.0804919654525855
0.912655870403562 0.0997772572253344
0.935908212832042 0.117823588216287
0.959160555260522 0.138469325883766
0.982412897689002 0.160750636632092
1.00566524011748 0.188361756082941
1.02891758254596 0.222448134121826
1.05216992497444 0.260743695064262
1.07542226740292 0.300762980287017
1.0986746098314 0.337469581435058
1.12192695225988 0.370037213353101
1.14517929468836 0.394909959530293
1.16843163711684 0.416810070800635
1.19168397954532 0.43198573845403
1.2149363219738 0.448312136088606
1.23818866440228 0.462513509724576
1.26144100683076 0.477889044455543
1.28469334925924 0.491416664856758
1.30794569168772 0.501581549044798
1.3311980341162 0.510205987883391
1.35445037654468 0.517839177564629
1.37770271897316 0.526036489403291
1.40095506140164 0.534335763904527
1.42420740383012 0.540317797801234
1.4474597462586 0.542874203576019
1.47071208868708 0.541565357335694
1.49396443111556 0.539532936149904
1.51721677354404 0.536922911442895
1.54046911597252 0.536150712476096
1.563721458401 0.536406064212152
1.58697380082948 0.537961594914769
1.61022614325796 0.53589544223546
1.63347848568644 0.531136951222951
1.65673082811492 0.524665923493313
1.6799831705434 0.521443592960709
1.70323551297188 0.521861240004659
1.72648785540036 0.523314869626362
1.74974019782884 0.526023946167109
1.77299254025732 0.528187159475958
1.7962448826858 0.531112791470113
1.81949722511428 0.532511961883547
1.84274956754276 0.534813878951134
1.86600190997124 0.538276033586884
1.88925425239972 0.54139779693967
1.9125065948282 0.542879701054658
1.93575893725668 0.541485022862704
1.95901127968516 0.540000237062675
1.98226362211364 0.537336549749367
2.00551596454212 0.533410577780893
2.0287683069706 0.528588991219265
2.05202064939908 0.524197964090799
2.07527299182756 0.521968511189539
2.09852533425604 0.518142519553662
2.12177767668452 0.511747337217064
2.145030019113 0.503072072041356
2.16828236154148 0.49409983487751
2.19153470396996 0.486647646148239
2.21478704639844 0.479135651862062
2.23803938882692 0.470754729152522
2.2612917312554 0.46344686442917
2.28454407368387 0.455366901654561
2.30779641611235 0.449042125100124
2.33104875854083 0.440710481347218
2.35430110096931 0.434353346719376
2.37755344339779 0.428507827995051
2.40080578582627 0.423829043673429
2.42405812825475 0.418346594352076
2.44731047068323 0.412100186256147
2.47056281311171 0.406080243193468
2.49381515554019 0.402423975218719
2.51706749796867 0.399767206234885
2.54031984039715 0.397806569133179
2.56357218282563 0.394458132654464
2.58682452525411 0.392544861224266
2.61007686768259 0.39172466162162
2.63332921011107 0.394861163211383
2.65658155253955 0.396703681567166
2.67983389496803 0.393364038500377
2.70308623739651 0.386132653229712
2.72633857982499 0.37834245080011
2.74959092225347 0.373763829455557
2.77284326468195 0.369459099870318
2.79609560711043 0.366097069168033
2.81934794953891 0.363288610541975
2.84260029196739 0.359179908732141
2.86585263439587 0.354081112729253
2.88910497682435 0.34775352333056
2.91235731925283 0.342225508906699
2.93560966168131 0.337576742722235
2.95886200410979 0.332587335614144
2.98211434653827 0.327649913464717
3.00536668896675 0.322273562522092
3.02861903139523 0.318507894675443
3.05187137382371 0.314505741780772
3.07512371625219 0.313246186841758
3.09837605868067 0.311731514669309
3.12162840110915 0.309208602207609
3.14488074353763 0.305154336873496
3.16813308596611 0.302073303848067
3.19138542839459 0.301263605915123
3.21463777082307 0.300516580294679
3.23789011325155 0.301373428059332
3.26114245568003 0.300227445765676
3.28439479810851 0.299113462877298
3.30764714053699 0.294095050283878
3.33089948296547 0.291023649902677
3.35415182539395 0.286012546088516
3.37740416782243 0.285650583156836
3.40065651025091 0.286331648535817
3.42390885267939 0.291648107786687
3.44716119510787 0.29630554172207
3.47041353753635 0.300271354841219
3.49366587996483 0.301892905644961
3.51691822239331 0.30086862714568
3.54017056482179 0.300199696532437
3.56342290725027 0.298636720993222
3.58667524967875 0.298730681768779
3.60992759210723 0.30060020880785
3.63317993453571 0.30175799355103
3.65643227696419 0.302119295037666
3.67968461939267 0.302665785576721
3.70293696182115 0.303964230512108
3.72618930424963 0.305069119075441
3.74944164667811 0.30216430219999
3.77269398910659 0.300762497728572
3.79594633153507 0.299420765261398
3.81919867396355 0.300097427565381
3.84245101639203 0.296702215978509
3.86570335882051 0.295168919385747
3.88895570124899 0.295941848797928
3.91220804367747 0.298605021231784
};
\addlegendentry{1.14}
\addplot [thick, color7, forget plot]
table {%
0 0
};
\addplot [thick, color8, forget plot]
table {%
0 0
};
\addplot [thick, color0, forget plot]
table {%
0 0
};
\addplot [very thick, color1, dash pattern=on 1pt off 3pt on 3pt off 3pt]
table {%
-0.715008099590029 -0.977078629671483
-0.691755757161549 -0.97958080339942
-0.668503414733069 -0.985601704394626
-0.645251072304589 -0.987946543068235
-0.621998729876109 -0.989499379551297
-0.59874638744763 -0.985874382661249
-0.57549404501915 -0.978045845387372
-0.55224170259067 -0.9704056686802
-0.52898936016219 -0.965587190919786
-0.50573701773371 -0.963691008893558
-0.48248467530523 -0.96174159152292
-0.45923233287675 -0.963959236728428
-0.435979990448271 -0.964352269961657
-0.412727648019791 -0.96397380245266
-0.389475305591311 -0.960097640368888
-0.366222963162831 -0.956660165579275
-0.342970620734351 -0.95176621368964
-0.319718278305871 -0.947998538689399
-0.296465935877391 -0.949597180396349
-0.273213593448911 -0.955080348344585
-0.249961251020432 -0.967428019236179
-0.226708908591952 -0.975369077189662
-0.203456566163472 -0.979795562070762
-0.180204223734992 -0.976493417024315
-0.156951881306512 -0.979128924626232
-0.133699538878032 -0.990370499303324
-0.110447196449552 -1.00578611422895
-0.0871948540210724 -1.01364763139406
-0.0639425115925925 -1.01052575509055
-0.0406901691641126 -1.00530553651926
-0.0174378267356328 -1.00761502481176
0.00581451569284712 -1.01158624225495
0.029066858121327 -1.01241068375298
0.0523192005498069 -1.00876101569774
0.0755715429782867 -1.00844566177959
0.0988238854067666 -1.00776727200637
0.122076227835246 -1.00439900138615
0.145328570263726 -1.00074920220013
0.168580912692206 -0.997922108607218
0.191833255120686 -0.999630568143434
0.215085597549166 -1.00105722898436
0.238337939977646 -1.00344231432908
0.261590282406126 -1.00617757503332
0.284842624834606 -1.00673105807078
0.308094967263085 -1.00554222567397
0.331347309691565 -1.0009606597347
0.354599652120045 -0.998575670916274
0.377851994548525 -0.995359550915517
0.401104336977005 -0.989265451199567
0.424356679405485 -0.972039978997573
0.447609021833965 -0.941269549277157
0.470861364262445 -0.899669157538095
0.494113706690925 -0.8483618267555
0.517366049119404 -0.799356976648618
0.540618391547884 -0.740585809169429
0.563870733976364 -0.670787776559718
0.587123076404844 -0.573019934660669
0.610375418833324 -0.460893711048907
0.633627761261804 -0.341806287942957
0.656880103690284 -0.224759924670631
0.680132446118764 -0.110780951422376
0.703384788547243 -0.00650862228605564
0.726637130975723 0.0831763873843505
0.749889473404203 0.149275562107913
0.773141815832683 0.199767116022437
0.796394158261163 0.231384222706741
0.819646500689643 0.257494022612978
0.842898843118123 0.274798871133969
0.866151185546603 0.28957916748907
0.889403527975082 0.29436013379691
0.912655870403562 0.297492984134088
0.935908212832042 0.301773627050199
0.959160555260522 0.316733588906394
0.982412897689002 0.33618366723787
1.00566524011748 0.361807734050309
1.02891758254596 0.388630819002161
1.05216992497444 0.418061390574556
1.07542226740292 0.449333600399112
1.0986746098314 0.482596244870907
1.12192695225988 0.519797287979957
1.14517929468836 0.556384959253138
1.16843163711684 0.59237583454429
1.19168397954532 0.624986773130316
1.2149363219738 0.654309537348644
1.23818866440228 0.678656383114245
1.26144100683076 0.699123735030452
1.28469334925924 0.716487070753594
1.30794569168772 0.732538222134643
1.3311980341162 0.744982149004585
1.35445037654468 0.755335287693409
1.37770271897316 0.762419324513672
1.40095506140164 0.768102695996431
1.42420740383012 0.771859892510325
1.4474597462586 0.775214367048276
1.47071208868708 0.7769009447336
1.49396443111556 0.776174472713303
1.51721677354404 0.773605397327158
1.54046911597252 0.771244347446773
1.563721458401 0.771262446797746
1.58697380082948 0.771802696050891
1.61022614325796 0.77144579483653
1.63347848568644 0.768285184112834
1.65673082811492 0.763304660398785
1.6799831705434 0.756950026205803
1.70323551297188 0.749230145953663
1.72648785540036 0.743036999090147
1.74974019782884 0.737974584938204
1.77299254025732 0.733610100284806
1.7962448826858 0.727556043917607
1.81949722511428 0.721275041445726
1.84274956754276 0.716897282391562
1.86600190997124 0.714375090297548
1.88925425239972 0.713461801132411
1.9125065948282 0.71404319450306
1.93575893725668 0.717099909431118
1.95901127968516 0.721660796060832
1.98226362211364 0.727760938036545
2.00551596454212 0.734180282734987
2.0287683069706 0.739910699565131
2.05202064939908 0.744285977512234
2.07527299182756 0.746191264244061
2.09852533425604 0.744066612181905
2.12177767668452 0.738399927085018
2.145030019113 0.729730420377003
2.16828236154148 0.720844642410071
2.19153470396996 0.711210132338518
2.21478704639844 0.700750749201165
2.23803938882692 0.68779160009817
2.2612917312554 0.672995777310897
2.28454407368387 0.657848434115973
2.30779641611235 0.644347592098129
2.33104875854083 0.631139132809281
2.35430110096931 0.619173163726775
2.37755344339779 0.607122810377072
2.40080578582627 0.595879764358395
2.42405812825475 0.581814854247275
2.44731047068323 0.565612740315391
2.47056281311171 0.547837240731501
2.49381515554019 0.531858290070958
2.51706749796867 0.514835568941879
2.54031984039715 0.498048916018084
2.56357218282563 0.481300417053374
2.58682452525411 0.469134874385946
2.61007686768259 0.459873101607446
2.63332921011107 0.453536174234707
2.65658155253955 0.446367078556365
2.67983389496803 0.435415176952394
2.70308623739651 0.422832509632681
2.72633857982499 0.409898702187813
2.74959092225347 0.399618166148384
2.77284326468195 0.38939644474169
2.79609560711043 0.381726300089033
2.81934794953891 0.375866006557058
2.84260029196739 0.370754419972328
2.86585263439587 0.363543661075521
2.88910497682435 0.355726450248625
2.91235731925283 0.349036552974675
2.93560966168131 0.345262656254808
2.95886200410979 0.342957056299458
2.98211434653827 0.340417144563624
3.00536668896675 0.338335996031321
3.02861903139523 0.336283151880347
3.05187137382371 0.33628399273791
3.07512371625219 0.334174204993341
3.09837605868067 0.332207554385036
3.12162840110915 0.328199423749661
3.14488074353763 0.325523121670432
3.16813308596611 0.321068794949941
3.19138542839459 0.316497715859496
3.21463777082307 0.313761396473528
3.23789011325155 0.315225667600875
3.26114245568003 0.317711050052281
3.28439479810851 0.317948300382429
3.30764714053699 0.314755170320352
3.33089948296547 0.30982502130869
3.35415182539395 0.305750702398492
3.37740416782243 0.304222722628796
3.40065651025091 0.305126646846198
3.42390885267939 0.307064860631057
3.44716119510787 0.307470502873076
3.47041353753635 0.306017383713972
3.49366587996483 0.303146135057134
3.51691822239331 0.301445946699928
3.54017056482179 0.301079588938977
3.56342290725027 0.303360634042707
3.58667524967875 0.306689559492358
3.60992759210723 0.313883316061585
3.63317993453571 0.321572795792867
3.65643227696419 0.328598226606474
3.67968461939267 0.334509767314882
3.70293696182115 0.337146942188109
3.72618930424963 0.338298083325384
3.74944164667811 0.337708952591705
3.77269398910659 0.340114637288488
3.79594633153507 0.34436902276235
3.81919867396355 0.347630683438963
3.84245101639203 0.349750834186465
3.86570335882051 0.352386931131545
3.88895570124899 0.354726417405886
3.91220804367747 0.356394188719418
};
\addlegendentry{1.30}
\addplot [thick, color2, forget plot]
table {%
0 0
};
\addplot [thick, color3, forget plot]
table {%
0 0
};
\addplot [thick, color4, forget plot]
table {%
0 0
};
\addplot [very thick, color5, dotted]
table {%
-0.715008099590029 -0.88979125161774
-0.691755757161549 -0.892220967810349
-0.668503414733069 -0.894589737606647
-0.645251072304589 -0.895815685455443
-0.621998729876109 -0.896256388327975
-0.59874638744763 -0.892807647075323
-0.57549404501915 -0.879936557559564
-0.55224170259067 -0.861290882510926
-0.52898936016219 -0.847217038825931
-0.50573701773371 -0.839015239182227
-0.48248467530523 -0.834106200946019
-0.45923233287675 -0.829864540437983
-0.435979990448271 -0.826206594816326
-0.412727648019791 -0.825873636463577
-0.389475305591311 -0.831668798278163
-0.366222963162831 -0.845075922594902
-0.342970620734351 -0.867896030774758
-0.319718278305871 -0.895656960191854
-0.296465935877391 -0.92710188894004
-0.273213593448911 -0.952753631799764
-0.249961251020432 -0.975108126785965
-0.226708908591952 -0.988141272845483
-0.203456566163472 -0.994386943136516
-0.180204223734992 -0.99256842850825
-0.156951881306512 -1.00032424366818
-0.133699538878032 -1.01849085952205
-0.110447196449552 -1.02827432726476
-0.0871948540210724 -1.02523409889523
-0.0639425115925925 -1.0132249917962
-0.0406901691641126 -1.01337098410087
-0.0174378267356328 -1.02289210899431
0.00581451569284712 -1.02603003522319
0.029066858121327 -1.02417780995708
0.0523192005498069 -1.01440766446099
0.0755715429782867 -1.01281509008316
0.0988238854067666 -1.01174123245684
0.122076227835246 -1.00754994275872
0.145328570263726 -1.0029392776016
0.168580912692206 -0.996874268504063
0.191833255120686 -0.997648693638575
0.215085597549166 -0.999579654744412
0.238337939977646 -1.00705379678619
0.261590282406126 -1.01407387327719
0.284842624834606 -1.01562828521168
0.308094967263085 -1.01405151329446
0.331347309691565 -1.00656333952473
0.354599652120045 -0.998499325420531
0.377851994548525 -0.984843135335391
0.401104336977005 -0.966880281511962
0.424356679405485 -0.936352693490889
0.447609021833965 -0.887529132311542
0.470861364262445 -0.822636563246772
0.494113706690925 -0.749153867008316
0.517366049119404 -0.681890877342534
0.540618391547884 -0.606879209529142
0.563870733976364 -0.509605473290977
0.587123076404844 -0.37368152663601
0.610375418833324 -0.232629273808413
0.633627761261804 -0.0993730469721514
0.656880103690284 0.0253652889466152
0.680132446118764 0.147063884256866
0.703384788547243 0.253576855551573
0.726637130975723 0.325096782671985
0.749889473404203 0.37059738492818
0.773141815832683 0.401030336569183
0.796394158261163 0.419585037355861
0.819646500689643 0.432236036732662
0.842898843118123 0.436435997377745
0.866151185546603 0.439877701364192
0.889403527975082 0.435128593513668
0.912655870403562 0.428405869865049
0.935908212832042 0.42253321323994
0.959160555260522 0.425784869741386
0.982412897689002 0.434796194482745
1.00566524011748 0.453149445282974
1.02891758254596 0.483433983643317
1.05216992497444 0.519906033184121
1.07542226740292 0.560068803968081
1.0986746098314 0.598689134401831
1.12192695225988 0.638658826547952
1.14517929468836 0.672532086999325
1.16843163711684 0.700085709053271
1.19168397954532 0.721350057545673
1.2149363219738 0.741087394060622
1.23818866440228 0.757397368058816
1.26144100683076 0.775215334591505
1.28469334925924 0.79168140529241
1.30794569168772 0.809200295910447
1.3311980341162 0.825632824116741
1.35445037654468 0.84234470831406
1.37770271897316 0.855821169660417
1.40095506140164 0.86711692984176
1.42420740383012 0.877272322287673
1.4474597462586 0.887980924896218
1.47071208868708 0.896799763913191
1.49396443111556 0.903772610207984
1.51721677354404 0.908124273825619
1.54046911597252 0.911739900574228
1.563721458401 0.917144826852722
1.58697380082948 0.923047445631114
1.61022614325796 0.926424440626004
1.63347848568644 0.924445689709863
1.65673082811492 0.921945639901063
1.6799831705434 0.91942339835953
1.70323551297188 0.915713132015826
1.72648785540036 0.908333247167009
1.74974019782884 0.903236658371748
1.77299254025732 0.900036831457728
1.7962448826858 0.898628323744307
1.81949722511428 0.89614582780073
1.84274956754276 0.893734709829091
1.86600190997124 0.893713018941491
1.88925425239972 0.892350787481287
1.9125065948282 0.890142640848432
1.93575893725668 0.885785833052152
1.95901127968516 0.881302769651607
1.98226362211364 0.876029265584173
2.00551596454212 0.869523496463877
2.0287683069706 0.861915953574612
2.05202064939908 0.854884429151733
2.07527299182756 0.848177008004787
2.09852533425604 0.841598977813033
2.12177767668452 0.835505230712305
2.145030019113 0.830355927149848
2.16828236154148 0.827699078527986
2.19153470396996 0.825325128212115
2.21478704639844 0.823932081759809
2.23803938882692 0.821835715543557
2.2612917312554 0.82176754089166
2.28454407368387 0.822177682320418
2.30779641611235 0.822788101715366
2.33104875854083 0.819975433094104
2.35430110096931 0.817039538254726
2.37755344339779 0.813718000724973
2.40080578582627 0.811706059395055
2.42405812825475 0.806509566836453
2.44731047068323 0.800039047020195
2.47056281311171 0.792563083062585
2.49381515554019 0.786496916817653
2.51706749796867 0.778258108884116
2.54031984039715 0.767755482578763
2.56357218282563 0.756265361618816
2.58682452525411 0.745687186327809
2.61007686768259 0.737713603429258
2.63332921011107 0.736221782021421
2.65658155253955 0.734916908810743
2.67983389496803 0.724915560545646
2.70308623739651 0.707847974488021
2.72633857982499 0.690115108400615
2.74959092225347 0.67908652040141
2.77284326468195 0.667193690047933
2.79609560711043 0.655970216134206
2.81934794953891 0.645338160708114
2.84260029196739 0.635739269094977
2.86585263439587 0.627180546943969
2.88910497682435 0.62019249347739
2.91235731925283 0.615170870430214
2.93560966168131 0.610773628892022
2.95886200410979 0.604283991590454
2.98211434653827 0.59496092102068
3.00536668896675 0.585080942668013
3.02861903139523 0.576882282701257
3.05187137382371 0.570492347700426
3.07512371625219 0.563906487188263
3.09837605868067 0.558043937232303
3.12162840110915 0.552419101082704
3.14488074353763 0.548593720918035
3.16813308596611 0.544491705363775
3.19138542839459 0.540273379432341
3.21463777082307 0.534128645063635
3.23789011325155 0.528511497982813
3.26114245568003 0.5251527931806
3.28439479810851 0.523319753596588
3.30764714053699 0.523439239672995
3.33089948296547 0.522175773277129
3.35415182539395 0.523956913572249
3.37740416782243 0.526279177335567
3.40065651025091 0.529198855117023
3.42390885267939 0.530867992365699
3.44716119510787 0.530505124324181
3.47041353753635 0.530309546229111
3.49366587996483 0.529658178157269
3.51691822239331 0.531643259145024
3.54017056482179 0.535604964860391
3.56342290725027 0.538309493659123
3.58667524967875 0.540141969198521
3.60992759210723 0.541401973132807
3.63317993453571 0.541993755706088
3.65643227696419 0.542039950288603
3.67968461939267 0.540478410698615
3.70293696182115 0.539018628030796
3.72618930424963 0.537111319028094
3.74944164667811 0.535729208809937
3.77269398910659 0.534362444293745
3.79594633153507 0.531287958173033
3.81919867396355 0.526721204232424
3.84245101639203 0.524918393162488
3.86570335882051 0.526682929656423
3.88895570124899 0.534877854027685
3.91220804367747 0.539331945071253
};
\addlegendentry{1.47}
\addplot [thick, color6, forget plot]
table {%
0 0
};
\addplot [thick, white!49.8039215686275!black, forget plot]
table {%
0 0
};
\addplot [thick, color7, forget plot]
table {%
0 0
};
\addplot [very thick, color8, dash pattern=on 5pt off 5pt]
table {%
-0.715008099590029 -0.923508363973699
-0.691755757161549 -0.926397870419952
-0.668503414733069 -0.931503037000065
-0.645251072304589 -0.930280252634477
-0.621998729876109 -0.926632568225809
-0.59874638744763 -0.915470405011588
-0.57549404501915 -0.909981802340605
-0.55224170259067 -0.902243951831318
-0.52898936016219 -0.895122067782638
-0.50573701773371 -0.886900800303959
-0.48248467530523 -0.88093586168562
-0.45923233287675 -0.88489716594742
-0.435979990448271 -0.889394655532694
-0.412727648019791 -0.894730891423215
-0.389475305591311 -0.890824396746929
-0.366222963162831 -0.886208447034379
-0.342970620734351 -0.883316086742696
-0.319718278305871 -0.884215652923656
-0.296465935877391 -0.89014299710459
-0.273213593448911 -0.905581558933751
-0.249961251020432 -0.929148137181327
-0.226708908591952 -0.949213165014402
-0.203456566163472 -0.954417702276113
-0.180204223734992 -0.954134557451638
-0.156951881306512 -0.962498152635949
-0.133699538878032 -0.991340222692563
-0.110447196449552 -1.01581661618742
-0.0871948540210724 -1.02244625417808
-0.0639425115925925 -1.00812019457197
-0.0406901691641126 -1.00739891937996
-0.0174378267356328 -1.01544588049812
0.00581451569284712 -1.0247034535223
0.029066858121327 -1.01886093210234
0.0523192005498069 -1.0114620325448
0.0755715429782867 -1.00766997518835
0.0988238854067666 -1.00754759571327
0.122076227835246 -1.00447088725795
0.145328570263726 -1.00028286479857
0.168580912692206 -0.990636517282085
0.191833255120686 -0.986509896581063
0.215085597549166 -0.988828419307542
0.238337939977646 -1.00198555184077
0.261590282406126 -1.01638859179884
0.284842624834606 -1.02017140692334
0.308094967263085 -1.01992459374933
0.331347309691565 -1.0105916783032
0.354599652120045 -1.0007697755305
0.377851994548525 -0.985259800483425
0.401104336977005 -0.967844357385256
0.424356679405485 -0.939848570035188
0.447609021833965 -0.899803705433583
0.470861364262445 -0.856860336127548
0.494113706690925 -0.826677550961187
0.517366049119404 -0.808133296493379
0.540618391547884 -0.78519940476079
0.563870733976364 -0.734661146546002
0.587123076404844 -0.647354858827806
0.610375418833324 -0.543815366747285
0.633627761261804 -0.434617736773853
0.656880103690284 -0.312944149979246
0.680132446118764 -0.181587050929745
0.703384788547243 -0.0467771082136257
0.726637130975723 0.0624414031939897
0.749889473404203 0.150509802399478
0.773141815832683 0.215425262380624
0.796394158261163 0.25488920337071
0.819646500689643 0.279776079595446
0.842898843118123 0.291927175147779
0.866151185546603 0.297391912940347
0.889403527975082 0.29267200235773
0.912655870403562 0.290854160666361
0.935908212832042 0.297348388884151
0.959160555260522 0.311040798483214
0.982412897689002 0.323719164160426
1.00566524011748 0.346247628025679
1.02891758254596 0.376953275297462
1.05216992497444 0.418759052581305
1.07542226740292 0.458766044118237
1.0986746098314 0.502841248776954
1.12192695225988 0.53826541407854
1.14517929468836 0.569361247288875
1.16843163711684 0.589768293566266
1.19168397954532 0.609245999330572
1.2149363219738 0.626667963398245
1.23818866440228 0.642811157889311
1.26144100683076 0.655647680037805
1.28469334925924 0.665863326948063
1.30794569168772 0.674696348743715
1.3311980341162 0.685209743727603
1.35445037654468 0.695138569221641
1.37770271897316 0.704268526965405
1.40095506140164 0.71061134323567
1.42420740383012 0.713733988667835
1.4474597462586 0.716291498412295
1.47071208868708 0.719837907920232
1.49396443111556 0.725545919785111
1.51721677354404 0.73200691512852
1.54046911597252 0.737116160968133
1.563721458401 0.746101346824364
1.58697380082948 0.755504279096159
1.61022614325796 0.763429678127561
1.63347848568644 0.766125198563584
1.65673082811492 0.769514468083648
1.6799831705434 0.776210077275693
1.70323551297188 0.783724035033292
1.72648785540036 0.790985287653424
1.74974019782884 0.796097575781963
1.77299254025732 0.79936785832274
1.7962448826858 0.798425316527442
1.81949722511428 0.797616377101454
1.84274956754276 0.798079990987046
1.86600190997124 0.799403501535207
1.88925425239972 0.799048717993083
1.9125065948282 0.797389277640385
1.93575893725668 0.794776550113889
1.95901127968516 0.791113578086857
1.98226362211364 0.787211757101552
2.00551596454212 0.78307162589986
2.0287683069706 0.77797506667188
2.05202064939908 0.771582008060678
2.07527299182756 0.764950176738677
2.09852533425604 0.757144893218098
2.12177767668452 0.74628660268147
2.145030019113 0.73478467409548
2.16828236154148 0.721601411009051
2.19153470396996 0.70933188732785
2.21478704639844 0.69689825729322
2.23803938882692 0.686279282468544
2.2612917312554 0.676103374397975
2.28454407368387 0.666025249687516
2.30779641611235 0.656049112464206
2.33104875854083 0.647902277480884
2.35430110096931 0.641431312204997
2.37755344339779 0.638389619999517
2.40080578582627 0.637247132260146
2.42405812825475 0.638646135493114
2.44731047068323 0.641887935063451
2.47056281311171 0.647594241639135
2.49381515554019 0.656023255865056
2.51706749796867 0.663866333653792
2.54031984039715 0.670637072329234
2.56357218282563 0.675656602145655
2.58682452525411 0.683302335233759
2.61007686768259 0.693404875376385
2.63332921011107 0.708316955057619
2.65658155253955 0.721033826733333
2.67983389496803 0.725469032453107
2.70308623739651 0.721470368288413
2.72633857982499 0.715360566199679
2.74959092225347 0.713226673547463
2.77284326468195 0.710657818463427
2.79609560711043 0.707662732780811
2.81934794953891 0.703347613838833
2.84260029196739 0.69768676370742
2.86585263439587 0.690834917306643
2.88910497682435 0.682758991202309
2.91235731925283 0.674259564271551
2.93560966168131 0.664544351869361
2.95886200410979 0.654533004942071
2.98211434653827 0.646661656839302
3.00536668896675 0.639563548303585
3.02861903139523 0.63375119410595
3.05187137382371 0.630436712264597
3.07512371625219 0.628432337481233
3.09837605868067 0.626734193768391
3.12162840110915 0.623074951962281
3.14488074353763 0.62033010032185
3.16813308596611 0.617867906922762
3.19138542839459 0.614966700396771
3.21463777082307 0.610722077504056
3.23789011325155 0.606263819928792
3.26114245568003 0.601715477666804
3.28439479810851 0.596570658639485
3.30764714053699 0.590943109564356
3.33089948296547 0.584987821314848
3.35415182539395 0.579059867983914
3.37740416782243 0.579009591166703
3.40065651025091 0.581498315410307
3.42390885267939 0.584076750267711
3.44716119510787 0.579987270249481
3.47041353753635 0.571667260569027
3.49366587996483 0.565469663612027
3.51691822239331 0.559619253177564
3.54017056482179 0.554575652593646
3.56342290725027 0.549930856253903
3.58667524967875 0.549209335717474
3.60992759210723 0.552413229489669
3.63317993453571 0.553716259682225
3.65643227696419 0.552987123777989
3.67968461939267 0.550622744566017
3.70293696182115 0.550319361772293
3.72618930424963 0.55114817351586
3.74944164667811 0.551042727865843
3.77269398910659 0.551013823035366
3.79594633153507 0.551146591055312
3.81919867396355 0.549680995793566
3.84245101639203 0.547385678923019
3.86570335882051 0.545342427632539
3.88895570124899 0.547663683840868
3.91220804367747 0.549426380691004
};
\addlegendentry{1.63}
\end{axis}

\end{tikzpicture}

%% file: tikz_matplotlib/media_ux_max.tikz
\begin{tikzpicture}

\definecolor{color0}{rgb}{0.12156862745098,0.466666666666667,0.705882352941177}
\definecolor{color1}{rgb}{1,0.498039215686275,0.0549019607843137}
\definecolor{color2}{rgb}{0.172549019607843,0.627450980392157,0.172549019607843}
\definecolor{color3}{rgb}{0.83921568627451,0.152941176470588,0.156862745098039}
\definecolor{color4}{rgb}{0.580392156862745,0.403921568627451,0.741176470588235}

\begin{axis}[thick,
legend cell align={left},
legend style={fill opacity=0.8, draw opacity=1, text opacity=1,legend columns=-1,at={(0.5,1.2)},anchor=north, draw=white!80!black},
tick align=outside,
tick pos=left,
x grid style={white!69.0196078431373!black},
xlabel={\(\displaystyle f^+\)},
xmajorgrids,
xmin=0.945087176144696, xmax=1.66204986149584,
xtick style={color=black},
y grid style={white!69.0196078431373!black},
ylabel={\(\displaystyle u_x-U\)},
ymajorgrids,
ymin=0, ymax=1,
ytick style={color=black}
]
\addlegendimage{empty legend}
\addlegendentry{\hspace{-0cm}\textbf{$\bm U_R$}}	
\addplot [thick, color0, mark=*, mark size=5, mark options={solid,fill opacity=0}, only marks]
table {%
0.977676389115203 0.374261505390057
};
\addlegendentry{$0.29$}
\addplot [thick, color0, mark=*, mark size=5, mark options={solid,fill opacity=0}, forget plot]
table {%
1.14062245396774 0.550265813242177
};
\addplot [thick, color0, mark=*, mark size=5, mark options={solid,fill opacity=0}, forget plot]
table {%
1.30356851882027 0.774637611295671
};
\addplot [thick, color0, mark=*, mark size=5, mark options={solid,fill opacity=0}, forget plot]
table {%
1.4665145836728 0.919965987889056
};
\addplot [thick, color0, mark=*, mark size=5, mark options={solid,fill opacity=0}, forget plot]
table {%
1.62946064852534 0.791721531354684
};
\addplot [thick, color1, mark=square, mark size=5, mark options={solid,fill opacity=0}, only marks]
table {%
0.977676389115203 0.177099137687274
};
\addlegendentry{$0.37$}
\addplot [thick, color1, mark=square, mark size=5, mark options={solid,fill opacity=0}, forget plot]
table {%
1.14062245396774 0.242308033241384
};
\addplot [thick, color1, mark=square, mark size=5, mark options={solid,fill opacity=0}, forget plot]
table {%
1.30356851882027 0.34981715276466
};
\addplot [thick, color1, mark=square, mark size=5, mark options={solid,fill opacity=0}, forget plot]
table {%
1.4665145836728 0.415484594209327
};
\addplot [thick, color1, mark=square, mark size=5, mark options={solid,fill opacity=0}, forget plot]
table {%
1.62946064852534 0.39265024320569
};
\addplot [thick, color2, mark=triangle, mark size=5, mark options={solid,rotate=270,fill opacity=0}, only marks]
table {%
0.977676389115203 0.1120862273563
};
\addlegendentry{$0.46$}
\addplot [thick, color2, mark=triangle, mark size=5, mark options={solid,rotate=270,fill opacity=0}, forget plot]
table {%
1.14062245396774 0.153078459119872
};
\addplot [thick, color2, mark=triangle, mark size=5, mark options={solid,rotate=270,fill opacity=0}, forget plot]
table {%
1.30356851882027 0.173022482481016
};
\addplot [thick, color2, mark=triangle, mark size=5, mark options={solid,rotate=270,fill opacity=0}, forget plot]
table {%
1.4665145836728 0.176476071839074
};
\addplot [thick, color2, mark=triangle, mark size=5, mark options={solid,rotate=270,fill opacity=0}, forget plot]
table {%
1.62946064852534 0.187319949402079
};
\addplot [thick, color3, mark=diamond, mark size=5, mark options={solid,fill opacity=0}, only marks]
table {%
0.977676389115203 0.0381466130303075
};
\addlegendentry{$0.54$}
\addplot [thick, color3, mark=diamond, mark size=5, mark options={solid,fill opacity=0}, forget plot]
table {%
1.14062245396774 0.0787467576093992
};
\addplot [thick, color3, mark=diamond, mark size=5, mark options={solid,fill opacity=0}, forget plot]
table {%
1.30356851882027 0.111623070615624
};
\addplot [thick, color3, mark=diamond, mark size=5, mark options={solid,fill opacity=0}, forget plot]
table {%
1.4665145836728 0.116959186499871
};
\addplot [thick, color3, mark=diamond, mark size=5, mark options={solid,fill opacity=0}, forget plot]
table {%
1.62946064852534 0.10244749908297
};
\addplot [thick, color4, mark=triangle, mark size=5, mark options={solid,rotate=90,fill opacity=0}, only marks]
table {%
0.977676389115203 0.00499304510107773
};
\addlegendentry{$0.59$}
\addplot [thick, color4, mark=triangle, mark size=5, mark options={solid,rotate=90,fill opacity=0}, forget plot]
table {%
1.14062245396774 0.0569989500027297
};
\addplot [thick, color4, mark=triangle, mark size=5, mark options={solid,rotate=90,fill opacity=0}, forget plot]
table {%
1.30356851882027 0.0586809043686444
};
\addplot [thick, color4, mark=triangle, mark size=5, mark options={solid,rotate=90,fill opacity=0}, forget plot]
table {%
1.4665145836728 0.0897056915118598
};
\addplot [thick, color4, mark=triangle, mark size=5, mark options={solid,rotate=90,fill opacity=0}, forget plot]
table {%
1.62946064852534 0.0620107687174102
};
\end{axis}

\end{tikzpicture}

%% file: tikz_matplotlib/media_uystd_0.29.tikz
\begin{tikzpicture}

\definecolor{color0}{rgb}{0.12156862745098,0.466666666666667,0.705882352941177}
\definecolor{color1}{rgb}{1,0.498039215686275,0.0549019607843137}
\definecolor{color2}{rgb}{0.172549019607843,0.627450980392157,0.172549019607843}
\definecolor{color3}{rgb}{0.83921568627451,0.152941176470588,0.156862745098039}
\definecolor{color4}{rgb}{0.580392156862745,0.403921568627451,0.741176470588235}
\definecolor{color5}{rgb}{0.549019607843137,0.337254901960784,0.294117647058824}
\definecolor{color6}{rgb}{0.890196078431372,0.466666666666667,0.76078431372549}
\definecolor{color7}{rgb}{0.737254901960784,0.741176470588235,0.133333333333333}
\definecolor{color8}{rgb}{0.0901960784313725,0.745098039215686,0.811764705882353}

\begin{axis}[thick,
legend cell align={left},
legend style={fill opacity=0.8, draw opacity=1, text opacity=1, draw=white!80!black,legend columns=-1,at={(0.5,1.2)},anchor=north},
tick align=outside,
tick pos=left,
x grid style={white!69.0196078431373!black},
xlabel={\(\displaystyle x/L\)},
xmajorgrids,
xmin=1, xmax=4.07032397219113,
xtick style={color=black},
y grid style={white!69.0196078431373!black},
ylabel={\(\displaystyle \langle {u^\prime}^2_y \rangle^{1/2}\)},
ymajorgrids,
ymin=0, ymax=1,
ytick style={color=black}
]
\addlegendimage{empty legend}
\addlegendentry{\hspace{-0cm}\textbf{$\bm f^+$}}	
\addplot [thick, color0, forget plot]
table {%
0.5 0.5
};
\addplot [thick, color1, forget plot]
table {%
0.5 0.5
};
\addplot [thick, color2, forget plot]
table {%
0.5 0.5
};
\addplot [very thick, color3]
table {%
-0.715008099590029 0.440793652047435
-0.621998729876109 0.434637389979808
-0.52898936016219 0.408949221352341
-0.435979990448271 0.363260530039849
-0.342970620734351 0.334817731296674
-0.249961251020432 0.292247967980288
-0.156951881306512 0.224942829095126
-0.0639425115925925 0.230634691689997
0.029066858121327 0.106043270058239
0.122076227835246 0.0274979465018991
0.215085597549166 0.0430455604610649
0.308094967263085 0.0772299834529927
0.401104336977005 0.120976577610452
0.494113706690925 0.204350616781525
0.587123076404844 0.26889314141397
0.680132446118764 0.343738308220407
0.773141815832683 0.413131553903451
0.866151185546603 0.475412566643262
0.959160555260522 0.531275055219841
1.05216992497444 0.506744427787937
1.14517929468836 0.516033270911455
1.23818866440228 0.551865933099559
1.3311980341162 0.577950870536177
1.42420740383012 0.582785225994901
1.51721677354404 0.59895356956031
1.61022614325796 0.592905763064488
1.70323551297188 0.598478789065783
1.7962448826858 0.610226068859399
1.88925425239972 0.608491616309337
1.98226362211364 0.62041737296323
2.07527299182756 0.622687406679232
2.16828236154148 0.62925899301078
2.2612917312554 0.623450642863326
2.35430110096931 0.617695327509997
2.44731047068323 0.610062214813331
2.54031984039715 0.597630868025096
2.63332921011107 0.582172809716803
2.72633857982499 0.567701121274303
2.81934794953891 0.551249965627981
2.91235731925283 0.540150045432664
3.00536668896675 0.527473150272555
3.09837605868067 0.521137220600752
3.19138542839459 0.510009607670443
3.28439479810851 0.497491728099833
3.37740416782243 0.485303787010118
3.47041353753635 0.476755264645603
3.56342290725027 0.464136443135186
3.65643227696419 0.456619625770466
3.74944164667811 0.443889661879048
3.84245101639203 0.439898691336937
};
\addlegendentry{0.98}
\addplot [thick, color4, forget plot]
table {%
0.5 0.5
};
\addplot [thick, color5, forget plot]
table {%
0.5 0.5
};
\addplot [thick, color6, forget plot]
table {%
0.5 0.5
};
\addplot [very thick, white!49.8039215686275!black, dashed]
table {%
-0.715008099590029 0.51727737202266
-0.621998729876109 0.496528312264028
-0.52898936016219 0.458114777732973
-0.435979990448271 0.414835832260315
-0.342970620734351 0.347501800771564
-0.249961251020432 0.308879438934295
-0.156951881306512 0.232205051625697
-0.0639425115925925 0.245500087207931
0.029066858121327 0.115082475989222
0.122076227835246 0.0280480782798044
0.215085597549166 0.0466609851855353
0.308094967263085 0.0836303977153424
0.401104336977005 0.137431366627433
0.494113706690925 0.224435203542975
0.587123076404844 0.298157239577647
0.680132446118764 0.411313728966947
0.773141815832683 0.493389271058686
0.866151185546603 0.600223416658913
0.959160555260522 0.678112967757623
1.05216992497444 0.655896329505502
1.14517929468836 0.661136015319331
1.23818866440228 0.694124399822724
1.3311980341162 0.718693786751647
1.42420740383012 0.736499332906425
1.51721677354404 0.762018123663676
1.61022614325796 0.784058044020028
1.70323551297188 0.794147850479295
1.7962448826858 0.805827791169986
1.88925425239972 0.813332959861087
1.98226362211364 0.813471464387693
2.07527299182756 0.81450331743127
2.16828236154148 0.811735637078182
2.2612917312554 0.796496883975339
2.35430110096931 0.786227860191928
2.44731047068323 0.76585225807915
2.54031984039715 0.750996614778153
2.63332921011107 0.731617218969818
2.72633857982499 0.718979742060455
2.81934794953891 0.707909189468287
2.91235731925283 0.678100855952468
3.00536668896675 0.659144671425726
3.09837605868067 0.640404458881835
3.19138542839459 0.615524664740607
3.28439479810851 0.599981537068577
3.37740416782243 0.581213573795401
3.47041353753635 0.559141644707366
3.56342290725027 0.540695484881819
3.65643227696419 0.53041318033587
3.74944164667811 0.523533300821791
3.84245101639203 0.507853722420869
};
\addlegendentry{1.14}
\addplot [thick, color7, forget plot]
table {%
0.5 0.5
};
\addplot [thick, color8, forget plot]
table {%
0.5 0.5
};
\addplot [thick, color0, forget plot]
table {%
0.5 0.5
};
\addplot [very thick, color1, dash pattern=on 1pt off 3pt on 3pt off 3pt]
table {%
-0.715008099590029 0.506523117699155
-0.621998729876109 0.48297164568311
-0.52898936016219 0.463104367794761
-0.435979990448271 0.440973751993963
-0.342970620734351 0.366124068028415
-0.249961251020432 0.317328753474124
-0.156951881306512 0.244011475143872
-0.0639425115925925 0.257837384968974
0.029066858121327 0.120497808930524
0.122076227835246 0.0306026060722691
0.215085597549166 0.0538977330436522
0.308094967263085 0.0979366054341462
0.401104336977005 0.15167545201649
0.494113706690925 0.227045066397474
0.587123076404844 0.31480276903898
0.680132446118764 0.452134990968356
0.773141815832683 0.602954353184657
0.866151185546603 0.750599524781947
0.959160555260522 0.833112134778995
1.05216992497444 0.825063709948841
1.14517929468836 0.808224472369473
1.23818866440228 0.866280609616331
1.3311980341162 0.8903807900418
1.42420740383012 0.911920592073285
1.51721677354404 0.929420025410814
1.61022614325796 0.939975212854709
1.70323551297188 0.939720555809026
1.7962448826858 0.936369608609004
1.88925425239972 0.934570044069283
1.98226362211364 0.925595384640326
2.07527299182756 0.907537914873564
2.16828236154148 0.895574321125337
2.2612917312554 0.871941664438547
2.35430110096931 0.83983756297541
2.44731047068323 0.802479573219241
2.54031984039715 0.755817490436678
2.63332921011107 0.712328487353284
2.72633857982499 0.684028149489003
2.81934794953891 0.641811125565627
2.91235731925283 0.611632874870202
3.00536668896675 0.584691526162348
3.09837605868067 0.568098182943882
3.19138542839459 0.551569230809843
3.28439479810851 0.543225097157244
3.37740416782243 0.52789815434246
3.47041353753635 0.52028973291079
3.56342290725027 0.507440767318399
3.65643227696419 0.484370628726293
3.74944164667811 0.470488111234352
3.84245101639203 0.445869376066455
};
\addlegendentry{1.30}
\addplot [thick, color2, forget plot]
table {%
0.5 0.5
};
\addplot [thick, color3, forget plot]
table {%
0.5 0.5
};
\addplot [thick, color4, forget plot]
table {%
0.5 0.5
};
\addplot [very thick, color5, dotted]
table {%
-0.715008099590029 0.641746536406141
-0.621998729876109 0.637008892162932
-0.52898936016219 0.592227131713542
-0.435979990448271 0.547724434752674
-0.342970620734351 0.452368433146784
-0.249961251020432 0.371489759621459
-0.156951881306512 0.283509383225105
-0.0639425115925925 0.30513012484518
0.029066858121327 0.150043874156691
0.122076227835246 0.0378785953783895
0.215085597549166 0.0660487082294723
0.308094967263085 0.11287601536741
0.401104336977005 0.169310300405057
0.494113706690925 0.21435472621112
0.587123076404844 0.317909072035173
0.680132446118764 0.467601627993986
0.773141815832683 0.652305773680422
0.866151185546603 0.804603823099782
0.959160555260522 0.922938485670722
1.05216992497444 0.882426538196063
1.14517929468836 0.882072318548702
1.23818866440228 0.937434976230422
1.3311980341162 0.955253859335625
1.42420740383012 0.975373859520106
1.51721677354404 0.981807477353254
1.61022614325796 0.985425610528171
1.70323551297188 0.986698029325042
1.7962448826858 0.976503242821054
1.88925425239972 0.9615359839151
1.98226362211364 0.942713278106642
2.07527299182756 0.919676663554532
2.16828236154148 0.897865652377276
2.2612917312554 0.884575305859311
2.35430110096931 0.864206788882145
2.44731047068323 0.839717580179716
2.54031984039715 0.810960921094182
2.63332921011107 0.78843299634895
2.72633857982499 0.767989498109476
2.81934794953891 0.730749918592315
2.91235731925283 0.698636936560479
3.00536668896675 0.669055138265053
3.09837605868067 0.656628752149498
3.19138542839459 0.628836790332559
3.28439479810851 0.611884629716197
3.37740416782243 0.588122090548977
3.47041353753635 0.571759235496916
3.56342290725027 0.551686058126927
3.65643227696419 0.535590044111621
3.74944164667811 0.512370989821701
3.84245101639203 0.501565133448966
};
\addlegendentry{1.47}
\addplot [thick, color6, forget plot]
table {%
0.5 0.5
};
\addplot [thick, white!49.8039215686275!black, forget plot]
table {%
0.5 0.5
};
\addplot [thick, color7, forget plot]
table {%
0.5 0.5
};
\addplot [very thick, color8, dash pattern=on 5pt off 5pt]
table {%
-0.715008099590029 0.475285959612164
-0.621998729876109 0.446640948002304
-0.52898936016219 0.411485830398501
-0.435979990448271 0.386155364923342
-0.342970620734351 0.372864706629364
-0.249961251020432 0.363783622181721
-0.156951881306512 0.282584074923123
-0.0639425115925925 0.28811465970726
0.029066858121327 0.139781249212932
0.122076227835246 0.0385760388133888
0.215085597549166 0.0711022625431388
0.308094967263085 0.11903515386612
0.401104336977005 0.167589544368321
0.494113706690925 0.199624766399604
0.587123076404844 0.248740472075493
0.680132446118764 0.367227370588476
0.773141815832683 0.570933455629833
0.866151185546603 0.717621901328817
0.959160555260522 0.832056501444439
1.05216992497444 0.825132322821169
1.14517929468836 0.838893945940916
1.23818866440228 0.88745739304584
1.3311980341162 0.914039799012694
1.42420740383012 0.931081349577385
1.51721677354404 0.933656517369295
1.61022614325796 0.933923122845638
1.70323551297188 0.931711084630481
1.7962448826858 0.93912797778441
1.88925425239972 0.9404213759271
1.98226362211364 0.936850787080214
2.07527299182756 0.929596365764303
2.16828236154148 0.929468670969049
2.2612917312554 0.917463913694644
2.35430110096931 0.894614691417451
2.44731047068323 0.883307474485017
2.54031984039715 0.874480164094601
2.63332921011107 0.856352973229394
2.72633857982499 0.833846804537009
2.81934794953891 0.810874420877369
2.91235731925283 0.774158890702029
3.00536668896675 0.741402508350672
3.09837605868067 0.711402750338499
3.19138542839459 0.67940874737858
3.28439479810851 0.638641275993387
3.37740416782243 0.617946197346131
3.47041353753635 0.601040587225584
3.56342290725027 0.576370660634839
3.65643227696419 0.559289076961178
3.74944164667811 0.547203892648759
3.84245101639203 0.519459544217604
};
\addlegendentry{1.63}
\end{axis}

\end{tikzpicture}

%% file: tikz_matplotlib/media_uystd_max.tikz
\begin{tikzpicture}

\definecolor{color0}{rgb}{0.12156862745098,0.466666666666667,0.705882352941177}
\definecolor{color1}{rgb}{1,0.498039215686275,0.0549019607843137}
\definecolor{color2}{rgb}{0.172549019607843,0.627450980392157,0.172549019607843}
\definecolor{color3}{rgb}{0.83921568627451,0.152941176470588,0.156862745098039}
\definecolor{color4}{rgb}{0.580392156862745,0.403921568627451,0.741176470588235}

\begin{axis}[thick,
legend cell align={left},
legend style={fill opacity=0.8, draw opacity=1, text opacity=1,legend columns=-1,at={(0.5,1.2)},anchor=north,  draw=white!80!black},
tick align=outside,
tick pos=left,
x grid style={white!69.0196078431373!black},
xlabel={\(\displaystyle f^+\)},
xmajorgrids,
xmin=0.945087176144696, xmax=1.66204986149584,
xtick style={color=black},
y grid style={white!69.0196078431373!black},
ylabel={\(\displaystyle \langle {u^\prime}^2_y \rangle^{1/2}\)},
ymajorgrids,
ymin=0, ymax=1,
ytick style={color=black}
]
\addlegendimage{empty legend}
\addlegendentry{\hspace{-0cm}\textbf{$\bm U_R$}}	
\addplot [thick, color0, mark=*, mark size=5, mark options={solid,fill opacity=0}, only marks]
table {%
0.977676389115203 0.624106663731838
};
\addlegendentry{$0.29$}
\addplot [thick, color0, mark=*, mark size=5, mark options={solid,fill opacity=0}, forget plot]
table {%
1.14062245396774 0.815315923461693
};
\addplot [thick, color0, mark=*, mark size=5, mark options={solid,fill opacity=0}, forget plot]
table {%
1.30356851882027 0.942330810900164
};
\addplot [thick, color0, mark=*, mark size=5, mark options={solid,fill opacity=0}, forget plot]
table {%
1.4665145836728 0.986794704576999
};
\addplot [thick, color0, mark=*, mark size=5, mark options={solid,fill opacity=0}, forget plot]
table {%
1.62946064852534 0.942535269626532
};
\addplot [thick, color1, mark=square, mark size=5, mark options={solid,fill opacity=0}, only marks]
table {%
0.977676389115203 0.394531923884546
};
\addlegendentry{$0.37$}
\addplot [thick, color1, mark=square, mark size=5, mark options={solid,fill opacity=0}, forget plot]
table {%
1.14062245396774 0.538151869879936
};
\addplot [thick, color1, mark=square, mark size=5, mark options={solid,fill opacity=0}, forget plot]
table {%
1.30356851882027 0.672030442935084
};
\addplot [thick, color1, mark=square, mark size=5, mark options={solid,fill opacity=0}, forget plot]
table {%
1.4665145836728 0.72454463893877
};
\addplot [thick, color1, mark=square, mark size=5, mark options={solid,fill opacity=0}, forget plot]
table {%
1.62946064852534 0.689902343169374
};
\addplot [thick, color2, mark=triangle, mark size=5, mark options={solid,rotate=270,fill opacity=0}, only marks]
table {%
0.977676389115203 0.315612403589976
};
\addlegendentry{$0.46$}
\addplot [thick, color2, mark=triangle, mark size=5, mark options={solid,rotate=270,fill opacity=0}, forget plot]
table {%
1.14062245396774 0.370759848009681
};
\addplot [thick, color2, mark=triangle, mark size=5, mark options={solid,rotate=270,fill opacity=0}, forget plot]
table {%
1.30356851882027 0.495425669840644
};
\addplot [thick, color2, mark=triangle, mark size=5, mark options={solid,rotate=270,fill opacity=0}, forget plot]
table {%
1.4665145836728 0.552328366576195
};
\addplot [thick, color2, mark=triangle, mark size=5, mark options={solid,rotate=270,fill opacity=0}, forget plot]
table {%
1.62946064852534 0.56053068958442
};
\addplot [thick, color3, mark=diamond, mark size=5, mark options={solid,fill opacity=0}, only marks]
table {%
0.977676389115203 0.275367776904129
};
\addlegendentry{$0.54$}
\addplot [thick, color3, mark=diamond, mark size=5, mark options={solid,fill opacity=0}, forget plot]
table {%
1.14062245396774 0.303230330734925
};
\addplot [thick, color3, mark=diamond, mark size=5, mark options={solid,fill opacity=0}, forget plot]
table {%
1.30356851882027 0.390506403708265
};
\addplot [thick, color3, mark=diamond, mark size=5, mark options={solid,fill opacity=0}, forget plot]
table {%
1.4665145836728 0.448126921883812
};
\addplot [thick, color3, mark=diamond, mark size=5, mark options={solid,fill opacity=0}, forget plot]
table {%
1.62946064852534 0.455072054192248
};
\addplot [thick, color4, mark=triangle, mark size=5, mark options={solid,rotate=90,fill opacity=0}, only marks]
table {%
0.977676389115203 0.331351037095494
};
\addlegendentry{$0.59$}
\addplot [thick, color4, mark=triangle, mark size=5, mark options={solid,rotate=90,fill opacity=0}, forget plot]
table {%
1.14062245396774 0.353970184584635
};
\addplot [thick, color4, mark=triangle, mark size=5, mark options={solid,rotate=90,fill opacity=0}, forget plot]
table {%
1.30356851882027 0.334447276945883
};
\addplot [thick, color4, mark=triangle, mark size=5, mark options={solid,rotate=90,fill opacity=0}, forget plot]
table {%
1.4665145836728 0.383861154040618
};
\addplot [thick, color4, mark=triangle, mark size=5, mark options={solid,rotate=90,fill opacity=0}, forget plot]
table {%
1.62946064852534 0.379697141883467
};
\end{axis}

\end{tikzpicture}

%% file: tikz_matplotlib/profiles_UR029fr146.tikz
\begin{tikzpicture}

\definecolor{color0}{rgb}{0.12156862745098,0.466666666666667,0.705882352941177}
\definecolor{color1}{rgb}{1,0.498039215686275,0.0549019607843137}
\definecolor{color2}{rgb}{0.172549019607843,0.627450980392157,0.172549019607843}

\begin{axis}[ultra thick,
label style={font=\large},
tick label style={font=\large},
legend cell align={left},
legend style={at={(0.55,0.2)},anchor=west,fill opacity=0.8, draw opacity=1, text opacity=1, draw=white!80!black,font=\large},
tick align=outside,
tick pos=left,
x grid style={white!69.0196078431373!black},
xlabel={\(\displaystyle y/L\)},
xmajorgrids,
xmin=-1.82297231682709, xmax=1.98808660720076,
xtick style={color=black},
y grid style={white!69.0196078431373!black},
ymajorgrids,
ymin=-1.27742526070337, ymax=0.668144472960279,
ytick style={color=black}
]
\addplot [semithick, color0, mark=square, mark size=2.5, mark options={solid,fill opacity=0}, only marks]
table {%
-1.64974236573492 -0.0468952762209341
-1.62649002330644 -0.0467881214524728
-1.60323768087796 -0.0452573243011926
-1.57998533844948 -0.0441765946143782
-1.556732996021 -0.0438015938990991
-1.53348065359252 -0.0439283442173741
-1.51022831116404 -0.0449100185093776
-1.48697596873556 -0.0449560825036146
-1.46372362630708 -0.0453916151812762
-1.4404712838786 -0.0449564397370156
-1.41721894145012 -0.0434510433406986
-1.39396659902164 -0.0410599990203211
-1.37071425659316 -0.0381397351742491
-1.34746191416468 -0.0356135614386134
-1.3242095717362 -0.0341865056471135
-1.30095722930772 -0.033093192665519
-1.27770488687924 -0.0319832046340207
-1.25445254445076 -0.0297046295759417
-1.23120020202228 -0.0289244361903782
-1.2079478595938 -0.028301817429426
-1.18469551716532 -0.0279893798765728
-1.16144317473684 -0.0255937058921483
-1.13819083230836 -0.0236084638046048
-1.11493848987988 -0.0220541364950157
-1.0916861474514 -0.0205523209342705
-1.06843380502292 -0.0197175048553809
-1.04518146259444 -0.0190091376158561
-1.02192912016596 -0.0177305249470265
-0.998676777737481 -0.0150202251877273
-0.975424435309001 -0.012408412868111
-0.952172092880522 -0.0105645738228875
-0.928919750452042 -0.00949385421085219
-0.905667408023562 -0.00804161681587746
-0.882415065595082 -0.00677161064896572
-0.859162723166602 -0.00518017745576725
-0.835910380738122 -0.00439769670869735
-0.812658038309642 -0.00259510804819975
-0.789405695881162 -0.000121107671517037
-0.766153353452683 0.00349106037975191
-0.742901011024203 0.00803842977394641
-0.719648668595723 0.0146550921094394
-0.696396326167243 0.0243724855483511
-0.673143983738763 0.035555154487791
-0.649891641310283 0.046593337592298
-0.626639298881803 0.0576071986165059
-0.603386956453323 0.0696499851128288
-0.580134614024844 0.0817588000761097
-0.556882271596364 0.0916419225161361
-0.533629929167884 0.0974397933123603
-0.510377586739404 0.0986609758254209
-0.487125244310924 0.0951573385630165
-0.463872901882444 0.0866096043852004
-0.440620559453964 0.0735356659536716
-0.417368217025484 0.056379312112891
-0.394115874597004 0.0348100396792945
-0.370863532168525 0.00682390995899814
-0.347611189740045 -0.0303561534672547
-0.324358847311565 -0.0817955405856623
-0.301106504883085 -0.149674365991358
-0.277854162454605 -0.237138850084085
-0.254601820026125 -0.34089208571257
-0.231349477597645 -0.460959466697892
-0.208097135169165 -0.588158996694151
-0.184844792740686 -0.712044275176468
-0.161592450312206 -0.823594623460869
-0.138340107883726 -0.92440685906791
-0.115087765455246 -1.01719316873616
-0.0918354230267661 -1.10017900154713
-0.0685830805982862 -1.16108418549704
-0.0453307381698063 -1.18899027280957
-0.0220783957413265 -1.18072607400975
0.00117394668715341 -1.1402450225923
0.0244262891156333 -1.07980820329051
0.0476786315441132 -1.00607835375712
0.070930973972593 -0.92762779187991
0.0941833164010729 -0.842839137023829
0.117435658829553 -0.748948813751572
0.140688001258033 -0.648779102248137
0.163940343686513 -0.549149970236776
0.187192686114992 -0.45613294424714
0.210445028543472 -0.371598584797455
0.233697370971952 -0.298799054681547
0.256949713400432 -0.237446468892723
0.280202055828912 -0.186358817248959
0.303454398257392 -0.139902181457399
0.326706740685872 -0.097843248032313
0.349959083114352 -0.0599321598428061
0.373211425542831 -0.0305942121489187
0.396463767971311 -0.00590144094119184
0.419716110399791 0.0162701479032378
0.442968452828271 0.037348095136852
0.466220795256751 0.0533397125722565
0.489473137685231 0.0626556966792076
0.512725480113711 0.0654953055316079
0.535977822542191 0.0656257878453817
0.55923016497067 0.0645329564196147
0.58248250739915 0.0636270935961799
0.60573484982763 0.057549726611377
0.62898719225611 0.0456779192084616
0.65223953468459 0.0266755136566033
0.67549187711307 0.00771368595720029
0.69874421954155 -0.00575691835115598
0.721996561970029 -0.0110462621852674
0.745248904398509 -0.0126833399990765
0.768501246826989 -0.015238488908286
0.791753589255469 -0.0186264976045833
0.815005931683949 -0.0213725273313447
0.838258274112429 -0.0231245118148496
0.861510616540909 -0.0245487593727589
0.884762958969389 -0.0263601853445692
0.908015301397869 -0.0277928338275691
0.931267643826348 -0.0296142957702227
0.954519986254828 -0.0311619298342313
0.977772328683308 -0.0350592283928506
1.00102467111179 -0.0377561558021595
1.02427701354027 -0.0387163963351415
1.04752935596875 -0.0379381709337556
1.07078169839723 -0.0381568544094001
1.09403404082571 -0.0398041681538916
1.11728638325419 -0.0416900860762064
1.14053872568267 -0.0434326170168506
1.16379106811115 -0.0466230697554336
1.18704341053963 -0.0505186823180925
1.21029575296811 -0.0537276023308732
1.23354809539659 -0.0570584304541876
1.25680043782507 -0.0587188600769918
1.28005278025355 -0.0604741685454762
1.30330512268203 -0.0613291036666277
1.32655746511051 -0.0685979487340099
1.34980980753899 -0.0739082542634761
1.37306214996747 -0.0787925440335498
1.39631449239595 -0.073435729935779
1.41956683482443 -0.0686930438639252
1.44281917725291 -0.0670027050051159
1.46607151968139 -0.0704648817130308
1.48932386210987 -0.0735423473292506
1.51257620453835 -0.0746029282527376
1.53582854696683 -0.0739698965865471
1.55908088939531 -0.0760126857002252
1.58233323182379 -0.0764544563047471
1.60558557425226 -0.0751627169190282
1.62883791668074 -0.0718183693337138
1.65209025910922 -0.0702957051369665
1.6753426015377 -0.0701894892302891
1.69859494396618 -0.0686692872725858
1.72184728639466 -0.0624381267094167
1.74509962882314 -0.0538493661408554
1.76835197125162 -0.0425202828623282
1.7916043136801 -0.0277960479315322
1.81485665610858 -0.0195288933456267
};
\addlegendentry{$\langle u_x\rangle\left(1-\langle u_x\rangle\right)$}
\addplot [semithick, color1, mark=*, mark size=2.5, mark options={solid,fill opacity=0}, only marks]
table {%
-1.64974236573492 0.00364273936368524
-1.62649002330644 0.00260772571673605
-1.60323768087796 0.00207286961654408
-1.57998533844948 0.00203496078927278
-1.556732996021 0.00196419183612067
-1.53348065359252 0.00192346886239641
-1.51022831116404 0.00205568017450576
-1.48697596873556 0.00226050499334899
-1.46372362630708 0.00225707146757527
-1.4404712838786 0.00206689433357669
-1.41721894145012 0.00200729585047764
-1.39396659902164 0.00199924747457329
-1.37071425659316 0.00200564342148329
-1.34746191416468 0.0019818831306879
-1.3242095717362 0.00191773992447939
-1.30095722930772 0.00193786551221055
-1.27770488687924 0.00191872083465946
-1.25445254445076 0.00191500817294537
-1.23120020202228 0.0019478730723189
-1.2079478595938 0.00191456917152801
-1.18469551716532 0.00176830591657922
-1.16144317473684 0.00178860300902955
-1.13819083230836 0.0019038742749992
-1.11493848987988 0.00197675164395118
-1.0916861474514 0.00204682768784179
-1.06843380502292 0.00218986654119739
-1.04518146259444 0.00240553484140442
-1.02192912016596 0.00268015783569734
-0.998676777737481 0.00285909745551262
-0.975424435309001 0.00314848894071272
-0.952172092880522 0.00351597677517378
-0.928919750452042 0.00401622706020024
-0.905667408023562 0.00461744004528895
-0.882415065595082 0.00568802127362963
-0.859162723166602 0.00692356479733573
-0.835910380738122 0.00881404794239997
-0.812658038309642 0.0109787212583908
-0.789405695881162 0.0135584429782085
-0.766153353452683 0.0166691813951489
-0.742901011024203 0.020682064743177
-0.719648668595723 0.0251313533867696
-0.696396326167243 0.0302383905561234
-0.673143983738763 0.0352223994059234
-0.649891641310283 0.0405810350946183
-0.626639298881803 0.0464425352214045
-0.603386956453323 0.0539037008160153
-0.580134614024844 0.0630724761856315
-0.556882271596364 0.0734881668748629
-0.533629929167884 0.0853987178043772
-0.510377586739404 0.099720965756948
-0.487125244310924 0.11702150386648
-0.463872901882444 0.136868695863778
-0.440620559453964 0.158600933220223
-0.417368217025484 0.181979530098274
-0.394115874597004 0.207313121938024
-0.370863532168525 0.235964211802427
-0.347611189740045 0.267218303297345
-0.324358847311565 0.30340427193087
-0.301106504883085 0.342050284831806
-0.277854162454605 0.383935209728187
-0.254601820026125 0.421509504605306
-0.231349477597645 0.453197149272609
-0.208097135169165 0.478100303744706
-0.184844792740686 0.50062897410158
-0.161592450312206 0.51787806760733
-0.138340107883726 0.530838568885217
-0.115087765455246 0.540666050606925
-0.0918354230267661 0.550482129065668
-0.0685830805982862 0.557430749790623
-0.0453307381698063 0.563632801428494
-0.0220783957413265 0.569435616054447
0.00117394668715341 0.576062166383117
0.0244262891156333 0.579709485066477
0.0476786315441132 0.579416077791829
0.070930973972593 0.575708209882547
0.0941833164010729 0.568676540125604
0.117435658829553 0.55618552461292
0.140688001258033 0.537058613069238
0.163940343686513 0.510244868776624
0.187192686114992 0.480875953237424
0.210445028543472 0.448398304869827
0.233697370971952 0.415605267844491
0.256949713400432 0.380960677453395
0.280202055828912 0.345045939053076
0.303454398257392 0.308868772039494
0.326706740685872 0.273247627927217
0.349959083114352 0.240909988240904
0.373211425542831 0.208905059274529
0.396463767971311 0.179173204058522
0.419716110399791 0.150372112032236
0.442968452828271 0.124408020289224
0.466220795256751 0.102094052273504
0.489473137685231 0.0836229353431103
0.512725480113711 0.0694528908643263
0.535977822542191 0.0582259661430814
0.55923016497067 0.0484684754216796
0.58248250739915 0.0404187724393983
0.60573484982763 0.0335637540128581
0.62898719225611 0.028242659375269
0.65223953468459 0.0235193133601844
0.67549187711307 0.019531331029641
0.69874421954155 0.0162825289607149
0.721996561970029 0.0138468014054532
0.745248904398509 0.0112541170454245
0.768501246826989 0.00893931269681083
0.791753589255469 0.0070791489180524
0.815005931683949 0.00590755106547932
0.838258274112429 0.00524311547744376
0.861510616540909 0.00482502984549579
0.884762958969389 0.00459773692114205
0.908015301397869 0.00435421159894171
0.931267643826348 0.00405179141439556
0.954519986254828 0.00401115731682543
0.977772328683308 0.00389307770530401
1.00102467111179 0.00388948854741115
1.02427701354027 0.00374288010212666
1.04752935596875 0.00392943557262057
1.07078169839723 0.00425369525333397
1.09403404082571 0.00452714991286119
1.11728638325419 0.00492881700161333
1.14053872568267 0.00483135178595011
1.16379106811115 0.00488214456904636
1.18704341053963 0.00467879292016065
1.21029575296811 0.00496326657717019
1.23354809539659 0.00549222784424778
1.25680043782507 0.00559569893863278
1.28005278025355 0.00552836015173899
1.30330512268203 0.00589697892944445
1.32655746511051 0.00587851188685832
1.34980980753899 0.00597307614436388
1.37306214996747 0.00649207078239554
1.39631449239595 0.00685305693483784
1.41956683482443 0.00626336200132631
1.44281917725291 0.00621416315947784
1.46607151968139 0.00675274492156898
1.48932386210987 0.00715972640606795
1.51257620453835 0.00751866429254554
1.53582854696683 0.00757882224427356
1.55908088939531 0.0082443827073562
1.58233323182379 0.00856059938464346
1.60558557425226 0.00833146243134775
1.62883791668074 0.00859751363305931
1.65209025910922 0.00849511315648037
1.6753426015377 0.00835880506559714
1.69859494396618 0.00794807707364002
1.72184728639466 0.00809153433989501
1.74509962882314 0.00799510727753952
1.76835197125162 0.00792906807138458
1.7916043136801 0.0101653319509015
1.81485665610858 0.0143409294894582
};
\addlegendentry{$\langle u^\prime_y\rangle^2$}
\addplot [semithick, color2, mark=diamond, mark size=2.5, mark options={solid,fill opacity=0}, only marks]
table {%
-1.64974236573492 -0.0032647734757158
-1.62649002330644 -0.00241658752116619
-1.60323768087796 -0.00203662709591876
-1.57998533844948 -0.0019742428675194
-1.556732996021 -0.00199160918975209
-1.53348065359252 -0.00202979500978477
-1.51022831116404 -0.00190998145699904
-1.48697596873556 -0.00182754440486147
-1.46372362630708 -0.00177695393198967
-1.4404712838786 -0.00172991600860699
-1.41721894145012 -0.00187714359185614
-1.39396659902164 -0.00199843286860022
-1.37071425659316 -0.00184471464583138
-1.34746191416468 -0.00176819572414969
-1.3242095717362 -0.00181605419954703
-1.30095722930772 -0.00176413557680843
-1.27770488687924 -0.00178535223047193
-1.25445254445076 -0.00187647490292447
-1.23120020202228 -0.00201368918161711
-1.2079478595938 -0.00193602584928477
-1.18469551716532 -0.00185672220346027
-1.16144317473684 -0.00180595519259808
-1.13819083230836 -0.00186964268553616
-1.11493848987988 -0.00195966737949368
-1.0916861474514 -0.00210448566120489
-1.06843380502292 -0.00219536458187559
-1.04518146259444 -0.00225842803178978
-1.02192912016596 -0.00248617679546271
-0.998676777737481 -0.00277399563918357
-0.975424435309001 -0.00319549262584381
-0.952172092880522 -0.00362652586386259
-0.928919750452042 -0.00411150148452903
-0.905667408023562 -0.00459486451506888
-0.882415065595082 -0.005338116479219
-0.859162723166602 -0.00636375425296226
-0.835910380738122 -0.00769031647417127
-0.812658038309642 -0.00919097124101446
-0.789405695881162 -0.011374350210142
-0.766153353452683 -0.0141794764494908
-0.742901011024203 -0.0175986189975815
-0.719648668595723 -0.0217307207232989
-0.696396326167243 -0.0267933296305684
-0.673143983738763 -0.0324501669974004
-0.649891641310283 -0.037626865615718
-0.626639298881803 -0.0416637577929947
-0.603386956453323 -0.0441739196353184
-0.580134614024844 -0.0463101828422902
-0.556882271596364 -0.0491553736782051
-0.533629929167884 -0.0543330969175678
-0.510377586739404 -0.0615536612059129
-0.487125244310924 -0.0702549667748113
-0.463872901882444 -0.0793978825771508
-0.440620559453964 -0.0888135488166479
-0.417368217025484 -0.0989263592519472
-0.394115874597004 -0.11012120400638
-0.370863532168525 -0.119718823937224
-0.347611189740045 -0.124920500287807
-0.324358847311565 -0.122491301830866
-0.301106504883085 -0.113052404059663
-0.277854162454605 -0.097279407702144
-0.254601820026125 -0.0774439772782752
-0.231349477597645 -0.0592541822656216
-0.208097135169165 -0.0458938320464497
-0.184844792740686 -0.0379760128464618
-0.161592450312206 -0.0344667698791383
-0.138340107883726 -0.0346311415590983
-0.115087765455246 -0.035932689858635
-0.0918354230267661 -0.0376186561679923
-0.0685830805982862 -0.0397891203847776
-0.0453307381698063 -0.042134427410212
-0.0220783957413265 -0.0445115033729923
0.00117394668715341 -0.0479252827082662
0.0244262891156333 -0.0522224357527412
0.0476786315441132 -0.0577072593716789
0.070930973972593 -0.0644677009385672
0.0941833164010729 -0.0711766907013447
0.117435658829553 -0.0784982744921026
0.140688001258033 -0.0857499282101489
0.163940343686513 -0.0935855260832059
0.187192686114992 -0.10208072516924
0.210445028543472 -0.111648618892958
0.233697370971952 -0.118684731653953
0.256949713400432 -0.122607274856582
0.280202055828912 -0.121945876211264
0.303454398257392 -0.121868963426331
0.326706740685872 -0.12306931760163
0.349959083114352 -0.124750841887664
0.373211425542831 -0.123589611224397
0.396463767971311 -0.120857085965522
0.419716110399791 -0.116904271009428
0.442968452828271 -0.113216678351629
0.466220795256751 -0.107436830197333
0.489473137685231 -0.100007566964697
0.512725480113711 -0.0890067837166106
0.535977822542191 -0.0767895630235096
0.55923016497067 -0.065655034704284
0.58248250739915 -0.0563559815038939
0.60573484982763 -0.0479737470143949
0.62898719225611 -0.0409972588729454
0.65223953468459 -0.0330955799123603
0.67549187711307 -0.0255680541987741
0.69874421954155 -0.0190439010741301
0.721996561970029 -0.0147128185511213
0.745248904398509 -0.0118503815214589
0.768501246826989 -0.00957103653259765
0.791753589255469 -0.00783654252873846
0.815005931683949 -0.00678254399899046
0.838258274112429 -0.0059463683411757
0.861510616540909 -0.00504987222713042
0.884762958969389 -0.0043762567607639
0.908015301397869 -0.00391098563292744
0.931267643826348 -0.00358929247017279
0.954519986254828 -0.00333054049959247
0.977772328683308 -0.00339205918453525
1.00102467111179 -0.00332079454475194
1.02427701354027 -0.00312153939557423
1.04752935596875 -0.0030116289183015
1.07078169839723 -0.00314816703020888
1.09403404082571 -0.0036938396323211
1.11728638325419 -0.00377323527969736
1.14053872568267 -0.00378204334500486
1.16379106811115 -0.00376208240847907
1.18704341053963 -0.00341532631297808
1.21029575296811 -0.00355706941828076
1.23354809539659 -0.00389793575907789
1.25680043782507 -0.00442789010738089
1.28005278025355 -0.00473778313914588
1.30330512268203 -0.00469243174298453
1.32655746511051 -0.00446939709991409
1.34980980753899 -0.00465148272194097
1.37306214996747 -0.00543850490247916
1.39631449239595 -0.00568881310455056
1.41956683482443 -0.00566117097969562
1.44281917725291 -0.00529570209802079
1.46607151968139 -0.00507629239617507
1.48932386210987 -0.00476368170983246
1.51257620453835 -0.00494741014136496
1.53582854696683 -0.00511389087016589
1.55908088939531 -0.00523526962123373
1.58233323182379 -0.00465506349355669
1.60558557425226 -0.0047901543279831
1.62883791668074 -0.00480046060972657
1.65209025910922 -0.00475875887557775
1.6753426015377 -0.0048071079103596
1.69859494396618 -0.00453436635154035
1.72184728639466 -0.00422489641735812
1.74509962882314 -0.0043859919361967
1.76835197125162 -0.0050643856965206
1.7916043136801 -0.00652612380796902
1.81485665610858 -0.00885662716106922
};
\addlegendentry{$-\langle u^\prime_x\rangle^2$}
\end{axis}

\end{tikzpicture}

%% file: tikz_matplotlib/drag_Trefftz_1.tikz
\begin{tikzpicture}

\definecolor{color0}{rgb}{0.12156862745098,0.466666666666667,0.705882352941177}
\definecolor{color1}{rgb}{1,0.498039215686275,0.0549019607843137}
\definecolor{color2}{rgb}{0.172549019607843,0.627450980392157,0.172549019607843}
\definecolor{color3}{rgb}{0.83921568627451,0.152941176470588,0.156862745098039}
\definecolor{color4}{rgb}{0.580392156862745,0.403921568627451,0.741176470588235}

\begin{axis}[thick,
legend style={
	fill opacity=0.8,
	draw opacity=1,
	text opacity=1,
	at={(0.03,0.03)},
	anchor=south west,
	draw=white!80!black
},
tick align=outside,
tick pos=left,
unbounded coords=jump,
x grid style={white!69.0196078431373!black},
xlabel={\(\displaystyle f^+\)},
xmajorgrids,
xmin=0.945087176144696, xmax=1.66204986149584,
xtick style={color=black},
y grid style={white!69.0196078431373!black},
ylabel={\(\displaystyle C_D\)},
ymajorgrids,
ymin=-0.700609554573197, ymax=-0.0165559158647663,
ytick style={color=black}
]
\path [draw=color0, semithick]
(axis cs:0.977676389115203,-0.227116236237486)
--(axis cs:0.977676389115203,-0.207167401613214);

\path [draw=color0, semithick]
(axis cs:1.14062245396774,-0.359191979923292)
--(axis cs:1.14062245396774,-0.319774895257931);

\path [draw=color0, semithick]
(axis cs:1.30356851882027,-0.554159981653062)
--(axis cs:1.30356851882027,-0.517620657742803);

\path [draw=color0, semithick]
(axis cs:1.4665145836728,-0.669516207359178)
--(axis cs:1.4665145836728,-0.590737080506158);

\path [draw=color0, semithick]
(axis cs:1.62946064852534,-0.448532657553124)
--(axis cs:1.62946064852534,-0.410141812730333);

\path [draw=color1, semithick]
(axis cs:0.977676389115203,-0.110690160764305)
--(axis cs:0.977676389115203,-0.0953206199275496);

\path [draw=color1, semithick]
(axis cs:1.14062245396774,-0.127911311639318)
--(axis cs:1.14062245396774,-0.11260471753306);

\path [draw=color1, semithick]
(axis cs:1.30356851882027,-0.247654149238841)
--(axis cs:1.30356851882027,-0.216609179965119);

\path [draw=color1, semithick]
(axis cs:1.4665145836728,-0.277174777652286)
--(axis cs:1.4665145836728,-0.258586166734726);

\path [draw=color1, semithick]
(axis cs:1.62946064852534,-0.198041529294227)
--(axis cs:1.62946064852534,-0.188501777833932);

\path [draw=color2, semithick]
;

\path [draw=color2, semithick]
(axis cs:1.14062245396774,-0.129858840741662)
--(axis cs:1.14062245396774,-0.110673662165718);

\path [draw=color2, semithick]
(axis cs:1.30356851882027,-0.133115674623791)
--(axis cs:1.30356851882027,-0.105179440692384);

\path [draw=color2, semithick]
(axis cs:1.4665145836728,-0.133710348922275)
--(axis cs:1.4665145836728,-0.102470905313982);

\path [draw=color2, semithick]
(axis cs:1.62946064852534,-0.141975663198765)
--(axis cs:1.62946064852534,-0.102255129032901);

\path [draw=color3, semithick]
;

\path [draw=color3, semithick]
;

\path [draw=color3, semithick]
(axis cs:1.30356851882027,-0.0876502075076221)
--(axis cs:1.30356851882027,-0.0742716905271296);

\path [draw=color3, semithick]
(axis cs:1.4665145836728,-0.0930695313326109)
--(axis cs:1.4665145836728,-0.0684123394330022);

\path [draw=color3, semithick]
(axis cs:1.62946064852534,-0.0876245018747165)
--(axis cs:1.62946064852534,-0.0493766330737095);

\path [draw=color4, semithick]
;

\path [draw=color4, semithick]
;

\path [draw=color4, semithick]
(axis cs:1.30356851882027,-0.0546881368366023)
--(axis cs:1.30356851882027,-0.0546881368366023);

\path [draw=color4, semithick]
(axis cs:1.4665145836728,-0.0774503087823711)
--(axis cs:1.4665145836728,-0.0621649952340057);

\path [draw=color4, semithick]
(axis cs:1.62946064852534,-0.0731956628067834)
--(axis cs:1.62946064852534,-0.0476492630787859);

\addplot [semithick, color0, mark=-, mark size=4, mark options={solid}, only marks,forget plot]
table {%
0.977676389115203 -0.227116236237486
1.14062245396774 -0.359191979923292
1.30356851882027 -0.554159981653062
1.4665145836728 -0.669516207359178
1.62946064852534 -0.448532657553124
};
\addplot [semithick, color0, mark=-, mark size=4, mark options={solid}, only marks, forget plot]
table {%
0.977676389115203 -0.207167401613214
1.14062245396774 -0.319774895257931
1.30356851882027 -0.517620657742803
1.4665145836728 -0.590737080506158
1.62946064852534 -0.410141812730333
};
\addplot [thick, color0, dashed,forget plot]
table {%
1.04062245396774 -0.243212945632298
1.07161393789182 -0.267058278946612
1.10260542181591 -0.296538895891521
1.13359690573999 -0.330385102839626
1.16458838966407 -0.367327206163516
1.19557987358816 -0.40609551223579
1.22657135751224 -0.445420327429053
1.25756284143633 -0.484031958115896
1.28855432536041 -0.520660710668908
1.3195458092845 -0.554036891460695
1.35053729320858 -0.582890806863851
1.38152877713266 -0.605952763250974
1.41252026105675 -0.621953066994656
1.44351174498083 -0.629622024467496
1.47450322890492 -0.627689942042098
1.505494712829 -0.61488712609105
1.53648619675309 -0.589943882986953
1.56747768067717 -0.551590519102392
1.59846916460125 -0.498557340809981
1.62946064852534 -0.429574654482307
};
\addplot [semithick, color1, mark=-, mark size=4, mark options={solid}, only marks, forget plot]
table {%
0.977676389115203 -0.110690160764305
1.14062245396774 -0.127911311639318
1.30356851882027 -0.247654149238841
1.4665145836728 -0.277174777652286
1.62946064852534 -0.198041529294227
};
\addplot [semithick, color1, mark=-, mark size=4, mark options={solid}, only marks, forget plot]
table {%
0.977676389115203 -0.0953206199275496
1.14062245396774 -0.11260471753306
1.30356851882027 -0.216609179965119
1.4665145836728 -0.258586166734726
1.62946064852534 -0.188501777833932
};
\addplot [thick, color1, dashed, forget plot]
table {%
1.04062245396774 -0.0972033998227406
1.07161393789182 -0.102706195745745
1.10260542181591 -0.112143749968409
1.13359690573999 -0.124810924487678
1.16458838966407 -0.140002581300502
1.19557987358816 -0.157013582403825
1.22657135751224 -0.175138789794599
1.25756284143633 -0.193673065469774
1.28855432536041 -0.211911271426293
1.3195458092845 -0.229148269661107
1.35053729320858 -0.244678922171164
1.38152877713266 -0.257798090953416
1.41252026105675 -0.267800638004801
1.44351174498083 -0.273981425322274
1.47450322890492 -0.275635314902786
1.505494712829 -0.272057168743284
1.53648619675309 -0.262541848840707
1.56747768067717 -0.246384217192016
1.59846916460125 -0.222879135794148
1.62946064852534 -0.191321466644064
};
\addplot [thick, color2, mark=-, mark size=4, mark options={solid}, only marks, forget plot]
table {%
0.977676389115203 nan
1.14062245396774 -0.129858840741662
1.30356851882027 -0.133115674623791
1.4665145836728 -0.133710348922275
1.62946064852534 -0.141975663198765
};
\addplot [thick, color2, mark=-, mark size=4, mark options={solid}, only marks, forget plot]
table {%
0.977676389115203 nan
1.14062245396774 -0.110673662165718
1.30356851882027 -0.105179440692384
1.4665145836728 -0.102470905313982
1.62946064852534 -0.102255129032901
};
\addplot [thick, color2, dashed, forget plot]
table {%
1.04062245396774 -0.118817759954268
1.07161393789182 -0.119536254274995
1.10260542181591 -0.119998486557634
1.13359690573999 -0.120238994524147
1.16458838966407 -0.120292315896496
1.19557987358816 -0.120192988396642
1.22657135751224 -0.119975549746547
1.25756284143633 -0.119674537668172
1.28855432536041 -0.119324489883478
1.3195458092845 -0.118959944114428
1.35053729320858 -0.118615438082982
1.38152877713266 -0.118325509511103
1.41252026105675 -0.118124696120751
1.44351174498083 -0.118047535633889
1.47450322890492 -0.118128565772477
1.505494712829 -0.118402324258478
1.53648619675309 -0.118903348813852
1.56747768067717 -0.119666177160562
1.59846916460125 -0.120725347020568
1.62946064852534 -0.122115396115833
};
\addplot [thick, color3, mark=-, mark size=4, mark options={solid}, only marks, forget plot]
table {%
0.977676389115203 nan
1.14062245396774 nan
1.30356851882027 -0.0876502075076221
1.4665145836728 -0.0930695313326109
1.62946064852534 -0.0876245018747165
};
\addplot [thick, color3, mark=-, mark size=4, mark options={solid}, only marks, forget plot]
table {%
0.977676389115203 nan
1.14062245396774 nan
1.30356851882027 -0.0742716905271296
1.4665145836728 -0.0684123394330022
1.62946064852534 -0.0493766330737095
};
\addplot [thick, color3, dashed, forget plot]
table {%
nan nan
nan nan
nan nan
nan nan
nan nan
nan nan
nan nan
nan nan
nan nan
nan nan
nan nan
nan nan
nan nan
nan nan
nan nan
nan nan
nan nan
nan nan
nan nan
1.62946064852534 -0.0685005674742121
};
\addplot [thick, color4, mark=-, mark size=4, mark options={solid}, only marks, forget plot]
table {%
0.977676389115203 nan
1.14062245396774 nan
1.30356851882027 -0.0546881368366023
1.4665145836728 -0.0774503087823711
1.62946064852534 -0.0731956628067834
};
\addplot [thick, color4, mark=-, mark size=4, mark options={solid}, only marks, forget plot]
table {%
0.977676389115203 nan
1.14062245396774 nan
1.30356851882027 -0.0546881368366023
1.4665145836728 -0.0621649952340057
1.62946064852534 -0.0476492630787859
};
\addplot [thick, color4, dashed, forget plot]
table {%
nan nan
nan nan
nan nan
nan nan
nan nan
nan nan
nan nan
nan nan
nan nan
nan nan
nan nan
nan nan
nan nan
nan nan
nan nan
nan nan
nan nan
nan nan
nan nan
1.62946064852534 -0.0604224629427824
};
\addlegendimage{empty legend}
\addlegendentry{\hspace{0cm}\textbf{$\bm U_R$}}
\addplot [thick, color0, mark=o, mark size=6, mark options={solid,fill opacity=0}, only marks]
table {%
0.977676389115203 -0.21714181892535
1.14062245396774 -0.339483437590611
1.30356851882027 -0.535890319697933
1.4665145836728 -0.630126643932668
1.62946064852534 -0.429337235141728
};
\addlegendentry{0.29}
\addplot [thick, color1, mark=square, mark size=6, mark options={solid,fill opacity=0}, only marks]
table {%
0.977676389115203 -0.103005390345927
1.14062245396774 -0.120258014586189
1.30356851882027 -0.23213166460198
1.4665145836728 -0.267880472193506
1.62946064852534 -0.19327165356408
};
\addlegendentry{0.37}
\addplot [thick, color2, mark=triangle, mark size=6, mark options={solid,rotate=270,fill opacity=0}, only marks]
table {%
0.977676389115203 nan
1.14062245396774 -0.12026625145369
1.30356851882027 -0.119147557658087
1.4665145836728 -0.118090627118128
1.62946064852534 -0.122115396115833
};
\addlegendentry{0.46}
\addplot [thick, color3, mark=diamond, mark size=6, mark options={solid,fill opacity=0}, only marks]
table {%
0.977676389115203 nan
1.14062245396774 nan
1.30356851882027 -0.0809609490173759
1.4665145836728 -0.0807409353828065
1.62946064852534 -0.068500567474213
};
\addlegendentry{0.54}
\addplot [thick, color4, mark=triangle, mark size=6, mark options={solid,rotate=90,fill opacity=0}, only marks]
table {%
0.977676389115203 nan
1.14062245396774 nan
1.30356851882027 -0.0546881368366023
1.4665145836728 -0.0698076520081884
1.62946064852534 -0.0604224629427847
};
\addlegendentry{0.59}
\end{axis}

\end{tikzpicture}

%% file: tikz_matplotlib/CD_trefftz_2.tikz
\begin{tikzpicture}

\definecolor{color0}{rgb}{0.12156862745098,0.466666666666667,0.705882352941177}
\definecolor{color1}{rgb}{1,0.498039215686275,0.0549019607843137}
\definecolor{color2}{rgb}{0.172549019607843,0.627450980392157,0.172549019607843}
\definecolor{color3}{rgb}{0.83921568627451,0.152941176470588,0.156862745098039}
\definecolor{color4}{rgb}{0.580392156862745,0.403921568627451,0.741176470588235}

\begin{axis}[
tick align=outside,
tick pos=left,
x grid style={white!69.0196078431373!black},
xlabel={\(\displaystyle f^+\)},
xmajorgrids,
xmin=0.945087176144696, xmax=1.66204986149584,
xtick style={color=black},
y grid style={white!69.0196078431373!black},
ylabel={\(\displaystyle \tilde \ell_{Trefftz}\)},
ymajorgrids,
ymin=1.01743451660235, ymax=3.97949727116626,
ytick style={color=black}
]
\addplot [semithick, color0, mark=o, mark size=5, mark options={solid,fill opacity=0}, only marks]
table {%
0.977676389115203 1.84403691795954
1.14062245396774 1.58989420195793
1.30356851882027 1.45353344289745
1.4665145836728 1.30291527292317
1.62946064852534 1.15207373271889
};
\addplot [semithick, color1, mark=square, mark size=5, mark options={solid,fill opacity=0}, only marks]
table {%
0.977676389115203 2.40500240500241
1.14062245396774 2.04695719812499
1.30356851882027 1.79952564503997
1.4665145836728 1.61188280000161
1.62946064852534 1.44624453681126
};
\addplot [semithick, color2, mark=triangle, mark size=5, mark options={solid,rotate=270,fill opacity=0}, only marks]
table {%
0.977676389115203 3.00379535550759
1.14062245396774 2.57040123346358
1.30356851882027 2.30117820324006
1.4665145836728 2.01160551546828
1.62946064852534 1.79956735521473
};
\addplot [semithick, black, mark=o, mark size=7.5, mark options={solid,fill opacity=0}, only marks]
table {%
0.977676389115203 3.00379535550759
};
\addplot [semithick, color3, mark=diamond, mark size=5, mark options={solid,fill opacity=0}, only marks]
table {%
0.977676389115203 3.51049033474776
1.14062245396774 3.00606183067286
1.30356851882027 2.61702619192007
1.4665145836728 2.33105179629244
1.62946064852534 2.08931309149174
};
\addplot [semithick, black, mark=o, mark size=7.5, mark options={solid,fill opacity=0}, only marks]
table {%
0.977676389115203 3.51049033474776
};
\addplot [semithick, black, mark=o, mark size=7.5, mark options={solid,fill opacity=0}, only marks]
table {%
1.14062245396774 3.00606183067286
};
\addplot [semithick, color4, mark=triangle, mark size=5, mark options={solid,rotate=90,fill opacity=0}, only marks]
table {%
0.977676389115203 3.84485805504972
1.14062245396774 3.29559261861404
1.30356851882027 2.89248907362252
1.4665145836728 2.55111151928895
1.62946064852534 2.29600036736006
};
\addplot [semithick, black, mark=o, mark size=7.5, mark options={solid,fill opacity=0}, only marks]
table {%
0.977676389115203 3.84485805504972
};
\addplot [semithick, black, mark=o, mark size=7.5, mark options={solid,fill opacity=0}, only marks]
table {%
1.14062245396774 3.29559261861404
};
\end{axis}

\end{tikzpicture}

%% file: tikz_matplotlib/circulacion_piv2.tikz
\begin{tikzpicture}

\definecolor{color0}{rgb}{0.12156862745098,0.466666666666667,0.705882352941177}
\definecolor{color1}{rgb}{1,0.498039215686275,0.0549019607843137}
\definecolor{color2}{rgb}{0.172549019607843,0.627450980392157,0.172549019607843}
\definecolor{color3}{rgb}{0.83921568627451,0.152941176470588,0.156862745098039}
\definecolor{color4}{rgb}{0.580392156862745,0.403921568627451,0.741176470588235}
\definecolor{color5}{rgb}{0.549019607843137,0.337254901960784,0.294117647058824}

\begin{axis}[thick,
legend cell align={left},
legend style={
  fill opacity=0.8,
  draw opacity=1,
  text opacity=1,
  at={(0.03,0.97)},
  anchor=north west,
  draw=white!80!black
},
tick align=outside,
tick pos=left,
x grid style={white!69.0196078431373!black},
xlabel={\(\displaystyle f^+\)},
xmajorgrids,
xmin=0.6, xmax=1.8,
xtick style={color=black},
y grid style={white!69.0196078431373!black},
ylabel={\(\displaystyle \Gamma / (U_\infty L)\)},
ymajorgrids,
ymin=-0.107923407681409, ymax=1.33404901816074,
ytick style={color=black}
]
\addlegendimage{empty legend}
\addlegendentry{\hspace{0cm}\textbf{$\bm U_R$}}
\addplot [thick, color0, mark=*, mark size=5, mark options={solid}, only marks]
table {%
0.977676389115203 0.735106268201301
1.14062245396774 0.972050597489822
1.30356851882027 1.10034907064396
1.4665145836728 1.09388406058652
1.62946064852534 0.95511156644131
};
\addlegendentry{$0.29$}
\addplot [thick, color0, forget plot]
table {%
0.96 0.698025073022839
0.967474747474747 0.71271406975281
0.974949494949495 0.727136664412166
0.982424242424242 0.741292857000912
0.98989898989899 0.755182647519043
0.997373737373737 0.768806035966563
1.00484848484848 0.782163022343469
1.01232323232323 0.795253606649763
1.01979797979798 0.808077788885443
1.02727272727273 0.820635569050511
1.03474747474747 0.832926947144966
1.04222222222222 0.844951923168809
1.04969696969697 0.856710497122039
1.05717171717172 0.868202669004656
1.06464646464646 0.87942843881666
1.07212121212121 0.890387806558051
1.07959595959596 0.90108077222883
1.08707070707071 0.911507335828996
1.09454545454545 0.92166749735855
1.1020202020202 0.93156125681749
1.10949494949495 0.941188614205818
1.1169696969697 0.950549569523532
1.12444444444444 0.959644122770634
1.13191919191919 0.968472273947124
1.13939393939394 0.977034023053001
1.14686868686869 0.985329370088264
1.15434343434343 0.993358315052915
1.16181818181818 1.00112085794695
1.16929292929293 1.00861699877038
1.17676767676768 1.01584673752319
1.18424242424242 1.02281007420539
1.19171717171717 1.02950700881698
1.19919191919192 1.03593754135795
1.20666666666667 1.04210167182831
1.21414141414141 1.04799940022806
1.22161616161616 1.0536307265572
1.22909090909091 1.05899565081572
1.23656565656566 1.06409417300363
1.2440404040404 1.06892629312093
1.25151515151515 1.07349201116762
1.2589898989899 1.07779132714369
1.26646464646465 1.08182424104915
1.27393939393939 1.085590752884
1.28141414141414 1.08909086264823
1.28888888888889 1.09232457034185
1.29636363636364 1.09529187596486
1.30383838383838 1.09799277951726
1.31131313131313 1.10042728099904
1.31878787878788 1.10259538041021
1.32626262626263 1.10449707775077
1.33373737373737 1.10613237302071
1.34121212121212 1.10750126622004
1.34868686868687 1.10860375734876
1.35616161616162 1.10943984640687
1.36363636363636 1.11000953339436
1.37111111111111 1.11031281831124
1.37858585858586 1.11034970115751
1.38606060606061 1.11012018193317
1.39353535353535 1.10962426063821
1.4010101010101 1.10886193727264
1.40848484848485 1.10783321183645
1.4159595959596 1.10653808432966
1.42343434343434 1.10497655475225
1.43090909090909 1.10314862310423
1.43838383838384 1.1010542893856
1.44585858585859 1.09869355359635
1.45333333333333 1.09606641573649
1.46080808080808 1.09317287580602
1.46828282828283 1.09001293380493
1.47575757575758 1.08658658973323
1.48323232323232 1.08289384359092
1.49070707070707 1.078934695378
1.49818181818182 1.07470914509446
1.50565656565657 1.07021719274031
1.51313131313131 1.06545883831555
1.52060606060606 1.06043408182018
1.52808080808081 1.05514292325419
1.53555555555556 1.04958536261759
1.5430303030303 1.04376139991038
1.55050505050505 1.03767103513255
1.5579797979798 1.03131426828411
1.56545454545455 1.02469109936506
1.57292929292929 1.0178015283754
1.58040404040404 1.01064555531512
1.58787878787879 1.00322318018423
1.59535353535354 0.995534402982732
1.60282828282828 0.987579223710617
1.61030303030303 0.979357642367889
1.61777777777778 0.970869658954549
1.62525252525253 0.962115273470595
1.63272727272727 0.953094485916029
1.64020202020202 0.943807296290851
1.64767676767677 0.934253704595059
1.65515151515152 0.924433710828655
1.66262626262626 0.914347314991638
1.67010101010101 0.903994517084008
1.67757575757576 0.893375317105766
1.6850505050505 0.882489715056911
1.69252525252525 0.871337710937441
1.7 0.859919304747363
};
\addplot [thick, color0, mark=o, mark size=5, mark options={solid,fill opacity=0}, only marks, forget plot]
table {%
0.488838194557601 0.114173659313473
0.651784259410135 0.235942174760311
0.814730324262669 0.387169383337108
0.977676389115203 0.653335720033413
1.14062245396774 0.95073376632549
1.30356851882027 1.12081953981871
1.4665145836728 1.24379919459342
1.62946064852534 1.26850481698609
1.79240671337787 1.23169748177057
};
\addplot [thick, color1, mark=square*, mark size=5, mark options={solid}, only marks]
table {%
0.977676389115203 0.549716012281086
1.14062245396774 0.664393958086351
1.30356851882027 0.879789497981515
1.4665145836728 0.90931746468905
1.62946064852534 0.894322415687099
};
\addlegendentry{$0.37$}
\addplot [thick, color1, forget plot]
table {%
0.96 0.504771851470498
0.967474747474747 0.515141714429665
0.974949494949495 0.525377740558437
0.982424242424242 0.535479929856814
0.98989898989899 0.545448282324795
0.997373737373737 0.55528279796238
1.00484848484848 0.56498347676957
1.01232323232323 0.574550318746363
1.01979797979798 0.583983323892763
1.02727272727273 0.593282492208766
1.03474747474747 0.602447823694374
1.04222222222222 0.611479318349586
1.04969696969697 0.620376976174402
1.05717171717172 0.629140797168824
1.06464646464646 0.63777078133285
1.07212121212121 0.646266928666481
1.07959595959596 0.654629239169715
1.08707070707071 0.662857712842555
1.09454545454545 0.670952349684999
1.1020202020202 0.678913149697047
1.10949494949495 0.686740112878701
1.1169696969697 0.694433239229958
1.12444444444444 0.70199252875082
1.13191919191919 0.709417981441287
1.13939393939394 0.716709597301357
1.14686868686869 0.723867376331033
1.15434343434343 0.730891318530313
1.16181818181818 0.737781423899198
1.16929292929293 0.744537692437687
1.17676767676768 0.75116012414578
1.18424242424242 0.757648719023479
1.19171717171717 0.764003477070781
1.19919191919192 0.770224398287688
1.20666666666667 0.776311482674199
1.21414141414141 0.782264730230316
1.22161616161616 0.788084140956037
1.22909090909091 0.793769714851362
1.23656565656566 0.799321451916292
1.2440404040404 0.804739352150826
1.25151515151515 0.810023415554964
1.2589898989899 0.815173642128708
1.26646464646465 0.820190031872055
1.27393939393939 0.825072584785008
1.28141414141414 0.829821300867565
1.28888888888889 0.834436180119726
1.29636363636364 0.838917222541491
1.30383838383838 0.843264428132862
1.31131313131313 0.847477796893837
1.31878787878788 0.851557328824416
1.32626262626263 0.8555030239246
1.33373737373737 0.859314882194388
1.34121212121212 0.862992903633781
1.34868686868687 0.866537088242779
1.35616161616162 0.869947436021381
1.36363636363636 0.873223946969587
1.37111111111111 0.876366621087398
1.37858585858586 0.879375458374813
1.38606060606061 0.882250458831833
1.39353535353535 0.884991622458458
1.4010101010101 0.887598949254687
1.40848484848485 0.89007243922052
1.4159595959596 0.892412092355958
1.42343434343434 0.894617908661001
1.43090909090909 0.896689888135648
1.43838383838384 0.898628030779899
1.44585858585859 0.900432336593755
1.45333333333333 0.902102805577216
1.46080808080808 0.903639437730281
1.46828282828283 0.90504223305295
1.47575757575758 0.906311191545224
1.48323232323232 0.907446313207103
1.49070707070707 0.908447598038586
1.49818181818182 0.909315046039674
1.50565656565657 0.910048657210366
1.51313131313131 0.910648431550662
1.52060606060606 0.911114369060564
1.52808080808081 0.91144646974007
1.53555555555556 0.91164473358918
1.5430303030303 0.911709160607894
1.55050505050505 0.911639750796214
1.5579797979798 0.911436504154137
1.56545454545455 0.911099420681666
1.57292929292929 0.910628500378798
1.58040404040404 0.910023743245536
1.58787878787879 0.909285149281877
1.59535353535354 0.908412718487824
1.60282828282828 0.907406450863375
1.61030303030303 0.90626634640853
1.61777777777778 0.90499240512329
1.62525252525253 0.903584627007654
1.63272727272727 0.902043012061623
1.64020202020202 0.900367560285197
1.64767676767677 0.898558271678374
1.65515151515152 0.896615146241157
1.66262626262626 0.894538183973544
1.67010101010101 0.892327384875535
1.67757575757576 0.889982748947131
1.6850505050505 0.887504276188331
1.69252525252525 0.884891966599136
1.7 0.882145820179546
};
\addplot [thick, color1, mark=square, mark size=5, mark options={solid,fill opacity=0}, only marks, forget plot]
table {%
0.488838194557601 0.0597412894660873
0.651784259410135 0.15140587599969
0.814730324262669 0.262374888146545
0.977676389115203 0.427576330913833
1.14062245396774 0.630798661455741
1.30356851882027 0.787004959758949
1.4665145836728 0.896494128497957
1.62946064852534 0.943484746739188
1.79240671337787 0.918801673301881
};
\addplot [thick, color2, mark=triangle*, mark size=5, mark options={solid,rotate=270}, only marks]
table {%
0.977676389115203 0.334240914720801
1.14062245396774 0.493224960824863
1.30356851882027 0.599419422330518
1.4665145836728 0.677404631796694
1.62946064852534 0.65037073853954
};
\addlegendentry{$0.46$}
\addplot [thick, color2, forget plot]
table {%
0.96 0.308867916370854
0.967474747474747 0.318083288919246
0.974949494949495 0.327178342716913
0.982424242424242 0.336153077763854
0.98989898989899 0.345007494060069
0.997373737373737 0.35374159160556
1.00484848484848 0.362355370400324
1.01232323232323 0.370848830444363
1.01979797979798 0.379221971737676
1.02727272727273 0.387474794280264
1.03474747474747 0.395607298072127
1.04222222222222 0.403619483113263
1.04969696969697 0.411511349403674
1.05717171717172 0.419282896943361
1.06464646464646 0.426934125732321
1.07212121212121 0.434465035770556
1.07959595959596 0.441875627058065
1.08707070707071 0.449165899594849
1.09454545454545 0.456335853380907
1.1020202020202 0.46338548841624
1.10949494949495 0.470314804700847
1.1169696969697 0.477123802234728
1.12444444444444 0.483812481017885
1.13191919191919 0.490380841050315
1.13939393939394 0.49682888233202
1.14686868686869 0.503156604863
1.15434343434343 0.509364008643254
1.16181818181818 0.515451093672783
1.16929292929293 0.521417859951586
1.17676767676768 0.527264307479663
1.18424242424242 0.532990436257015
1.19171717171717 0.538596246283642
1.19919191919192 0.544081737559542
1.20666666666667 0.549446910084718
1.21414141414141 0.554691763859168
1.22161616161616 0.559816298882892
1.22909090909091 0.564820515155891
1.23656565656566 0.569704412678165
1.2440404040404 0.574467991449713
1.25151515151515 0.579111251470535
1.2589898989899 0.583634192740631
1.26646464646465 0.588036815260003
1.27393939393939 0.592319119028649
1.28141414141414 0.596481104046569
1.28888888888889 0.600522770313763
1.29636363636364 0.604444117830233
1.30383838383838 0.608245146595977
1.31131313131313 0.611925856610995
1.31878787878788 0.615486247875287
1.32626262626263 0.618926320388855
1.33373737373737 0.622246074151696
1.34121212121212 0.625445509163812
1.34868686868687 0.628524625425203
1.35616161616162 0.631483422935868
1.36363636363636 0.634321901695807
1.37111111111111 0.637040061705022
1.37858585858586 0.63963790296351
1.38606060606061 0.642115425471273
1.39353535353535 0.644472629228311
1.4010101010101 0.646709514234623
1.40848484848485 0.648826080490209
1.4159595959596 0.65082232799507
1.42343434343434 0.652698256749205
1.43090909090909 0.654453866752615
1.43838383838384 0.6560891580053
1.44585858585859 0.657604130507258
1.45333333333333 0.658998784258492
1.46080808080808 0.660273119258999
1.46828282828283 0.661427135508782
1.47575757575758 0.662460833007839
1.48323232323232 0.66337421175617
1.49070707070707 0.664167271753776
1.49818181818182 0.664840013000656
1.50565656565657 0.665392435496811
1.51313131313131 0.66582453924224
1.52060606060606 0.666136324236944
1.52808080808081 0.666327790480922
1.53555555555556 0.666398937974175
1.5430303030303 0.666349766716701
1.55050505050505 0.666180276708503
1.5579797979798 0.665890467949579
1.56545454545455 0.665480340439929
1.57292929292929 0.664949894179555
1.58040404040404 0.664299129168454
1.58787878787879 0.663528045406628
1.59535353535354 0.662636642894077
1.60282828282828 0.6616249216308
1.61030303030303 0.660492881616797
1.61777777777778 0.659240522852069
1.62525252525253 0.657867845336616
1.63272727272727 0.656374849070436
1.64020202020202 0.654761534053532
1.64767676767677 0.653027900285902
1.65515151515152 0.651173947767546
1.66262626262626 0.649199676498465
1.67010101010101 0.647105086478658
1.67757575757576 0.644890177708126
1.6850505050505 0.642554950186868
1.69252525252525 0.640099403914885
1.7 0.637523538892176
};
\addplot [thick, color2, mark=triangle, mark size=5, mark options={solid,rotate=270,fill opacity=0}, only marks, forget plot]
table {%
0.488838194557601 0.0127886908064179
0.651784259410135 0.0802320511003472
0.814730324262669 0.159936928658042
0.977676389115203 0.280361361614566
1.14062245396774 0.458228934974922
1.30356851882027 0.559304947286646
1.4665145836728 0.65438735639144
1.62946064852534 0.709090386783007
1.79240671337787 0.701686875225834
};
\addplot [thick, color3, mark=diamond*, mark size=5, mark options={solid}, only marks]
table {%
0.977676389115203 0.230420115993138
1.14062245396774 0.347614379293994
1.30356851882027 0.415189756022638
1.4665145836728 0.606853839872708
1.62946064852534 0.495117002598543
};
\addlegendentry{$0.54$}
\addplot [thick, color3, forget plot]
table {%
0.96 0.194449721595452
0.967474747474747 0.202629071253742
0.974949494949495 0.210708084376394
0.982424242424242 0.218686760963407
0.98989898989899 0.226565101014782
0.997373737373737 0.234343104530518
1.00484848484848 0.242020771510616
1.01232323232323 0.249598101955075
1.01979797979798 0.257075095863895
1.02727272727273 0.264451753237078
1.03474747474747 0.271728074074622
1.04222222222222 0.278904058376527
1.04969696969697 0.285979706142794
1.05717171717172 0.292955017373423
1.06464646464646 0.299829992068413
1.07212121212121 0.306604630227765
1.07959595959596 0.313278931851477
1.08707070707071 0.319852896939552
1.09454545454545 0.326326525491989
1.1020202020202 0.332699817508786
1.10949494949495 0.338972772989945
1.1169696969697 0.345145391935466
1.12444444444444 0.351217674345348
1.13191919191919 0.357189620219592
1.13939393939394 0.363061229558198
1.14686868686869 0.368832502361165
1.15434343434343 0.374503438628493
1.16181818181818 0.380074038360183
1.16929292929293 0.385544301556235
1.17676767676768 0.390914228216648
1.18424242424242 0.396183818341423
1.19171717171717 0.401353071930559
1.19919191919192 0.406421988984057
1.20666666666667 0.411390569501916
1.21414141414141 0.416258813484136
1.22161616161616 0.421026720930719
1.22909090909091 0.425694291841663
1.23656565656566 0.430261526216968
1.2440404040404 0.434728424056635
1.25151515151515 0.439094985360663
1.2589898989899 0.443361210129053
1.26646464646465 0.447527098361805
1.27393939393939 0.451592650058918
1.28141414141414 0.455557865220392
1.28888888888889 0.459422743846228
1.29636363636364 0.463187285936426
1.30383838383838 0.466851491490985
1.31131313131313 0.470415360509906
1.31878787878788 0.473878892993188
1.32626262626263 0.477242088940832
1.33373737373737 0.480504948352837
1.34121212121212 0.483667471229204
1.34868686868687 0.486729657569932
1.35616161616162 0.489691507375022
1.36363636363636 0.492553020644474
1.37111111111111 0.495314197378287
1.37858585858586 0.497975037576462
1.38606060606061 0.500535541238998
1.39353535353535 0.502995708365895
1.4010101010101 0.505355538957154
1.40848484848485 0.507615033012775
1.4159595959596 0.509774190532758
1.42343434343434 0.511833011517101
1.43090909090909 0.513791495965806
1.43838383838384 0.515649643878873
1.44585858585859 0.517407455256302
1.45333333333333 0.519064930098092
1.46080808080808 0.520622068404243
1.46828282828283 0.522078870174756
1.47575757575758 0.523435335409631
1.48323232323232 0.524691464108867
1.49070707070707 0.525847256272464
1.49818181818182 0.526902711900423
1.50565656565657 0.527857830992744
1.51313131313131 0.528712613549426
1.52060606060606 0.52946705957047
1.52808080808081 0.530121169055875
1.53555555555556 0.530674942005642
1.5430303030303 0.53112837841977
1.55050505050505 0.53148147829826
1.5579797979798 0.531734241641111
1.56545454545455 0.531886668448325
1.57292929292929 0.531938758719899
1.58040404040404 0.531890512455835
1.58787878787879 0.531741929656133
1.59535353535354 0.531493010320792
1.60282828282828 0.531143754449812
1.61030303030303 0.530694162043194
1.61777777777778 0.530144233100938
1.62525252525253 0.529493967623043
1.63272727272727 0.52874336560951
1.64020202020202 0.527892427060338
1.64767676767677 0.526941151975528
1.65515151515152 0.52588954035508
1.66262626262626 0.524737592198993
1.67010101010101 0.523485307507267
1.67757575757576 0.522132686279903
1.6850505050505 0.520679728516901
1.69252525252525 0.51912643421826
1.7 0.51747280338398
};
\addplot [thick, color3, mark=diamond, mark size=5, mark options={solid,fill opacity=0}, only marks, forget plot]
table {%
0.488838194557601 -0.0130635760358038
0.651784259410135 0.0448422028971108
0.814730324262669 0.112402786388467
0.977676389115203 0.193851235090395
1.14062245396774 0.336001057817464
1.30356851882027 0.431578071516923
1.4665145836728 0.526177724267605
1.62946064852534 0.583624123258458
1.79240671337787 0.572629248393012
};
\addplot [thick, color4, mark=triangle*, mark size=5, mark options={solid,rotate=90}, only marks]
table {%
0.977676389115203 0.11828852960847
1.14062245396774 0.205590036784776
1.30356851882027 0.308586148424784
1.4665145836728 0.544988941319155
1.62946064852534 0.461851254829283
};
\addlegendentry{$0.59$}
\addplot [thick, color4, forget plot]
table {%
0.96 0.0751767290449528
0.967474747474747 0.0827211759672755
0.974949494949495 0.090203254269551
0.982424242424242 0.0976229639517796
0.98989898989899 0.104980305013961
0.997373737373737 0.112275277456095
1.00484848484848 0.119507881278183
1.01232323232323 0.126678116480223
1.01979797979798 0.133785983062216
1.02727272727273 0.140831481024162
1.03474747474747 0.14781461036606
1.04222222222222 0.154735371087912
1.04969696969697 0.161593763189717
1.05717171717172 0.168389786671474
1.06464646464646 0.175123441533184
1.07212121212121 0.181794727774847
1.07959595959596 0.188403645396464
1.08707070707071 0.194950194398033
1.09454545454545 0.201434374779555
1.1020202020202 0.207856186541029
1.10949494949495 0.214215629682457
1.1169696969697 0.220512704203838
1.12444444444444 0.226747410105171
1.13191919191919 0.232919747386457
1.13939393939394 0.239029716047696
1.14686868686869 0.245077316088889
1.15434343434343 0.251062547510034
1.16181818181818 0.256985410311132
1.16929292929293 0.262845904492182
1.17676767676768 0.268644030053186
1.18424242424242 0.274379786994142
1.19171717171717 0.280053175315052
1.19919191919192 0.285664195015914
1.20666666666667 0.291212846096729
1.21414141414141 0.296699128557498
1.22161616161616 0.302123042398219
1.22909090909091 0.307484587618892
1.23656565656566 0.312783764219519
1.2440404040404 0.318020572200099
1.25151515151515 0.323195011560631
1.2589898989899 0.328307082301117
1.26646464646465 0.333356784421555
1.27393939393939 0.338344117921946
1.28141414141414 0.34326908280229
1.28888888888889 0.348131679062587
1.29636363636364 0.352931906702837
1.30383838383838 0.35766976572304
1.31131313131313 0.362345256123196
1.31878787878788 0.366958377903304
1.32626262626263 0.371509131063366
1.33373737373737 0.37599751560338
1.34121212121212 0.380423531523347
1.34868686868687 0.384787178823267
1.35616161616162 0.38908845750314
1.36363636363636 0.393327367562966
1.37111111111111 0.397503909002745
1.37858585858586 0.401618081822477
1.38606060606061 0.405669886022161
1.39353535353535 0.409659321601798
1.4010101010101 0.413586388561389
1.40848484848485 0.417451086900932
1.4159595959596 0.421253416620428
1.42343434343434 0.424993377719877
1.43090909090909 0.428670970199279
1.43838383838384 0.432286194058634
1.44585858585859 0.435839049297941
1.45333333333333 0.439329535917202
1.46080808080808 0.442757653916415
1.46828282828283 0.446123403295581
1.47575757575758 0.449426784054701
1.48323232323232 0.452667796193773
1.49070707070707 0.455846439712798
1.49818181818182 0.458962714611776
1.50565656565657 0.462016620890706
1.51313131313131 0.46500815854959
1.52060606060606 0.467937327588426
1.52808080808081 0.470804128007216
1.53555555555556 0.473608559805959
1.5430303030303 0.476350622984654
1.55050505050505 0.479030317543301
1.5579797979798 0.481647643481902
1.56545454545455 0.484202600800457
1.57292929292929 0.486695189498963
1.58040404040404 0.489125409577423
1.58787878787879 0.491493261035835
1.59535353535354 0.4937987438742
1.60282828282828 0.496041858092519
1.61030303030303 0.498222603690791
1.61777777777778 0.500340980669014
1.62525252525253 0.502396989027191
1.63272727272727 0.504390628765321
1.64020202020202 0.506321899883404
1.64767676767677 0.50819080238144
1.65515151515152 0.509997336259428
1.66262626262626 0.511741501517369
1.67010101010101 0.513423298155264
1.67757575757576 0.515042726173111
1.6850505050505 0.516599785570911
1.69252525252525 0.518094476348664
1.7 0.51952679850637
};
\addplot [thick, color4, mark=triangle, mark size=5, mark options={solid,rotate=90,fill opacity=0}, only marks, forget plot]
table {%
0.488838194557601 -0.0315712393343052
0.651784259410135 0.0162428803181606
0.814730324262669 0.0739098233877488
0.977676389115203 0.142923674789118
1.14062245396774 0.259185739756125
1.30356851882027 0.32924453618522
1.4665145836728 0.41835687430467
1.62946064852534 0.476817706008388
1.79240671337787 0.547855973511465
};
\addplot [thick, color5, mark=diamond*, mark size=5, mark options={solid}, only marks]
table {%
0.977676389115203 0.185815721237276
1.14062245396774 0.222967058232285
1.30356851882027 0.289007428626343
1.4665145836728 0.437313968956185
1.62946064852534 0.490850027586579
};
\addlegendentry{$0.66$}
\addplot [thick, color5, forget plot]
table {%
0.96 0.171460277251197
0.967474747474747 0.17366988512464
0.974949494949495 0.175914074158993
0.982424242424242 0.178192844354254
0.98989898989899 0.180506195710425
0.997373737373737 0.182854128227505
1.00484848484848 0.185236641905493
1.01232323232323 0.187653736744391
1.01979797979798 0.190105412744198
1.02727272727273 0.192591669904914
1.03474747474747 0.19511250822654
1.04222222222222 0.197667927709074
1.04969696969697 0.200257928352517
1.05717171717172 0.20288251015687
1.06464646464646 0.205541673122131
1.07212121212121 0.208235417248302
1.07959595959596 0.210963742535381
1.08707070707071 0.21372664898337
1.09454545454545 0.216524136592268
1.1020202020202 0.219356205362075
1.10949494949495 0.222222855292791
1.1169696969697 0.225124086384416
1.12444444444444 0.22805989863695
1.13191919191919 0.231030292050394
1.13939393939394 0.234035266624746
1.14686868686869 0.237074822360008
1.15434343434343 0.240148959256178
1.16181818181818 0.243257677313258
1.16929292929293 0.246400976531247
1.17676767676768 0.249578856910144
1.18424242424242 0.252791318449951
1.19171717171717 0.256038361150667
1.19919191919192 0.259319985012292
1.20666666666667 0.262636190034827
1.21414141414141 0.26598697621827
1.22161616161616 0.269372343562622
1.22909090909091 0.272792292067884
1.23656565656566 0.276246821734054
1.2440404040404 0.279735932561134
1.25151515151515 0.283259624549122
1.2589898989899 0.28681789769802
1.26646464646465 0.290410752007827
1.27393939393939 0.294038187478543
1.28141414141414 0.297700204110168
1.28888888888889 0.301396801902702
1.29636363636364 0.305127980856146
1.30383838383838 0.308893740970498
1.31131313131313 0.312694082245759
1.31878787878788 0.31652900468193
1.32626262626263 0.320398508279009
1.33373737373737 0.324302593036998
1.34121212121212 0.328241258955896
1.34868686868687 0.332214506035702
1.35616161616162 0.336222334276418
1.36363636363636 0.340264743678043
1.37111111111111 0.344341734240577
1.37858585858586 0.348453305964021
1.38606060606061 0.352599458848373
1.39353535353535 0.356780192893634
1.4010101010101 0.360995508099805
1.40848484848485 0.365245404466884
1.4159595959596 0.369529881994873
1.42343434343434 0.373848940683771
1.43090909090909 0.378202580533577
1.43838383838384 0.382590801544293
1.44585858585859 0.387013603715918
1.45333333333333 0.391470987048452
1.46080808080808 0.395962951541896
1.46828282828283 0.400489497196248
1.47575757575758 0.405050624011509
1.48323232323232 0.40964633198768
1.49070707070707 0.414276621124759
1.49818181818182 0.418941491422748
1.50565656565657 0.423640942881645
1.51313131313131 0.428374975501452
1.52060606060606 0.433143589282168
1.52808080808081 0.437946784223793
1.53555555555556 0.442784560326327
1.5430303030303 0.44765691758977
1.55050505050505 0.452563856014122
1.5579797979798 0.457505375599383
1.56545454545455 0.462481476345554
1.57292929292929 0.467492158252633
1.58040404040404 0.472537421320622
1.58787878787879 0.477617265549519
1.59535353535354 0.482731690939326
1.60282828282828 0.487880697490042
1.61030303030303 0.493064285201667
1.61777777777778 0.498282454074201
1.62525252525253 0.503535204107644
1.63272727272727 0.508822535301996
1.64020202020202 0.514144447657257
1.64767676767677 0.519500941173427
1.65515151515152 0.524892015850507
1.66262626262626 0.530317671688495
1.67010101010101 0.535777908687393
1.67757575757576 0.5412727268472
1.6850505050505 0.546802126167916
1.69252525252525 0.55236610664954
1.7 0.557964668292074
};
\addplot [thick, color5, mark=diamond, mark size=5, mark options={solid,fill opacity=0}, only marks, forget plot]
table {%
0.488838194557601 -0.0423792065067662
0.651784259410135 -0.00212864362197083
0.814730324262669 0.0454955685061092
0.977676389115203 0.10381640605695
1.14062245396774 0.182429109424235
1.30356851882027 0.269666318069076
1.4665145836728 0.344024004306531
1.62946064852534 0.392462594213513
1.79240671337787 0.431886613015992
};
\end{axis}

\end{tikzpicture}

%% file: tikzs/Circulation_Scheme.tikz
\begin{tikzpicture}[thick]
\pgfmathsetmacro{\radiovortex}{0.9}
\pgfmathsetmacro{\xvortex}{3.75}
\pgfmathsetmacro{\yvortex}{0.9}
\draw[dashed] (-2,-1) -- (-2,2);
\draw[dashed] (-1,-1) -- (-1,2);
\draw[->] (-1.9,-0.5) -- (-1.1,-0.5);
\draw[->] (-1.9,0.5) -- (-1.1,0.5);
\draw[->] (-1.9,1.5) -- (-1.1,1.5);
\node () at (-1.5,2) {$U$};

\draw[ultra thick](0,0).. controls(1,0.33) ..(2.7,1.8);


\begin{scope}[xshift=-0.85cm,yshift=-0.7cm]
\draw[dashed](2.7,1.8)--++(39:10mm);
\draw[dashed](2.7,1.8)--++(0:7mm);
\node[]at (3.15,2) {$\theta$} ;
\end{scope}

\centerarct[<->,black!70](0,0)(-20:20:1)(\small{$\omega_f$})

\draw[dashed](0,0)--(0,-.7);
\draw[black!70,<->] (0,-.7)--++(3.16,0)  node[midway,fill=white] {$L$};


\centerarc[color0,thick,-<](\xvortex,\radiovortex)(270:180:\radiovortex) 
\centerarc[color0,thick,->](\xvortex,\radiovortex)(180:90:\radiovortex) 
\centerarc[color0,thick](\xvortex,\radiovortex)(90:-90:\radiovortex) 

\draw[<->,color=black!70](\xvortex,\radiovortex*2)--++(0,-\radiovortex*2) node[midway,fill=white]{$\delta$};

\draw[black!70][->] (-.5,0)--++(1.2,0) ++ (.6,0) -- (5,0)node[right,below]{$x$};
\draw[black!70][->] (0,-1) -- (0,2) node[right]{$y$};
\node[color0] at (\xvortex,\yvortex+1.2){$\Gamma_U \sim  U\cos(\theta) \delta$};
\node[color0,fill=white,inner sep=1pt] at (\xvortex-1,\yvortex-0.6){$\Gamma_{u_\theta} \sim  \omega_f L  \delta$};
\end{tikzpicture}
		

%% file: tikz_matplotlib/freq_globales_f.tikz
\begin{tikzpicture}

\definecolor{color0}{rgb}{0.12156862745098,0.466666666666667,0.705882352941177}
\definecolor{color1}{rgb}{1,0.498039215686275,0.0549019607843137}

\begin{axis}[name=principal,thick,width=7cm,
legend cell align={left},
legend style={
  fill opacity=0.8,
  draw opacity=1,
  text opacity=1,
  at={(0.03,0.97)},
  anchor=north west,
  draw=white!80!black
},
tick align=outside,
tick pos=left,
x grid style={white!69.0196078431373!black},
xlabel={\(\displaystyle f^+\)},
xmajorgrids,
xmin=0.945087176144696, xmax=1.66204986149584,
xtick style={color=black},
y grid style={white!69.0196078431373!black},
ylabel={\(\displaystyle St\)},
ymajorgrids,
ymin=0.09, ymax=0.27,
ytick style={color=black}
]
\addlegendimage{empty legend}
\addlegendentry{$\bm U_R$}
\addplot [semithick, color0, mark=o, mark size=6, mark options={solid,fill opacity=0}, only marks]
table {%
0.977676389115203 0.149605646506382
1.14062245396774 0.14610423775836
1.30356851882027 0.186211283417518
1.4665145836728 0.224408469759572
1.62946064852534 0.257353542979595
};
\addlegendentry{$0.29$}
\addplot [semithick, color1, mark=square, mark size=6, mark options={solid,fill opacity=0}, only marks]
table {%
0.977676389115203 0.0986760647169751
1.14062245396774 0.114591559026165
1.30356851882027 0.110612685448867
1.4665145836728 0.139578885091592
1.62946064852534 0.0986760647169751
};
\addlegendentry{$0.37$}
%
\coordinate (insetPosition) at (rel axis cs:0.0,0.95);
\draw[color=color0,opacity=0.5,line width=9] (axis cs:0.97,0.205)--(axis cs:1.62,0.205); 
\draw[color=color1,opacity=0.5,line width=9] (axis cs:0.97,0.153)--(axis cs:1.62,0.153); 
\path[line0]
(axis cs:0.98,0.2055) -- (axis cs:1.6,0.2055) --(1.6,0.2054)--(0.98,0.2054)--cycle ;
\path[line1]
(axis cs:0.98,0.153) -- (axis cs:1.6,0.153) --(1.6,0.1529)--(0.98,0.1529)--cycle ;

\end{axis}

\begin{axis}[at={(insetPosition)}, thick,width=7cm, height=3cm, yshift=0.5cm,
label style={font=\small},
tick align=outside,
tick pos=left,
unbounded coords=jump,
xmajorgrids,
xmin=0.945087176144696, xmax=1.66204986149584,
xticklabels={,,},
y grid style={white!69.0196078431373!black},
ylabel={\(\displaystyle C_D\)},
ymajorgrids,
ymin=-0.700609554573197, ymax=-0.02,
ytick style={color=black}
]

\addplot [thick, color0, dashed,forget plot]
table {%
	1.04062245396774 -0.243212945632298
	1.07161393789182 -0.267058278946612
	1.10260542181591 -0.296538895891521
	1.13359690573999 -0.330385102839626
	1.16458838966407 -0.367327206163516
	1.19557987358816 -0.40609551223579
	1.22657135751224 -0.445420327429053
	1.25756284143633 -0.484031958115896
	1.28855432536041 -0.520660710668908
	1.3195458092845 -0.554036891460695
	1.35053729320858 -0.582890806863851
	1.38152877713266 -0.605952763250974
	1.41252026105675 -0.621953066994656
	1.44351174498083 -0.629622024467496
	1.47450322890492 -0.627689942042098
	1.505494712829 -0.61488712609105
	1.53648619675309 -0.589943882986953
	1.56747768067717 -0.551590519102392
	1.59846916460125 -0.498557340809981
	1.62946064852534 -0.429574654482307
};

\addplot [thick, color1, dashed, forget plot]
table {%
	1.04062245396774 -0.0972033998227406
	1.07161393789182 -0.102706195745745
	1.10260542181591 -0.112143749968409
	1.13359690573999 -0.124810924487678
	1.16458838966407 -0.140002581300502
	1.19557987358816 -0.157013582403825
	1.22657135751224 -0.175138789794599
	1.25756284143633 -0.193673065469774
	1.28855432536041 -0.211911271426293
	1.3195458092845 -0.229148269661107
	1.35053729320858 -0.244678922171164
	1.38152877713266 -0.257798090953416
	1.41252026105675 -0.267800638004801
	1.44351174498083 -0.273981425322274
	1.47450322890492 -0.275635314902786
	1.505494712829 -0.272057168743284
	1.53648619675309 -0.262541848840707
	1.56747768067717 -0.246384217192016
	1.59846916460125 -0.222879135794148
	1.62946064852534 -0.191321466644064
};

\addplot [thick, color0, mark=o, mark size=4, mark options={solid,fill opacity=0}, only marks]
table {%
	0.977676389115203 -0.21714181892535
	1.14062245396774 -0.339483437590611
	1.30356851882027 -0.535890319697933
	1.4665145836728 -0.630126643932668
	1.62946064852534 -0.429337235141728
};
\addplot [thick, color1, mark=square, mark size=4, mark options={solid,fill opacity=0}, only marks]
table {%
	0.977676389115203 -0.103005390345927
	1.14062245396774 -0.120258014586189
	1.30356851882027 -0.23213166460198
	1.4665145836728 -0.267880472193506
	1.62946064852534 -0.19327165356408
};

\end{axis}
\draw[color=color0,opacity=0.5,line width=7] (3.75,4.9)--++(0,-2.4); 
\draw[color=color1,opacity=0.5,line width=7] (3.95,5.6)--++(0,-4.4); 

\path[line1]
(3.95,5.5) --++ (0,-4.2) --++(-0.005,0)--++(0,4.2)--cycle ;
\path[line0]
(3.75,4.85) --++ (0,-2.2) --++(-0.005,0)--++(0,2.2)--cycle ;
\end{tikzpicture}

%% file: tikz_matplotlib/cluster_method_f.tikz
\begin{tikzpicture}

\node[] (inst1) at (0,0) {\includegraphics[width=3cm,trim=4cm 1.8cm 7cm 2.8cm,clip]{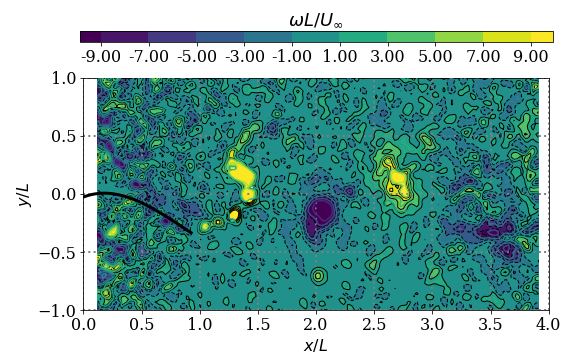}};

\node[right=0cm of inst1] (inst2) {\includegraphics[width=3cm,trim=4cm 1.8cm 7cm 2.8cm,clip]{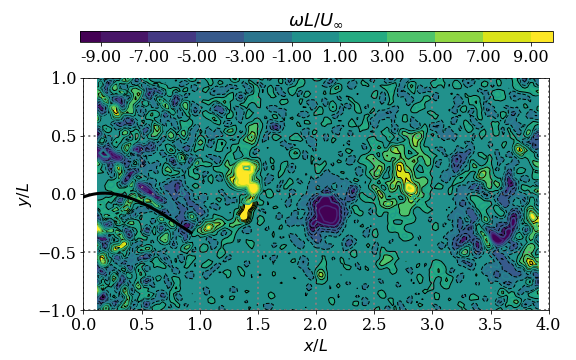}};
\node[right=0cm of inst2] (inst3) {\includegraphics[width=3cm,trim=4cm 1.8cm 7cm 2.8cm,clip]{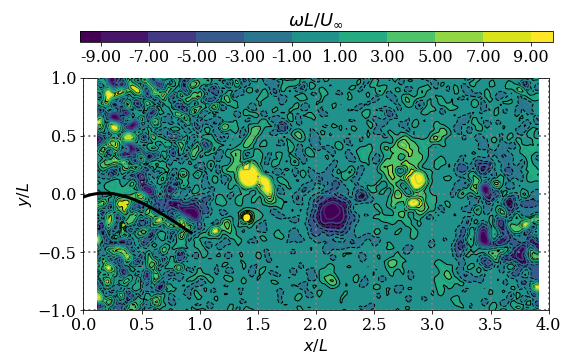}};
\node[below=0cm of inst2,yshift=-1cm] (inst4) {\includegraphics[width=3cm,trim=4cm 1.8cm 7cm 2.8cm,clip]{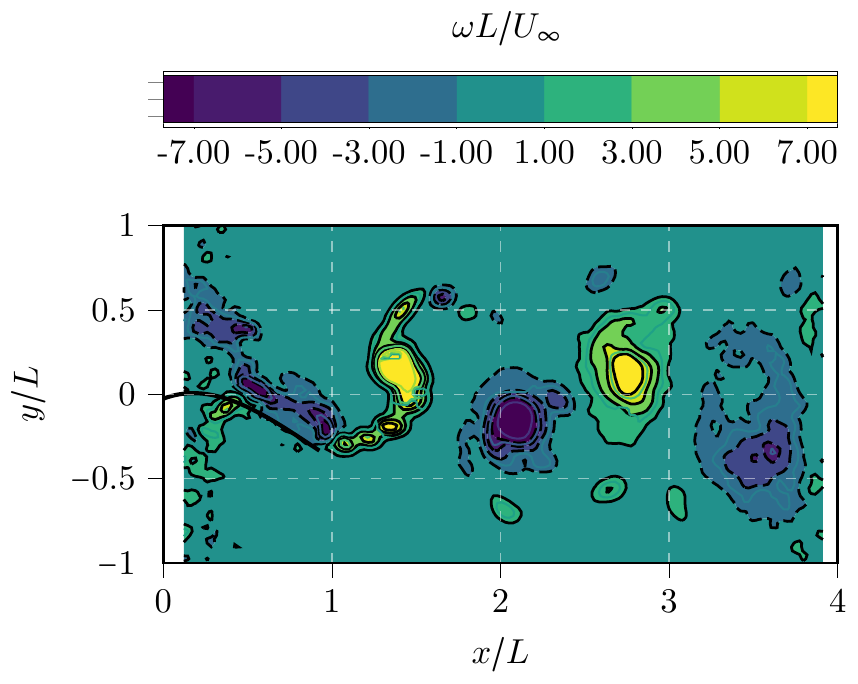}};
\node[below=0cm of inst1](text1){$\omega_{6}$};
\node[below=0cm of inst2](text2){$\omega_{22}$};
\node[below=0cm of inst3](text3){$\omega_{80}$};
\node[below=0cm of inst4](text3){$c_{1}$};

\draw[->,line width=3] ($(inst1)+(-1,-1)$) to [bend left=-45] (inst4);
\draw[->,line width=3] ($(inst3)+(1,-1)$) to [bend left=45] (inst4);
\draw[->,line width=3] ($(inst2)+(1,-1)$) to ($(inst4)+(1,)$);
	\node (species1)[left=of inst1,xshift=0.15cm,yshift=-1.65cm] [shape=rectangle,rounded corners=0.5cm,fill=gray,fill opacity=0.6,text opacity=0,thick] {
	
\begin{small}

\texttt{		\begin{tabular}{l|p{6cm}}
	\multicolumn{2}{l}{	\textbf{Algorithm Steps}}\\
	\hline\\
	\textbf{(0)}& Set $N_C$ centroids as random snapshots $c^{(0)}_k=\omega_j$. \\
	\textbf{(1)}& Compute the distance matrix $D_i^k$: $D_i^k=\mathrm{d}(\omega_i,c^{(0)}_k)$. \\
	\textbf{(2)}& Each $\omega_i$ is assigned to the $k$-centroid closest according to $D^k_i$.\\
	\textbf{(3)}& For an ensemble $\mathscr W$ of $N_s$ asigned snaphsots to a cluster, calculate the barycenter as $b_k =\sum_{i}\omega_i/N_s, i\in \mathscr{W}$.\\
	\textbf{(4)}& Centroids are displaced to new barycenters $c^{(1)}_k=b_k.$\\
	\textbf{(4)}&  If $|c^{(n)}_k - c^{(n-1)}_k|<tol$ no centroid is displaced, the method is stopped, otherwise it is iterated from \textbf{(1)}.\\	
	\end{tabular}
}
\end{small}
};

	\node (species1)[left=of inst1,yshift=-1.4cm] [shape=rectangle,draw,rounded corners=0.5cm,fill=color0!15,fill opacity=1,text opacity=1,thick] {

\begin{small}

\texttt{		\begin{tabular}{l|p{6cm}}
		\multicolumn{2}{l}{	\textbf{Algorithm Steps}}\\
\hline\\
	\textbf{(0)}& Set $N_C$ centroids as random snapshots $c^{(0)}_k=\omega_j$. \\
	\textbf{(1)}& Compute the distance matrix $D_i^k$: $D_i^k=\mathrm{d}(\omega_i,c^{(0)}_k)$. \\
\textbf{(2)}& Each $\omega_i$ is assigned to the $k$-centroid closest according to $D^k_i$.\\
\textbf{(3)}& For an ensemble $\mathscr W$ of $N_s$ asigned snaphsots to a cluster, calculate the barycenter as $b_k =\sum_{i}\omega_i/N_s, i\in \mathscr{W}$.\\
\textbf{(4)}& Centroids are displaced to new barycenters $c^{(1)}_k=b_k.$\\
\textbf{(4)}&  If $|c^{(n)}_k - c^{(n-1)}_k|<tol$ no centroid is displaced, the method is stopped, otherwise it is iterated from \textbf{(1)}.\\	
		\end{tabular}
}
\end{small}
	};

\end{tikzpicture}